\documentclass[%
 reprint,
 amsmath,amssymb,
 aps,
pra]{revtex4-2}

\usepackage{graphicx}
\usepackage{dcolumn}
\usepackage{bm}
\usepackage{upgreek} 
\usepackage{xcolor} 

\newcommand{\dx}{\mathrm{d}x}
\newcommand{\dy}{\mathrm{d}y}
\newcommand{\dt}{\mathrm{d}t}
\newcommand{\upd}{\mathrm{d}}

\begin{document}

\preprint{APS/123-QED}

\title{Membrane phononic integrated circuits}

\author{Timothy M.F. Hirsch}
\author{Nicolas P. Mauranyapin}
\author{Erick Romero}
\author{Glen I. Harris}
\author{Xiaoya Jin}
\author{Nishta Arora}
\author{Christiaan J. Bekker}
\author{Chao Meng}
\author{Warwick P. Bowen}
\email{Corresponding author: w.bowen@uq.edu.au}
\author{Christopher G. Baker}
\affiliation{School of Mathematics and Physics, The University of Queensland, QLD 4072, Australia}%

\date{\today}

\begin{abstract}
Phononic circuits constructed from high tensile stress membranes offer a range of desirable features such as high acoustic confinement, controllable nonlinearities, low mass, compact footprint, and ease of fabrication. This tutorial presents a systematic approach to modelling and designing phononic integrated circuits on this platform, beginning with acoustic confinement, wave propagation and dispersion, mechanical and actuation nonlinearities, as well as resonator dynamics. By adapting coupled mode theory from optoelectronics to suspended membranes, and validating this theory with several numerical techniques (finite element modelling, finite difference time domain simulations, and the transfer matrix method), we provide a comprehensive framework to engineer a broad variety of phononic circuit building blocks. As illustrative examples, we describe the implementation of several acoustic circuit elements including resonant and non-resonant variable-ratio power splitters, mode converters, mode (de)multiplexers, and in-line Fabry-P\'erot cavities based on evanescent tunnel barriers. These building blocks lay the foundation for phononic integrated circuits with applications in sensing, acoustic signal processing, and power-efficient and radiation-hard computing.

\end{abstract}

\maketitle

Over the last few decades, motivated by the compact wavelength and low dissipation of acoustic phonons, there has been a sustained interest in realising phononic integrated circuits to support the next generation of classical and quantum information processing technologies~\cite{roukesMechanicalComputionRedux2004, sklanSplashPopSizzle2015,fuPhononicIntegratedCircuitry2019,safavi-naeiniControllingPhononsPhotons2019,mayorGigahertzPhononicIntegrated2021,bienfaitPhononmediatedQuantumState2019,qiaoSplittingPhononsBuilding2023,yaoPerspectivesDevicesIntegrated2025}. Suspended membranes under high tensile stress are a promising candidate architecture to realise a general purpose integrated phononic circuit~\cite{chaElectricalTuningElastic2018,graczykowskiAcousticPhononPropagation2014,hatanakaPhononPropagationDynamics2015,romeroPropagationImagingMechanical2019,yaoPerspectivesDevicesIntegrated2025}. The strengths of the platform include potentially ultralow losses due to dissipation dilution, strain engineering and the high index contrast between the suspended material and the substrate~\cite{thompsonStrongDispersiveCoupling2008, sementilliNanomechanicalDissipationStrain2022,engelsenUltrahighqualityfactorMicroNanomechanical2024,tsaturyanUltracoherentNanomechanicalResonators2017,chakramDissipationUltrahighQuality2014,romeroPropagationImagingMechanical2019}; a readily attainable nonlinearity that can be used for mechanical logic~\cite{hatanakaBroadbandReconfigurableLogic2017,tadokoroHighlySensitiveImplementation2021,romeroAcousticallyDrivenSinglefrequency2024} or error correction~\cite{jinEngineeringErrorCorrecting2024}; and the availability of compact, low-loss evanescent couplers~\cite{mauranyapinTunnelingTransverseAcoustic2021}.

\begin{figure*}
    \centering
    \includegraphics[width=0.99\linewidth]{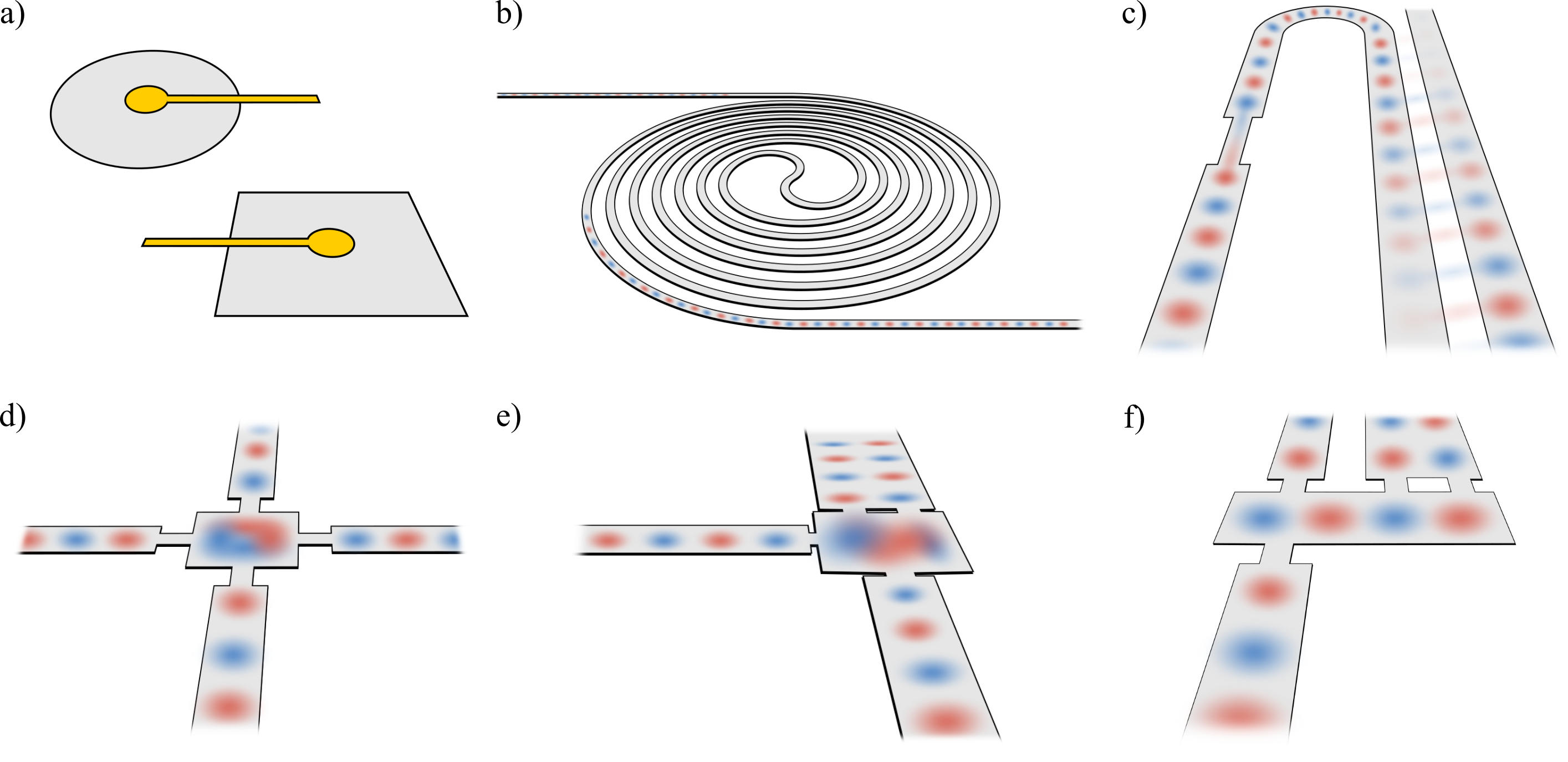}
    \caption[Phononic circuit components covered in this tutorial. ]{Phononic circuit components covered in this tutorial. (a) Resonators with electrostatic actuation and nonlinear dynamics. (b) Phononic waveguides with dispersive behaviour. (c) Parallel-waveguide evanescent coupling. (d) Resonance cavities and cross-talk free junctions. (e) Mode (demultiplexers). (f) Mode division power splitters.}
    \label{fig:intro-menu-figure}
\end{figure*}

Sections~\ref{Sec:Phononic-basics-equations-of-motion} to~\ref{sec:Phononic-basics-duffing} introduce the core concepts underpinning membrane phononic devices, including wave confinement, propagation, and nonlinear dynamics. Sections~\ref{sec:Coupling-evanescent-fields} to~\ref{sec:Coupling-examples-of-engineered-coupling} introduce more recent results centred around highly compact in-line evanescent couplers which, along with the accessible Duffing nonlinearity, are a key feature of the platform. We translate coupled mode theory (CMT)~\cite{hausCoupledmodeTheory1991} from the photonic to the phononic regime, and detail a range of numerical techniques (finite element modelling, finite difference time domain simulations, and the transfer matrix method), which allow us to precisely quantify the propagation and coupling of different spatial modes. Building on this capability, the tutorial concludes by numerically demonstrating several compact acoustic power splitting and mode (de)multiplexing devices (see Fig.~\ref{fig:intro-menu-figure}).

\section{The wave equation}
\label{Sec:Phononic-basics-equations-of-motion}
\begin{figure}[ht]
    \centering
    \includegraphics[width=0.99\linewidth]{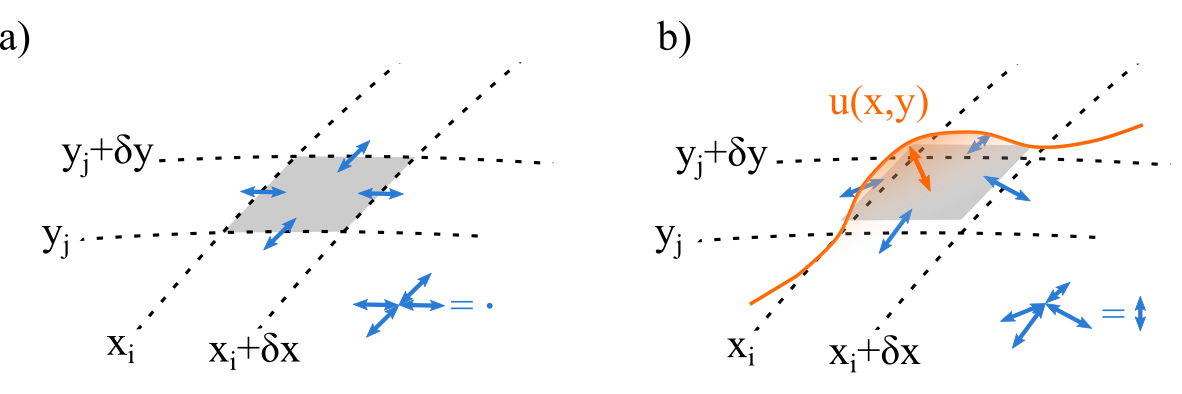}
    \caption[Continuum model of tension forces in a membrane]{Continuum model of tension forces in a membrane. (a) The $ij^\mathrm{th}$ membrane area segment (grey shading) experiences tension forces (blue arrows) between itself and its neighbours. At rest the tension forces are balanced. (b) When the area segment is suddenly moved as described by a displacement function $u(x,y)$ (orange), the tension forces become imbalanced.}
    \label{fig:Phononic-basics-membrane-tension-diagram}
\end{figure}

Membranes are the two-dimensional counterpart of strings. As illustrated in Fig.~\ref{fig:Phononic-basics-membrane-tension-diagram}, each infinitesimal area element of a membrane is connected to its neighbouring areas by a tension force. If an element is suddenly displaced from equilibrium, the imbalanced net tension forces act to return it to equilibrium. Over an infinite two-dimensional membrane, the aggregate behaviour of these tension forces---under approximations outlined shortly in Section~\ref{sec:Phononic-basics-rigidity-vs-tension}---is that the membrane follows the wave equation~\cite{schmidFundamentalsNanomechanicalResonators2016}:
\begin{equation}
\label{eqn:Phononic-basics-wave-eqn}
\frac{1}{c^2}\frac{\partial^2u}{\partial t^2}=\nabla^2u.
\end{equation}
Here $u(t,x,y)$ is the out-of-plane amplitude of vibration at time $t$ and position $(x,y)$ (see Fig.~\ref{fig:Phononic-basics-membrane-tension-diagram}). The wave speed is $c=\sqrt{\sigma/\rho}$, where $\rho$ is the volumetric density of the membrane material and $\sigma$ is the intrinsic tensile stress in the membrane. 

While membranes under tensile stress can be built from a variety of amorphous materials, chiefly silicon nitride~\cite{ghadimiRadiationInternalLoss2017,engelsenUltrahighqualityfactorMicroNanomechanical2024,zwicklHighQualityMechanical2008,verbridgeMegahertzNanomechanicalResonator2008} and silicon carbide~\cite{sementilliLowDissipationNanomechanicalDevices2025,xuHighStrengthAmorphousSilicon2024a}, and even strained crystalline materials such as graphene~\cite{garcia-sanchezImagingMechanicalVibrations2008,jungGHzNanomechanicalResonator2019} and silicon~\cite{beccariStrainedCrystallineNanomechanical2022}, this tutorial refers mainly to  highly stressed thin-film silicon nitride membranes like those used in our previous work~\cite{romeroPropagationImagingMechanical2019,mauranyapinTunnelingTransverseAcoustic2021,hirschDirectionalEmissionOnchip2024,romeroAcousticallyDrivenSinglefrequency2024}. Typical parameters for this material are $\rho=3200\,\mathrm{kg\cdot m^{-3}}$ and $\sigma\simeq1\,\mathrm{GPa}$, giving $c\simeq560\,\mathrm{m\cdot s^{-1}}$. Appendix~\ref{sec:Appendix-params+properties} lists further relevant physical parameters. 

\subsection{Rigidity vs tension justification}
\label{sec:Phononic-basics-rigidity-vs-tension}

\begin{figure}[ht]
    \includegraphics[width=0.99\linewidth]{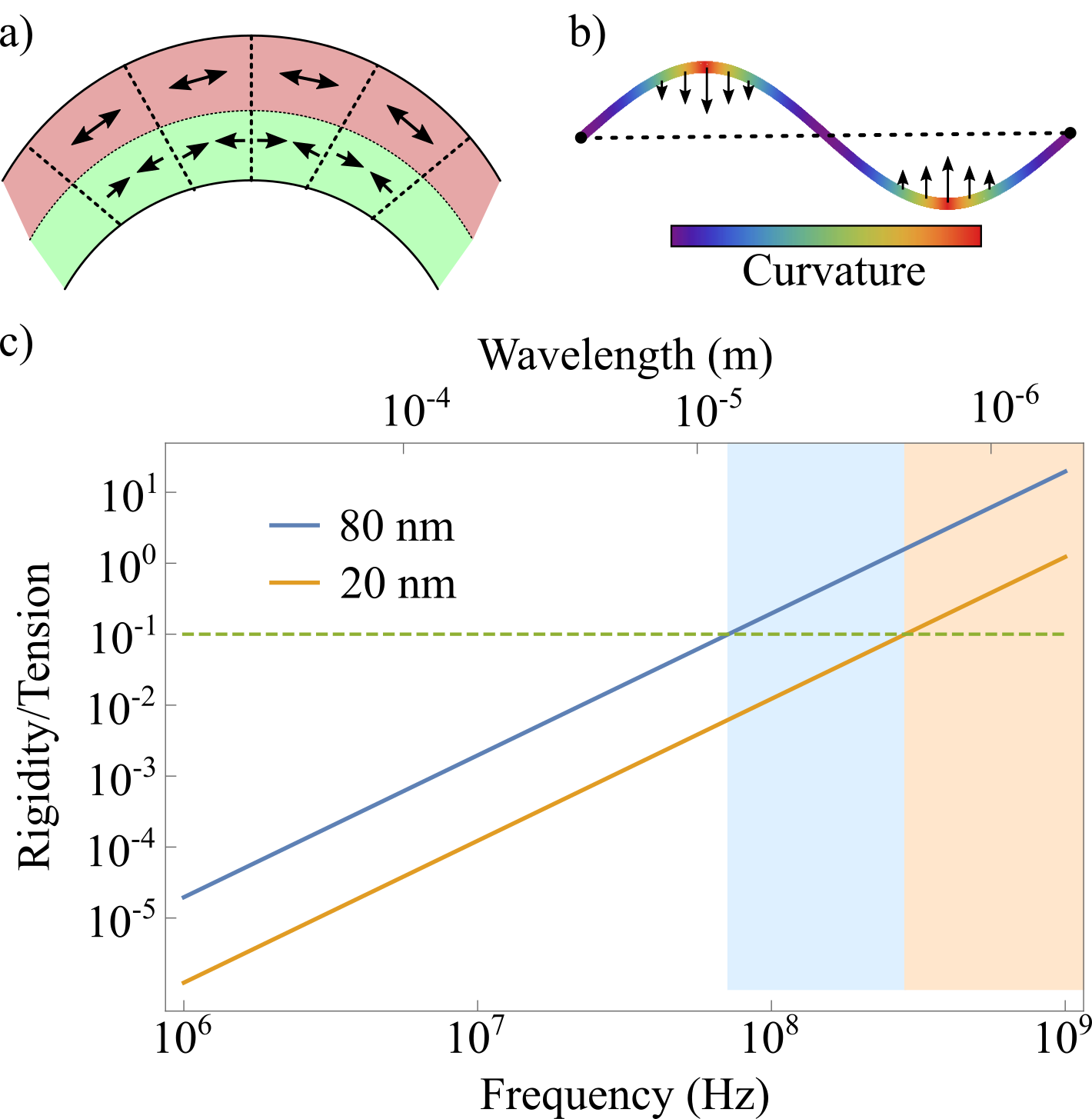}
    \caption[Rigidity, tension, and the ratio of plate-like to membrane-like restoring forces]{(a) In a beam, curvature causes zones of compression and tension (arrows) on the inward and outward sides of the curve respectively. The restoring force is proportional to the Young's modulus and acts to minimise the strain~\cite{ochsnerClassicalBeamTheories2021}. (b) In a string or membrane, areas of increased curvature (red shading) experience greater restoring force (arrows). In strings this force is proportional to the tension, and in membranes it is proportional to the tensile stress. (c) Ratio of plate-like to membrane-like restoring forces, from Eq.~\eqref{eqn:Phononic-basics-rigidity-vs-tension}, for silicon nitride membranes of thickness $100\,\mathrm{nm}$ (blue) and $30\,\mathrm{nm}$ (orange). Dashed line marks where tension is 10 times stronger. Blue and orange shadings show where the ratios cross the dashed line. The parameters used are here are: $Y=270\,\mathrm{GPa}$, $\sigma=1\,\mathrm{GPa}$, $\nu=0.27$ and $\rho=3200\,\mathrm{kg\cdot m^{-3}}$.}
    \label{fig:Phononic-basics-rigidity-vs-tension}
\end{figure}

Equation~\eqref{eqn:Phononic-basics-wave-eqn} is valid when the restoring force due to tensile stress is large compared with the force due to material rigidity. As illustrated in Fig.~\ref{fig:Phononic-basics-rigidity-vs-tension}(a-b), while both forces act to restore the membrane to its flat initial condition, their microscopic origins differ~\cite{schmidFundamentalsNanomechanicalResonators2016}. The equation of motion including both tensile stress and material rigidity is~\cite{schmidFundamentalsNanomechanicalResonators2016,ventselThinPlatesShells2001}:
\begin{equation}
    \label{eqn:Phononic-basics-rigidity+tension-eqn-of-motion}
    \rho\frac{\partial^2 u}{\partial t^2}=\sigma\nabla^2u + \frac{Yh^3}{12(1-\nu^2)}\nabla^4 u.
\end{equation}
Here the first term is the tension force that scales with the tensile stress. The second term represents flexural rigidity, and $h$ is the membrane thickness, $Y$ is the Young's modulus and $\nu$ the Poisson's ratio of the material.

Both terms in Eq.~\eqref{eqn:Phononic-basics-rigidity+tension-eqn-of-motion} act to `flatten' the membrane, minimising change in flexural displacement. However, because the rigidity scales with $\nabla^4$ and the tension with $\nabla^2$, at increasing wavenumbers (corresponding to increasing frequencies) the system will behave more like a plate than a membrane, as in the case of phononic waveguides fabricated from gallium arsenide~\cite{kurosuOnchipTemporalFocusing2018}. The ratio of plate-like to membrane-like restoring forces is given by:
\begin{equation}
    \label{eqn:Phononic-basics-rigidity-vs-tension}
    \frac{D\nabla^4u}{\sigma h \nabla^2u}=\frac{Yh^3k^2}{12(1-\nu^2)\sigma},
\end{equation}
where $k$ is the wavenumber. From Eq.~\eqref{eqn:Phononic-basics-rigidity-vs-tension} we can determine that in thin film silicon nitride, the transition from membrane-like to plate-like behaviour occurs at frequencies in excess of hundreds of megahertz. As shown in Fig.~\ref{fig:Phononic-basics-rigidity-vs-tension}(c), for an $80\,\mathrm{nm}$-thick membrane (as used in our previous work~\cite{romeroPropagationImagingMechanical2019,mauranyapinTunnelingTransverseAcoustic2021,romeroAcousticallyDrivenSinglefrequency2024}) with tensile stress of $1\,\mathrm{GPa}$, the tension is an order of magnitude stronger than the rigidity until around $70\,\mathrm{MHz}$ and becomes comparable around $200\,\mathrm{MHz}$. A thinner $20\,\mathrm{nm}$-thick membrane (as demonstrated in~\cite{norteMechanicalResonatorsQuantum2016}) the corresponding values are $300\,\mathrm{MHz}$ and $900\,\mathrm{MHz}$ (which corresponds to a wavelength of $\sim0.6\,\mathrm{\upmu m}$). This tutorial consider $~10-100\,\mathrm{\upmu m}$ sized silicon nitride membranes which have operating frequencies $<100\,\mathrm{MHz}$, so we will henceforth neglect the rigidity term in Eq.~\eqref{eqn:Phononic-basics-rigidity+tension-eqn-of-motion}.

\subsection{Boundary conditions}
\label{sec:Phononic-basics-boundary-conditions}

\begin{figure}[ht]
    \centering
    \includegraphics[width=0.99\linewidth]{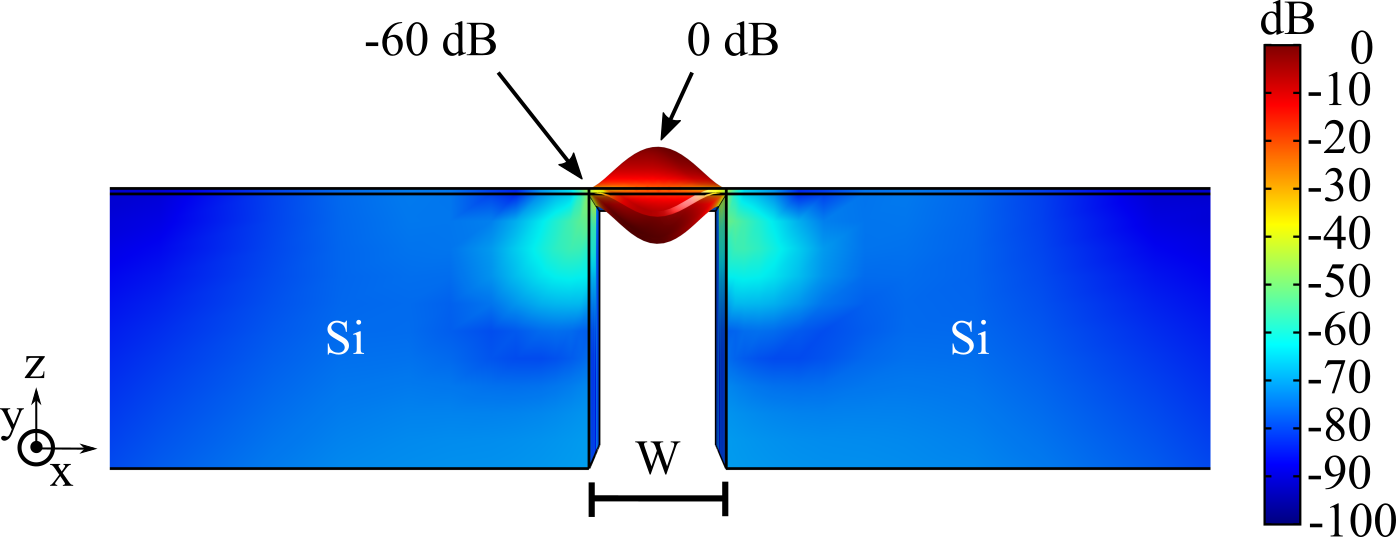}
    \caption[Finite element simulation of a suspended silicon nitride waveguide clamped to a silicon substrate, showing confinement of energy due to the impedance mismatch]{Finite element simulation of a suspended silicon nitride waveguide clamped to a silicon substrate, showing confinement of energy due to the impedance mismatch. The acoustic energy density (in arbitrary units) is plotted in a logarithmic scale. At the clamping points, simulations show the energy declines by a factor of $60\,\mathrm{dB}$ compared with the centre of the waveguide. In this simulation the nitride has stress $\sigma=1\,\mathrm{GPa}$ and is $3\,\mathrm{m}$ thick and $100\,\mathrm{\upmu m}$ wide. The substrate is $300\,\mathrm{\upmu m}$ thick.}
    \label{fig:Phononic-basics-Erick-clamping-FEM}
\end{figure}

When designing membrane phononic devices, one of the key physical facts to exploit is that there is typically very little motion at the boundary (where suspended material meets the substrate). This imposes Dirichlet boundary conditions on the solutions of the wave equation, with widespread implications for the physically permittable behaviour. For instance we will see that clamped edges lead to wave dispersion and the existence of resonant eigenmodes. 

Clamping losses---energy lost when the oscillating membrane tugs up and down on the substrate, generating travelling acoustic waves in the bulk---are reduced by the significant acoustic impedance mismatch between the membrane and substrate, equal to the ratio of wave speeds in the two media. Neglecting differences in thickness, the impedance $Z$ in a solid is $Z=\rho v_\mathrm{ph}$, where $\rho$ is the material density and $v_\mathrm{ph}$ is the phase velocity~\cite{auldAcousticFieldsWaves1990,safavi-naeiniControllingPhononsPhotons2019}. We saw from Eq.~\eqref{eqn:Phononic-basics-wave-eqn} that the speed of sound in a membrane is $c_\mathrm{mem}=\sqrt{\sigma/\rho_\mathrm{mem}}$, where $\rho_\mathrm{mem}$ is the membrane density. It can be shown that compressive/longitudinal waves in elastic, isotropic material travel at speed $c_\mathrm{sub}=\sqrt{Y/\rho_\mathrm{sub}}$, where $\rho_\mathrm{sub}$ is the volumetric density of the substrate~\cite{romerosanchezPhononicsEngineeringControl2019a}. The impedance mismatch is therefore:
\begin{equation}
    \label{eqn:Phononic-basics-impedance-mismatch-speed-ratio}
    \frac{Z_\mathrm{mem}}{Z_\mathrm{sub}}=\sqrt{\frac{\sigma}{Y}\frac{\rho_\mathrm{sub}}{\rho_\mathrm{mem}}}.
\end{equation}

Because it ignores thickness differences Eq.~\eqref{eqn:Phononic-basics-impedance-mismatch-speed-ratio} is a back-of-the-envelope estimate, but has illustrative value. For example, consider a silicon nitride membrane with stress of $\sigma=1\,\mathrm{GPa}$ and density $\rho_\mathrm{mem}=3200\,\mathrm{kg\cdot m^{-3}}$ on a silicon substrate with Young's modulus $Y=180\,\mathrm{GPa}$ and density $\rho_\mathrm{sub}=2650\,\mathrm{kg\cdot m^{-3}}$. According to Eq.~\eqref{eqn:Phononic-basics-impedance-mismatch-speed-ratio} the ratio of impedances is approximately $0.06$. If we also take into account the much larger thickness of the substrate compared to the membrane (typically on the order of $10^4$), the actual impedance mismatch is even larger~\cite{sementilliNanomechanicalDissipationStrain2022,photiadisAttachmentLossesHigh2004}. The fraction of energy in the membrane that is lost (per cycle of oscillation) into the substrate is proportional to the cube of the ratio of impedances~\cite{wilson-raeHighQNanomechanicsDestructive2011}.

Due to the high impedance mismatch we typically approximate the membrane as having zero amplitude of motion at the boundary where it meets the substrate. This is supported by numerical simulations, such as the one illustrated in Fig.~\ref{fig:Phononic-basics-Erick-clamping-FEM}. This shows a finite element simulation (COMSOL Multiphysics~\cite{COMSOLMultiphysics2020}) of the acoustic energy density in a realistic silicon nitride membrane of width $100\,\mathrm{\upmu m}$ and thickness $3\,\mathrm{\upmu m}$. The silicon substrate has depth $300\,\mathrm{\upmu m}$. The simulation shows strong energy confinement, with the amplitude of motion approximately $60\,\mathrm{dB}$ larger in the centre of the membrane than at the clamping points~\cite{romerosanchezPhononicsEngineeringControl2019a}.
The silicon nitride here is relatively thick compared to experimental samples because of computational limitations around using extreme aspect ratios. However, due to the high tensile stress, to first order the behaviour is still membrane-like instead of plate-like.
Corroborating this numerical simulation, experimental measurements of travelling waves in membrane waveguides record low losses of $0.4\,\mathrm{dB/cm}$~\cite{mauranyapinTunnelingTransverseAcoustic2021}, indicating motion at the membrane-substrate interface is minimal.


\section{Membrane device nomenclature}
\label{sec:Phononic-basics-nomenclature}

\begin{figure}[ht]
    \centering
    \includegraphics[width=0.99\linewidth]{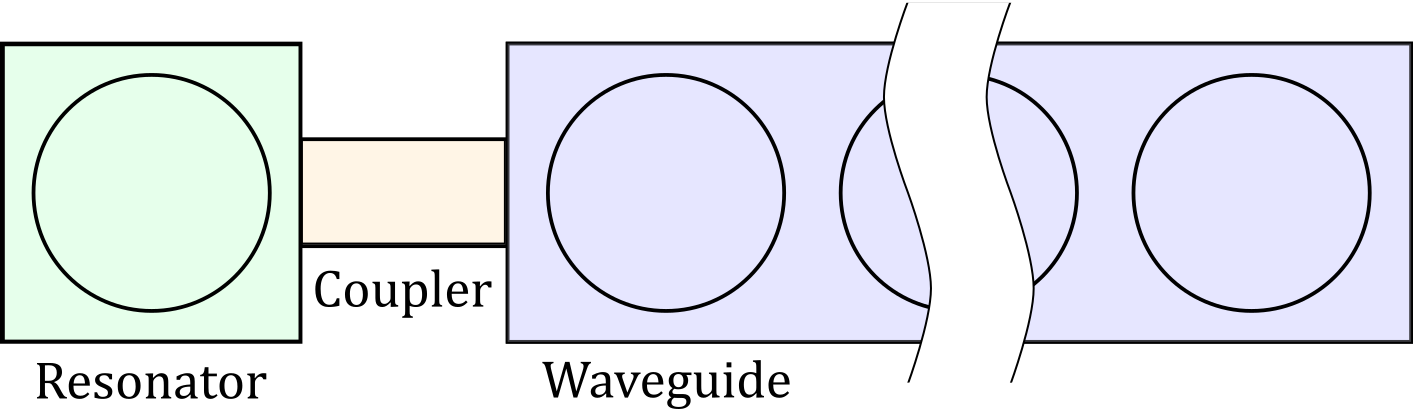}
    \caption[Pictorial examples of membrane resonators, couplers, and waveguides]{Pictorial example of a resonator (green), coupler (orange), and waveguide (blue). Circles illustrate the vibrational mode shapes, with the resonator and waveguide oscillating in their fundamental modes. The waveguide is comparatively much longer (indicated by the break).}
    \label{fig:Phononic-basics-waveguide-coupler-resonator-namescheme}
\end{figure}

As is the case in integrated photonics, in membrane phononic circuits there are some frequently-seen and visually recognisable circuit elements. Here we introduce three elements: waveguides, resonators, and evanescent couplers. Each is illustrated in Fig.~\ref{fig:Phononic-basics-waveguide-coupler-resonator-namescheme}. Waveguides confine and direct travelling acoustic waves and are studied in Section~\ref{sec:Phononic-basics-Dispersion}. Resonators store acoustic wave energy and will be explored in more detail in Section~\ref{sec:Phononic-basics-resonators}. Couplers connect circuit elements and will be detailed in Section~\ref{sec:Coupling-evanescent-coupling}.

Resonators, waveguides, and couplers can often be distinguished by their geometry---for example, waveguides tend to have larger aspect ratios than resonators, and couplers tend to have smaller lateral dimension than waveguides---but it is better to categorise phononic components by their function, as the fabrication and design scope for phononic circuits is very large.

\section{Waveguides and dispersion}
\label{sec:Phononic-basics-Dispersion}

One of the key tasks in a phononic circuit is transporting and directing the flow of acoustic energy. To do this we use phononic waveguides, which are the acoustic equivalent to wires in an electrical circuit or optical waveguides in a photonic circuit~\cite{wangPerspectivesPhononicWaveguides2024}. They are typically characterised by low transmission loss and large aspect ratios, and are used to route and store travelling waves---both classical and quantum~\cite{zivariNonclassicalMechanicalStates2022,eggleton_brillouin_2019,schmidt_convergence_2022}.

Some of the material platforms used for phononic waveguides include index-contrast slabs~\cite{fuPhononicIntegratedCircuitry2019,dengStronglyElectromechanicalCoupled2025,mayorGigahertzPhononicIntegrated2021,xuHighfrequencyTravelingwavePhononic2022,bicerGalliumNitridePhononic2022,fengGigahertzPhononicIntegrated2023}, phononic crystals~\cite{khelifGuidingBendingAcoustic2004,eichenfieldOptomechanicalCrystals2009,mohammadiChipComplexSignal2011,otsukaBroadbandEvolutionPhononiccrystalwaveguide2013,balramCoherentCouplingRadiofrequency2016,fangOpticalTransductionRouting2016}, and topological waveguides~\cite{zhangGigahertzTopologicalValley2022,hatanakaValleyPseudospinPolarized2024,xiSoftclampedTopologicalWaveguide2025}. In comparison with these platforms, suspended membrane waveguides have highly desirable properties such as low loss rates and a highly manipulable dispersion relationship~\cite{hatanakaPhononPropagationDynamics2015,chaElectricalTuningElastic2018,romeroPropagationImagingMechanical2019,wangHexagonalBoronNitride2019, kittlaus_-chip_2017}.

\begin{figure}[ht]
    \centering
    \includegraphics[width=0.99\linewidth]{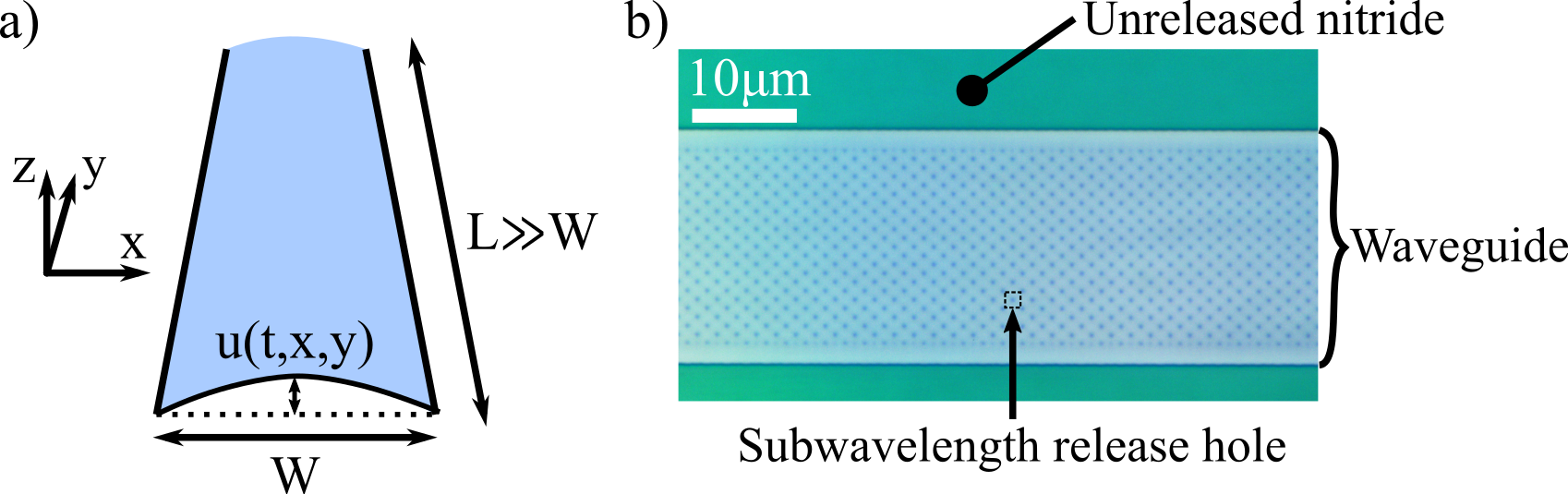}
    \caption[Illustration and optical micrograph of a suspended waveguide]{(a) Diagram of a waveguide of width $W$ and length $L\gg W$. The membrane (blue) has displacement $u(t,x,y)$ in the $z$-direction. (b) Optical microscope image of a silicon nitride membrane waveguide. Blue: released nitride waveguide. Green: unreleased nitride. The ``dots'' on the waveguide are submicron release holes used to access the sacrificial layer beneath the nitride~\cite{mauranyapinTunnelingTransverseAcoustic2021}.}
    \label{fig:Phononic-basics-waveguide-diagram+example}
\end{figure}

Consider the waveguide with width $W$ and length $L$ illustrated in Fig.~\ref{fig:Phononic-basics-waveguide-diagram+example}(a). In general the time-dependent membrane displacement is written as $u(t,x,y)$. Because waveguides have a large aspect ratio ($L\gg W$), to find $u(t,x,y)$ are motivated to look for a separable solution to Eq.~\eqref{eqn:Phononic-basics-wave-eqn} of the form:
\begin{equation}
    \label{eqn:Phononic-basics-waveguide-separable-solution}
    u(t,x,y)=Z_0\psi(x)e^{i(\Omega t-k_y y)}.
\end{equation}
This describes a wave of frequency $\Omega$ and wavenumber $k_y$ moving in the $+y$ direction, with a characteristic lateral modeshape $\psi(x)$. $Z_0$ is the maximum out-of-plane ($z$ axis) amplitude of oscillation and contains the dimensions of length, such that $\psi(x,y)$ is dimensionless and $\max(|\psi(x,y)|)=1$.

If we plug Eq.~\eqref{eqn:Phononic-basics-waveguide-separable-solution} into Eq.~\eqref{eqn:Phononic-basics-wave-eqn} and separate variables, we find:
\begin{equation}
    \label{eqn:Phononic-basics-dispersion-noclamp-deriv1}
    k_y^2-\frac{\Omega^2}{c^2}=\frac{\psi''(x)}{\psi(x)}.
\end{equation}
The left- and right-hands of the equation contain different variables and so must equal a constant. The only physically meaningful choice of constant is $-k_x^2$, as then the solution for $\psi(x)$ is simply:
\begin{equation}
    \label{eqn:Phononic-basics-dispersion-noclamp-deriv2}
\psi(x)=A\sin(k_x x)+B\cos(k_x x),    
\end{equation}
for constants $A$ and $B$. We will shortly solve for the constants and $k_x$ using the clamped boundary conditions of the waveguide.

Substituting Eq.~\eqref{eqn:Phononic-basics-dispersion-noclamp-deriv2} into Eq.~\eqref{eqn:Phononic-basics-dispersion-noclamp-deriv1} and rearranging, we arrive at the dispersion relationship:
\begin{equation}
    \Omega(k_x,k_y)=c\sqrt{k_x^2 + k_y^2}.
    \label{eqn:Phononic-basics-dispersion-noclamp}
\end{equation}

\begin{figure*}[ht]
    \centering
    \includegraphics[width=0.9\linewidth]{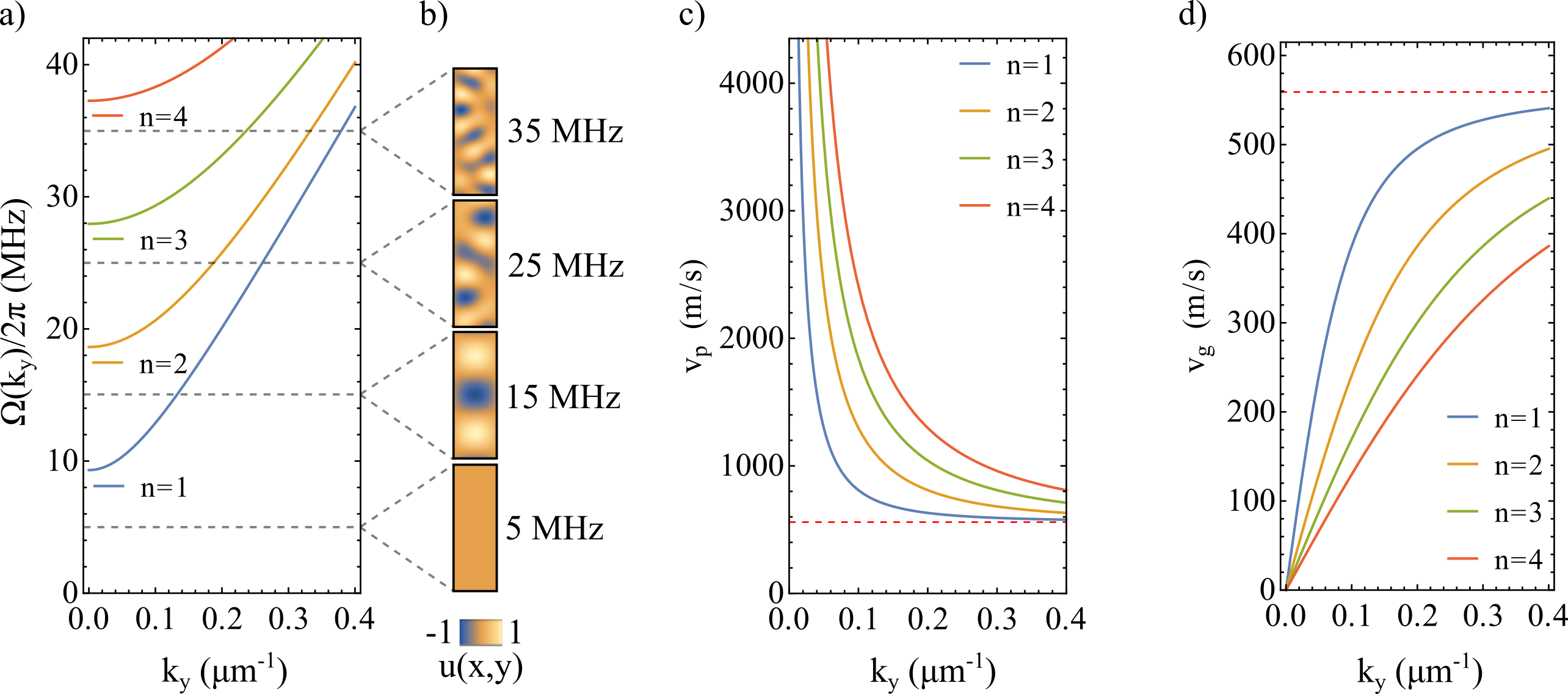}
    \caption[Dispersion relationship of suspended membrane waveguides]{(a) Dispersion relationship $\Omega(k)$ obtained from Eq.~\eqref{eqn:Phononic-basics-dispersion-clamp}, plotted for the first four modes. (b) Example displacements for different frequencies, found by summing over the modes that are above cutoff frequency with equal amplitude and random phase in each mode. (c) Phase velocity $v_p$ obtained from Eq.~\eqref{eqn:Phononic-basics-phase-velocity}. Dashed line indicates $c=\sqrt{\sigma/\rho}$. (d) Group velocity $v_g$ obtained from Eq.~\eqref{eqn:Phononic-basics-group-velocity}. Dashed line also indicates $c$. All values are calculated for a silicon nitride waveguide of width $W=30\,\mathrm{\upmu m}$, density $\rho=3200\,\mathrm{kg\cdot m^{-3}}$, and stress $\sigma=1\,\mathrm{GPa}$.}
    \label{fig:Phononic-basics-dispersion-relationship}
\end{figure*}

The clamped edges boundary condition from Section~\ref{sec:Phononic-basics-boundary-conditions} require that $\psi(0)=\psi(W)=0$. This immediately implies $B=0$, and because $\max(|\psi(x,y)|)=1$ we also conclude $A=1$. Finally, the condition $\psi(W)=0$ implies that $k_x=n\pi/W$ for integral $n$. The dispersion relationship then becomes:

\begin{equation}\label{eqn:Phononic-basics-dispersion-clamp}
    \Omega_n(k_y)=c\sqrt{k_y^2+\left(\frac{n\pi}{W}\right)^2}.
\end{equation}
The solutions to Eq.~\eqref{eqn:Phononic-basics-dispersion-clamp} are illustrated in Fig.~\ref{fig:Phononic-basics-dispersion-relationship}(a). By applying the clamped boundary conditions the space of solutions to a half-continuum: there is a continuum of longitudinal modes, but there are only discrete transverse modes. This is mathematically identical to the dispersion relationship of TEM modes in a microwave waveguide~\cite{montgomeryPrinciplesMicrowaveCircuits1987}.

At this point, because $k_x$ has been eliminated, we will drop the subscript on $k_y$.

From Eq.~\eqref{eqn:Phononic-basics-dispersion-clamp} we see that each transverse mode of the waveguide has a minimum possible frequency, found by setting $k\,(=k_y)=0$. This \emph{cutoff frequency}, $\Omega_{c,n}$ is a consequence of the restoring force introduced by the clamping on the sides of the waveguide, which causes even a wave with a zero wavenumber in the longitudinal direction to still oscillate. Because the transverse modeshape influences the restoring force, the cutoff frequency is mode-dependent. It has the value:
\begin{equation}
    \label{eqn:Phononic-basics-cutoff-frequency}
    \Omega_{c,n}=c\frac{n\pi}{W}.
\end{equation}
A useful feature of the cutoff frequency is that if the membrane is driven at a frequency strictly between $\Omega_{c,1}$ and $\Omega_{c,2}$ then only the fundamental mode of the waveguide will be excited~\cite{romeroPropagationImagingMechanical2019}. This is illustrated in Fig.~\ref{fig:Phononic-basics-dispersion-relationship}(b), and is analogous to the function of single mode optical fibers and microwave waveguides~\cite{hatanakaPhononPropagationDynamics2015,romerosanchezPhononicsEngineeringControl2019a,montgomeryPrinciplesMicrowaveCircuits1987}. Single mode operation eliminates modal dispersion and guarantees a consistent spatial distribution of wave energy~\cite{hausWavesFieldsOptoelectronics1984}.

The dispersion relationship~\eqref{eqn:Phononic-basics-dispersion-clamp} also describes the phase and group velocities of travelling waves. The phase velocity $v_p$ is:
\begin{equation}
    \label{eqn:Phononic-basics-phase-velocity}
    v_p\equiv\frac{\Omega(k)}{k}=c\sqrt{1+\left(\frac{n\pi}{Wk}\right)^2}=\frac{c}{\sqrt{1-\left(\frac{\Omega_{c,n}}{\Omega(k)}\right)^2}},
\end{equation}
and the group velocity $v_g$ is:
\begin{equation}
    \label{eqn:Phononic-basics-group-velocity}
    v_g\equiv\frac{d\Omega(k)}{dk}=\frac{c}{\sqrt{1+\left(\frac{n\pi}{Wk}\right)^2}}=c\sqrt{1-\left(\frac{\Omega_{c,n}}{\Omega(k)}\right)^2},
\end{equation}
with Equations~\eqref{eqn:Phononic-basics-phase-velocity} and~\eqref{eqn:Phononic-basics-group-velocity} leading to the identity $c=\sqrt{v_pv_g}$~\cite{ishimaruElectromagneticWavePropagation2017}.

Subfigures~\ref{fig:Phononic-basics-dispersion-relationship}(c-d) illustrate how the phase velocity and group velocity depend on the wave frequency. There are two regimes worth noting. As can be seen from Equations~\eqref{eqn:Phononic-basics-phase-velocity} and~\eqref{eqn:Phononic-basics-group-velocity}, as $k\rightarrow0$ and $\Omega\rightarrow\Omega_{c,n}$ from above, $v_p\rightarrow\infty$ and $v_g\rightarrow0$. In this situation the length of the waveguide is similar to the wavelength and it oscillates like a very large aspect ratio resonator. Since $v_g$ describes the speed at which energy is transferred in the waveguide~\cite{stevenwellingsonElectromagneticsII2023}, we have recovered the fact that no energy is transported (across long distances) when the waveguide is excited below the cutoff frequency. Conversely, as $k\rightarrow\infty$, the phase velocity and group velocity both converge to $c=\sqrt{\sigma/\rho}.$  This is because the clamped-edges boundary conditions are less relevant in the context of short wavelengths. The situation instead resembles the propagation of plane waves in a spatially infinite membrane, with the dispersionless plane wave dispersion relation: $\Omega(k)=ck$.

\section{Resonators}
\label{sec:Phononic-basics-resonators}

Within a phononic circuit resonators serve the crucial role of storing energy and enhancing the amplitude of mechanical motion. For example, this could be to increase the response to external force for the purpose of making a sensor~\cite{schmidFundamentalsNanomechanicalResonators2016}, or to introduce nonlinear behaviour for the purpose of mechanical computing~\cite{tadokoroHighlySensitiveImplementation2021,romeroAcousticallyDrivenSinglefrequency2024,hatanakaElectromechanicalMembraneResonator2012,dubcekInSensorPassiveSpeech2024}. This section explain the dynamics of membrane resonators, with an emphasis on the Duffing nonlinearity that often plays a significant and useful role.

\subsection{Resonator eigenmodes}

Whereas waveguides are bounded in one dimension, resonators have two dimensional boundary conditions. As we did in the case of waveguides, we can look for a separable solution for the resonator displacement that separates the time and spatial variables. Unlike the waveguide case, we cannot assume the solution will take the form of a travelling wave, so we will keep the $x$ and $y$ variables together:
\begin{equation}
    u(t,x,y)=Z_0e^{i\Omega t}\psi(x,y).
\end{equation}
As before, here $Z_0$ is the maximum amplitude and has dimensions of length. The modeshape $\psi(x,y)$ describes the spatial distribution of the out-of-plane displacement. Again here $\psi(x,y)$ is a dimensionless function normalised such that $\max_{x,y}|\psi(x,y)|=1$.

Plugging this solution into the wave Eq.~\eqref{eqn:Phononic-basics-wave-eqn}, we find:
\begin{equation}
    \frac{\Omega^2}{c^2}=-\frac{\nabla^2\psi(x,y)}{\psi(x,y)}.
\end{equation}
In a similar fashion to the waveguide case, we note that the left and right hand expressions feature different variables and hence must be constant.

\begin{figure}[ht]
    \centering
    \includegraphics[width=0.99\linewidth]{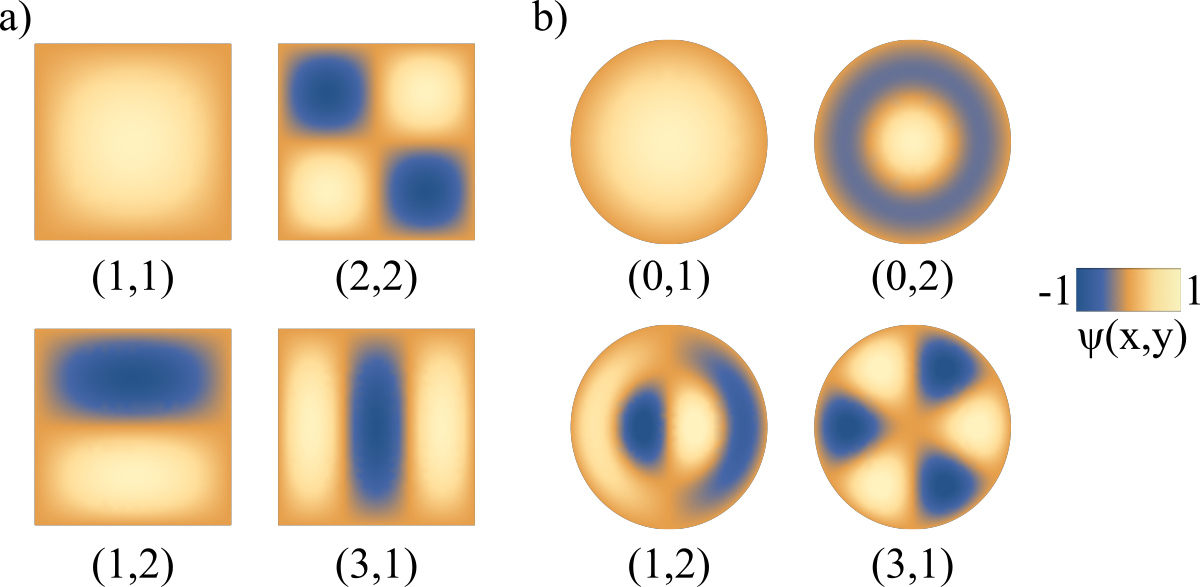}
    \caption[Example eigenmodes of square and circular membrane resonators]{(a) Modeshapes of a square membrane obtained from Eq.~\eqref{eqn:Phononic-basics-square-eigenmodes}. (b) Modeshapes of a circular membrane obtained from Eq.~\eqref{eqn:Phononic-basics-drum-eigenmodes}, classified by mode numbers $(m,n)$. The amplitudes are normalised to the range $[-1,1]$ in arbitrary units.}
    \label{fig:Phononic-basics-modeshapes-examples}
\end{figure}

There are some situations where we can solve for $\psi(x,y)$ analytically. For example, consider a rectangular membrane with side lengths $L_x$ and $L_y$. In that case we can try $\psi(x,y)=X(x)Y(y)$, separating variables again and introducing constants $k_x^2$ and $k_y^2$. Proceeding very similarly to the waveguides example and including the same the clamped boundary conditions:
\begin{equation}
    \label{eqn:Phononic-basics-square-eigenmodes}
    X(x)=\sin\left(\frac{n\pi}{L_x}x\right),\;\;Y(y)=\sin\left(\frac{m\pi}{L_y}y\right),
\end{equation}
for integers $n$ and $m$, which count the number of antinodes in the $X$ and $Y$ functions respectively. These solutions are illustrated in Fig.~\ref{fig:Phononic-basics-modeshapes-examples}(a).

As another example, for a circular membrane we can try separating into polar coordinates: $\psi(x,y)=R(r)\Theta(\theta)$. Separating variables and applying the boundary conditions will reveal that:
\begin{equation}
\label{eqn:Phononic-basics-drum-eigenmodes}
    u(r,\theta)=J_m(\lambda_{mn}r)\cos(m\theta+\varphi),
\end{equation}
for $m=0,1,2,\hdots$ and $n=1,2,3,\hdots$. Here $J_m$ is the Bessel function of the first kind of order $m$, and $\lambda_{mn}$ is the $n^\mathrm{th}$ root of $J_m$. The phaseshift $\varphi$ rotates the modeshape. These solutions are illustrated in Fig.~\ref{fig:Phononic-basics-modeshapes-examples}(b).

We generally may not be able to solve the modeshape analytically---and there generally may not be two integers that parametrise the modeshape as in Eqs.~\eqref{eqn:Phononic-basics-square-eigenmodes} and~\eqref{eqn:Phononic-basics-drum-eigenmodes}. However, we generally can say that, just as clamped boundary conditions in one dimension reduced the waveguide solution space to a half-continuum, clamped boundary conditions in two dimensions reduce the modeshape solution space to a discrete set. These solutions are called \emph{eigenmodes}.
 
A property of eigenmodes is that they form a basis that can be used to express any displacement:
\begin{equation}
    \label{eqn:Phononic-basics-eigenmode-expansion}
    u(t,x,y)= \sum_{m,n}Z_{0,mn}e^{i(\Omega_{mn}t+\varphi_{mn})}\psi_{mn}(x,y).
\end{equation}
Here $m$ and $n$ are the two mode numbers, $Z_{0,mn}$ is the amplitude of motion (with dimensions of length), $\Omega_{mn}$ is the \emph{eigenfrequency} of the $m,n^\mathrm{th}$ mode, $\varphi_{mn}$ is a phase shift used to match initial conditions, and $\psi_{mn}(x,y)$ is the normalised modeshape of the $m,n^\mathrm{th}$ mode. Each eigenmode harmonically oscillates at a particular eigenfrequency, just as each waveguide mode has a particular dispersion relationship.

Equation~\eqref{eqn:Phononic-basics-eigenmode-expansion} neglects dissipative and nonlinear processes, which introduce time dynamics beyond simple harmonic oscillation. It is possible to numerically solve two-dimensional continuum models with dissipation and nonlinearity, but this can be computationally expensive. However, a strength of Eq.~\eqref{eqn:Phononic-basics-eigenmode-expansion} is that it suggests we can model some regions of a two-dimensional material as systems of one-dimensional oscillators. This is the principle behind the lumped-element model, which we introduce in the next section.

\subsection{Lumped element model}
\label{sec:Phononic-basics-lumped-element-model}

When designing the resonators in a phononic circuit, it is often the case that the eigenmodes are known and the real interest is in nuanced time dynamics such as dissipation, nonlinearity, and the coupling between resonators. In these cases, instead of using the continuum model (i.e. the wave equation), it is often more insightful to note from Eq.~\eqref{eqn:Phononic-basics-eigenmode-expansion} that the motion can be described just by the amplitude of motion and knowledge of the eigenmode. This approach is called the lumped element model, which we explain briefly here. For more detail see~\cite{schmidFundamentalsNanomechanicalResonators2016} which we follow closely.

The lumped element model assumes that the membrane is oscillating in a particular eigenmode, so we can write the displacement as:
\begin{equation}
    \label{eqn:Phononic-basics-lumped-element-displacement-form}
    u(t,x,y)=z(t)\psi(x,y),
\end{equation}
where $z(t)$ describes the time dynamics and $\psi(x,y)$ describes the eigenmode shape. $z(t)$ has dimensions of length and $\psi(x,y)$ is dimensionless $\max_{x,y}|\psi(x,y)|=1$.  Comparing this with Eq.~\eqref{eqn:Phononic-basics-eigenmode-expansion}, we have restricted our attention to one eigenmode, merged the amplitude $Z_0$ with the time dynamics, and no longer assume the time dynamics correspond to harmonic oscillations.

In the simplest lumped element model $z(t)$ follows the equation of an undriven simple harmonic oscillator:
\begin{equation}
    \label{eqn:Phononic-basics-simple-harmonic-oscillator}
    m_\mathrm{eff}\frac{d^2z(t)}{dt^2}+k_\mathrm{eff}z(t)=0.
\end{equation}
Here $m_\mathrm{eff}$ and $k_\mathrm{eff}$ are effective values for the mass and spring constant, describing how the aggregated section of membrane behaves when vibrating at the eigenmode of interest. It is important to remember that $m_\mathrm{eff}$ and $k_\mathrm{eff}$ are only associated with that specific mode. Equation~\eqref{eqn:Phononic-basics-simple-harmonic-oscillator} takes solutions:
\begin{equation}
    z(t)=Z_0\cos(\Omega_0t+\varphi)
\end{equation}
where $Z_0$ is the maximum displacement, $\Omega_0=\sqrt{k_\mathrm{eff}/m_\mathrm{eff}}$ is the mode eigenfrequency and $\varphi$ is a phase shift used to match the initial conditions.

To use the lumped element model on a particular membrane eigenmode, we can first obtain the effective parameters $m_\mathrm{eff}$ and $k_\mathrm{eff}$ by comparing quantities such as kinetic energy or potential energy between the lumped element model and the wave equation~\cite{schmidFundamentalsNanomechanicalResonators2016}. As an example, consider a membrane of thickness $h$ and density $\rho$. The kinetic energy from the wave equation is:
\begin{equation}
    E_\mathrm{kin}^\mathrm{wave} = \frac{1}{2}\rho h \iint \left(\frac{\partial}{\partial t}z(t)\psi(x,y)\right)^2\,\dx\dy,
\end{equation}
where the double integrals are over the membrane area. This has maximum value:
\begin{align}
\max (E_\mathrm{kin}^\mathrm{wave})&=\frac{1}{2}\rho h\iint\max\left(\frac{\partial }{\partial t}z(t)\psi(x,y)\right)^2\,\dx\dy \nonumber\\
&=\frac{1}{2}\rho hZ_0^2\Omega_0^2\iint \psi(x,y)^2\,\dx\dy.
\end{align}
Meanwhile, in the lumped element model the maximum kinetic energy is:
\begin{equation}
\max(E_\mathrm{kin}^\mathrm{lumped})= \max\left(\frac{1}{2}m_\mathrm{eff}\left(\frac{\partial z(t)}{\partial t}\right)^2\right)=\frac{1}{2}Z_0^2\Omega_0^2m_\mathrm{eff}.
\end{equation}
Equating these two expressions we can solve for the effective mass:
\begin{equation}
m_\mathrm{eff}=\rho h \iint \psi(x,y)^2\,\dx\dy\label{eqn:Phononic-basics-m_eff}.
\end{equation}
$m_\mathrm{eff}$ can be interpreted as the total mass of the membrane that moves under the motion of the eigenmode. For example, if the eigenmode motion was  uniform and $\psi(x,y)=1$, then $m_\mathrm{eff}=\rho hA=m_\mathrm{tot}$, where $A$ is the area of the membrane and $m_\mathrm{tot}$ is the total mass of the membrane. Since $|\psi(x,y)|\leq1,$ $m_\mathrm{eff}\leq m_\mathrm{tot}$.

We can similarly obtain the effective spring constant, $k_\mathrm{eff}=\Omega_0^2m_\mathrm{eff}$, from the elastic potential energy:
\begin{widetext}
\begin{align}
    \label{eqn:Phononic-basics_k_eff-deriv-line1}
    \max(E_\mathrm{pot}^\mathrm{wave})&=\max\left(\frac{1}{2}\sigma h\iint \left[\left(\frac{\partial}{\partial x}z(t)\psi(x,y)\right)^2+\left(\frac{\partial}{\partial y}z(t)\psi(x,y)\right)^2\right]\,\dx\dy\right)\\
    &=
    \frac{1}{2}\sigma hZ_0^2\iint\left[\left(\frac{\partial \psi(x,y)}{\partial x}\right)^2+\left(\frac{\partial \psi(x,y)}{\partial y}\right)^2\right]\,\dx\dy, \nonumber\\
    \max(E_\mathrm{pot}^\mathrm{lumped})&= \frac{1}{2}k_\mathrm{eff}Z_0^2\nonumber\\
    \implies k_\mathrm{eff}&= \sigma h \iint\left[\left(\frac{\partial \psi(x,y)}{\partial x}\right)^2+\left(\frac{\partial \psi(x,y)}{\partial y}\right)^2\right]\,\dx\dy.
    \label{eqn:Phononic-basics-k_eff}
\end{align}
\end{widetext}
In this derivation, Eq.~\eqref{eqn:Phononic-basics_k_eff-deriv-line1} uses the two-dimensional version of the potential energy density in a string~\cite{schmidFundamentalsNanomechanicalResonators2016}. Equation~\eqref{eqn:Phononic-basics-k_eff} can be interpreted as the fact that the tensile restoring force acts to reduce deformation (strain) in the membrane. A mode with greater total strain therefore has a stronger effective spring constant.

By dividing Eq.~\eqref{eqn:Phononic-basics-k_eff} by Eq.~\eqref{eqn:Phononic-basics-m_eff} and simplifying we can obtain an equation relating the eigenfrequency $\Omega_0$ to the modeshape $\psi$:
\begin{equation}
    \Omega_0=\sqrt{\frac{\sigma}{\rho}}\sqrt{\frac{\iint\left[\left(\frac{\partial \psi(x,y)}{\partial x}\right)^2+\left(\frac{\partial \psi(x,y)}{\partial y}\right)^2\right]\,\dx\dy}{\iint \psi(x,y)^2\,\dx\dy}}.
\end{equation}
This equation constrains $\psi$ and provides some intuition for the relationship between modeshape and eigenfrequency. The dimensionless numerator scales with increasing strain and the denominator increases with mode area. This corresponds to the fact that eigenmodes confined to smaller areas with greater strain have higher eigenfrequencies.

\subsection{Damping}
Physically realisable phononic circuits will exhibit energy loss, both in terms of energy being lost from one circuit element to another and energy leaking out of the circuit to the environment. This loss can have an important effect on resonator dynamics.

To model energy loss we add a damping term and a forcing term to the lumped element model from Eq.~\eqref{eqn:Phononic-basics-simple-harmonic-oscillator}:
\begin{equation}
    \label{eqn:Phononic-basics-lumped-element-with-damping}
    m_\mathrm{eff}\frac{\upd^2z(t)}{\upd t^2}+\Gamma_\mathrm{eff}\frac{\upd z(t)}{\upd t}+k_\mathrm{eff}z(t)=F(t).
\end{equation}
Here, $\Gamma_\mathrm{eff}$ is the effective damping force coefficient, $F(t)$ is the drive force. It is worth reiterating that all of the effective quantities are defined with respect to a single eigenmode. 

$\Gamma_\mathrm{eff}$ can be found similar to how we found $k_\mathrm{eff}$ and $m_\mathrm{eff}$, by comparing the lumped element equation with damping, Eq.~\eqref{eqn:Phononic-basics-lumped-element-with-damping}, with the continuum wave equation with damping:
\begin{equation}
    \label{eqn:Phononic-basics-wave-equation-with-damping}
    \rho \frac{\partial^2u(t,x,y)}{\partial t^2}+\rho \gamma \frac{\partial u(t,x,y)}{\partial t}-\sigma \nabla^2u(t,x,y)=F(t,x,y).
\end{equation}
Here $\gamma$, with dimensions of inverse time, is called the \emph{damping rate}, and $F(t,x,y)$ is the external body force on the membrane. To find $\Gamma_\mathrm{eff}$ we first substitute the lumped element expression for the displacement, Eq.~\eqref{eqn:Phononic-basics-lumped-element-displacement-form} into Eq.~\eqref{eqn:Phononic-basics-wave-equation-with-damping}, then multiply by the eigenmodeshape $\psi(x,y)$ and integrate over all space. Comparing coefficients of $dz(t)/\dt$ then yields:
\begin{equation}
    \label{eqn:Phononic-basics-Gamma-vs-gamma}
    \Gamma_\mathrm{eff}=m_\mathrm{eff}\gamma.
\end{equation}
The general technique for finding effective parameters this way is called a Galerkin method~\cite{schmidFundamentalsNanomechanicalResonators2016}. To reduce the number of symbols we will henceforth exclusively use $\gamma$.

The presence of damping lowers the undriven oscillation frequency of the system to:
\begin{equation}
    \label{eqn:Phononic-basics-damped-resonance-frequency}
    \Omega' = \Omega_0\sqrt{1-\left(\frac{\gamma}{2\Omega_0}\right)^2},
\end{equation}
where $\Omega_0=\sqrt{k_\mathrm{eff}/m_\mathrm{eff}}$ is the undamped resonance frequency. For $\gamma/2>\Omega_0$ the system is overdamped and will quickly decay to zero amplitude without oscillating. For $\gamma/2=\Omega_0$ the system is critically damped---this corresponds to the quickest decay to zero amplitude for a damped oscillator. We usually consider the case where $\gamma/2\ll\Omega_0$, when the oscillator is called underdamped.

A slightly damped, undriven oscillator will oscillate at frequency $\Omega'$ with an exponentially decreasing envelope amplitude:
\begin{equation}
    \label{eqn:Phononic-basics-undriven-damped-decay}
    z(t)=z(0)\exp\left(-\frac{\gamma}{2}t\right)\exp(i\Omega't).
\end{equation}
Recognising that the energy in the oscillator is proportional to amplitude squared, we can see that $\gamma$ is the energy decay rate and $\gamma/2$ is the amplitude decay rate.

\subsection{Spectral response and quality factor}
Here we wish to understand how a damped and driven resonator responds to a single-frequency drive tone. That is, we want to understand the spectral response of the resonator. Consider again the damped oscillator from Eq.~\eqref{eqn:Phononic-basics-lumped-element-with-damping}, now with a sinusoidal drive at frequency $\Omega_d$:
\begin{equation}
    \label{eqn:Phononic-basics-lumped-element-damped-singlefreqdrive}
    m_\mathrm{eff}\frac{\upd^2z(t)}{\dt^2}+m_\mathrm{eff}\gamma\frac{\upd z(t)}{\dt}+k_\mathrm{eff}z(t)=F_0e^{i\Omega_d t}.
\end{equation}

The solution is an oscillation at the drive frequency, $z(t)=Z_0e^{i\Omega_d t}$, where the amplitude and phase offset with respect to the drive are bundled into the complex number $Z_0$. Substituting the solution into Eq.~\eqref{eqn:Phononic-basics-lumped-element-damped-singlefreqdrive} yields~\cite{schmidFundamentalsNanomechanicalResonators2016}:
\begin{equation}
    \label{eqn:Phononic-basics-spectral-magnitude}
    |Z_0|=\frac{F_0/k_\mathrm{eff}}{\sqrt{\left(1-\left(\frac{\Omega_d}{\Omega_0}\right)^2\right)^2+\left(\frac{\gamma}{\Omega_0}\right)^2\left(\frac{\Omega_d}{\Omega_0}\right)^2}},
\end{equation}
and
\begin{equation}
    \label{eqn:Phononic-basics-spectral-angle}
    \arg(Z_0)=\arctan\left(\frac{\frac{\gamma}{\Omega_0}\frac{\Omega_d}{\Omega_0}}{\left(\frac{\Omega_d}{\Omega_0}\right)^2-1}\right).
\end{equation}

\begin{figure*}
    \centering
    \includegraphics[width=0.99\linewidth]{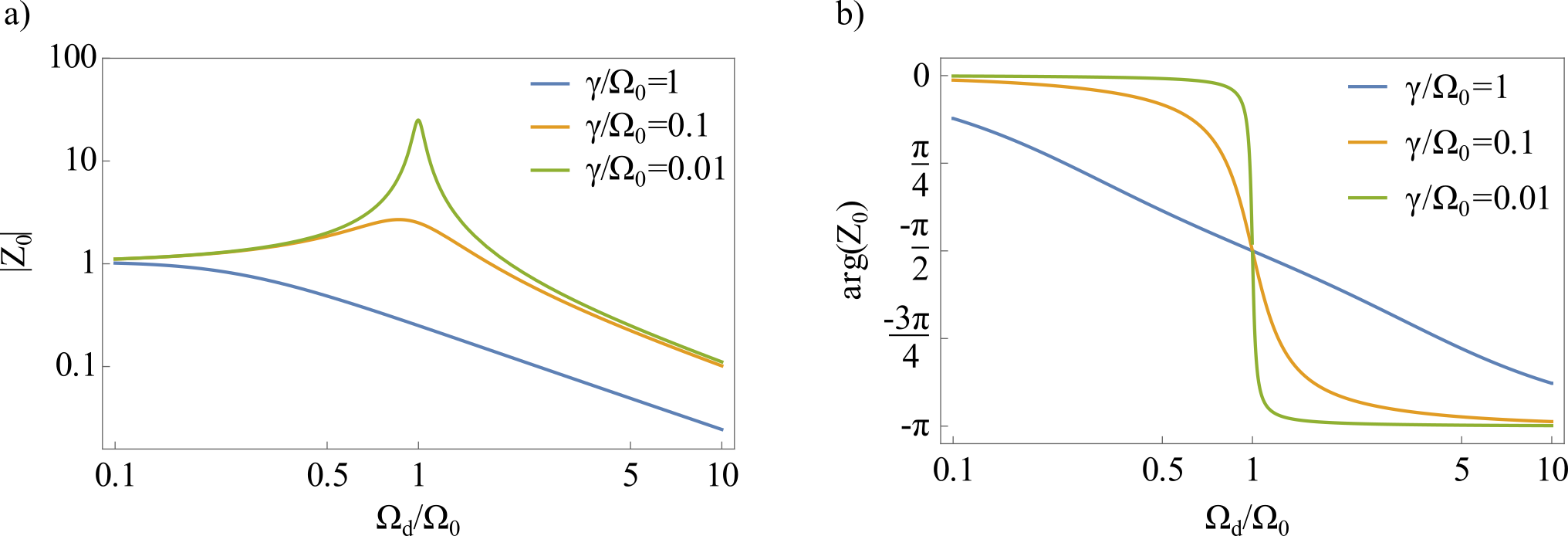}
    \caption[Spectral response from solving the linear, damped lumped element model]{(a) Magnitude of spectral response, for $F_0/k_\mathrm{eff}=1\,\mathrm{a.u.}$, as described in Eq.~\eqref{eqn:Phononic-basics-spectral-magnitude} (b) Angle of spectral response as described by Eq.~\eqref{eqn:Phononic-basics-spectral-angle}. The phase of the response shifts by $\pi$ through the resonance, which is a general behaviour observed in many physical systems that exhibit resonances.}
    \label{fig:Phononic-basics-spectral-response}
\end{figure*}

These two equations are illustrated in Fig.~\ref{fig:Phononic-basics-spectral-response} for different levels of damping. Equation~\eqref{eqn:Phononic-basics-spectral-magnitude} describes a Lorentzian peak at $\Omega_0$ with width defined by the damping.
Equation~\eqref{eqn:Phononic-basics-spectral-angle} describes the well-known $\pi$-phase shift that occurs as the drive frequency sweeps through the resonance, while $\arg(z)$ is negative for all frequencies such that that the response lags behind the drive~\cite{schmidFundamentalsNanomechanicalResonators2016}.

Figure~\ref{fig:Phononic-basics-spectral-response} also shows that the resonance is only visible in the underdamped limit where $\gamma/\Omega_0\ll1$. Henceforth in this paper we constrain the analysis to this limit.

In Eq.~\eqref{eqn:Phononic-basics-spectral-magnitude} and Eq.~\eqref{eqn:Phononic-basics-spectral-angle} we see the response is determined by two dimensionless ratios. The first ratio, $\Omega_d/\Omega_0$, describes how detuned the drive is from resonance. The second ratio, $\gamma/\Omega_0$, describes the width of the resonant peak. Specifically, $\gamma$ equals the full-width at half-maximum (FWHM) of the peak (measured in radians per second):
\begin{equation}
    \gamma=\mathrm{FWHM}.
\end{equation}

The ratio $\Omega_0/\gamma$ defines a number called the \emph{quality factor}, $Q$:
\begin{equation}
    \label{eqn:Phononic-basics-Q-linewidth-definition}
    Q=\frac{\Omega_0}{\gamma}=\frac{\Omega_0}{\mathrm{FWHM}}.
\end{equation}

The quality factor also equals the ratio of total energy stored in the oscillator, $E$, to the energy lost per oscillation, $\Delta E$:
\begin{equation}
    \label{eqn:Phononic-basics-Q-energy-definition}
    Q=2\pi \frac{E}{\Delta E}.
\end{equation}
Eqs.~\eqref{eqn:Phononic-basics-Q-linewidth-definition} and~\eqref{eqn:Phononic-basics-Q-energy-definition} are equivalent in the small damping limit, where we can use Eq.~\eqref{eqn:Phononic-basics-undriven-damped-decay} to calculate the ratio of total energy to energy lost per cycle:
\begin{align}
    \frac{E}{\Delta E}&=\frac{z(0)^2e^{-\gamma t}}{z(0)^2(e^{-\gamma t}-e^{-\gamma(t+2\pi/\Omega)})} \nonumber\\
    &= \frac{1}{1-e^{-2\pi\gamma/\Omega}} \nonumber\\
    &\simeq \frac{\Omega}{2\pi\gamma},
\end{align}
where in the last step we used the Taylor expansion under the justification that $\gamma/\Omega\ll1$. One need only then substitute the ratio of energies into Eq.~\eqref{eqn:Phononic-basics-Q-energy-definition} to obtain Eq.~\eqref{eqn:Phononic-basics-Q-linewidth-definition}.

The quality factor of a resonator can be determined experimentally using spectral or time domain measurements. Working in the spectral domain, measuring the linewidth of the resonance frequency yields the quality factor using Eq.~\eqref{eqn:Phononic-basics-Q-linewidth-definition}. Working in the time domain, measuring the rate of decay of energy yields the damping rate from Eq.~\eqref{eqn:Phononic-basics-undriven-damped-decay}; this is called a ringdown measurement~\cite{sementilliLowDissipationNanomechanicalDevices2025}. From the damping rate Eq.~\eqref{eqn:Phononic-basics-Q-linewidth-definition} yields the quality factor.

Different eigenmodes can have more or less damping and hence different quality factors. High quality factors are typically sought out in the field of nanomechanics as they correspond to reduced loss, improved sensitivity, and better frequency filtering ability~\cite{schmidFundamentalsNanomechanicalResonators2016,engelsenUltrahighqualityfactorMicroNanomechanical2024,sementilliNanomechanicalDissipationStrain2022,tsaturyanUltracoherentNanomechanicalResonators2017,aspelmeyerCavityOptomechanics2014}. However, in some situations it can be preferable to have a lower quality factor, such as engineering nanomechanical computing devices, where intrinsic losses remain low but the resonator is strongly coupled to its input and output waveguides to allow faster computation~\cite{tadokoroHighlySensitiveImplementation2021,romeroAcousticallyDrivenSinglefrequency2024}.

\begin{figure*}[ht]
    \centering
    \includegraphics[width=0.99\linewidth]{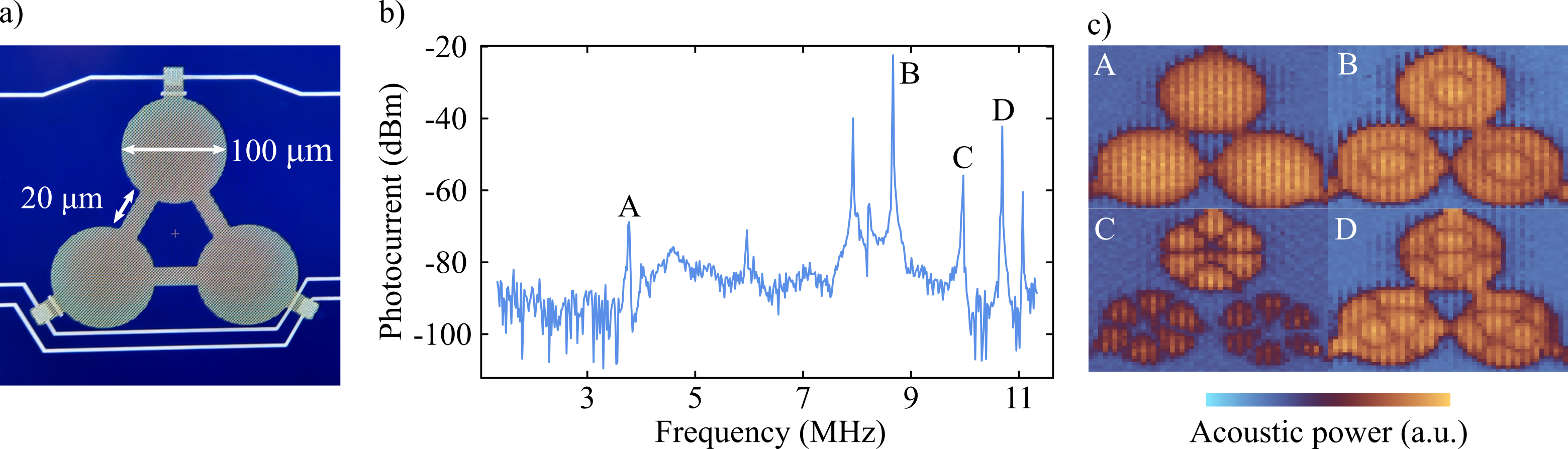}
    \caption[Experimental demonstration of how resonances correspond to distinct eigenmodes, using the example of a phononic error correcting device]{Experimental demonstration of how resonances correspond to distinct eigenmodes, using the example of a phononic error correcting device~\cite{jinNanomechanicalErrorCorrection2025}. (a) Optical microscope image of a phononic error-correcting device which consists of three coupled silicon nitride resonators. The resonators are $100\,\mathrm{\upmu m}$ diameter circular drums, coupled with evanescent tunnels. The device in this image has tunnels of length $20\,\mathrm{\upmu m}$. (b) Spectral response from optical Doppler interferometry, measured on the bottom left resonator when driving on the top resonator. These data were taken on a device with $11\,\mathrm{\upmu m}$ tunnels. (c) Experimentally obtained images of the modeshapes labelled in subfigure (b),   
    on the same device and again driving the top resonator. These images were taken by measuring the acoustic power spectral density around the drive frequency while driving at a particular resonance frequency. Higher-order drum modes can be identified by the number and position of their nodal lines (locations where the amplitude of vibration is zero). For instance, mode A corresponds to the $(m,n)=(0,1)$ circular drum mode, while mode D corresponds to the $(m,n)=(1,2)$ mode.}
    \label{fig:Phononic-basics-resonator-spectrum-example}
\end{figure*}

In this section we have considered just one resonance with effective parameters $k_\mathrm{eff}$, $m_\mathrm{eff}$, and $\gamma$. In practice a membrane will usually exhibit a range of resonances corresponding to the different eigenmodes. Fig.~\ref{fig:Phononic-basics-resonator-spectrum-example} shows an experimental example of this, where on a single device we can measure a large number of resonances, each corresponding to a particular spatial mode that can be experimentally driven and observed~\cite{jinNanomechanicalErrorCorrection2025}. In Fig.~\ref{fig:Phononic-basics-resonator-spectrum-example}(b) the resonance peak heights and widths vary because the different modes are more or less efficiently actuated by the electrode, and have larger or smaller damping rates. Differences in actuation efficiency can be observed for the mode labelled `C' in Fig.~\ref{fig:Phononic-basics-resonator-spectrum-example}(c). Section~\ref{sec:Coupling-coupled-mode-theory} goes into more detail about how the mode shape affects actuation efficiency and loss.



\section{Duffing nonlinearity}
\label{sec:Phononic-basics-duffing}
Nonlinearity enables many of the more useful and complex behaviours in phononic devices---for example it is essential for solitons and frequency combs~\cite{czaplewskiBifurcationGeneratedMechanical2018,shiVibrationalKerrSolitons2022,dejongMechanicalOvertoneFrequency2023,goryachevGenerationUltralowPower2020,mouharrarGenerationSolitonFrequency2024}, logic~\cite{tadokoroHighlySensitiveImplementation2021,hatanakaBroadbandReconfigurableLogic2017,yaoLogicmemoryDeviceMechanical2014,hafizMicroelectromechanicalReprogrammableLogic2016,romeroAcousticallyDrivenSinglefrequency2024}, and other phenomena such as four-wave mixing~\cite{ganesanFrequencyTransitionsPhononic2017,kurosuOnchipTemporalFocusing2018,kurosuMechanicalKerrNonlinearity2020,hatanakaBroadbandReconfigurableLogic2017} and bifurcations~\cite{mahboobHopfPerioddoublingBifurcations2016,czaplewskiBifurcationGeneratedMechanical2018}. In suspended membranes two key sources of nonlinearity are elongation stresses during large out-of-plane vibration, and electrostatic actuation nonlinearities. In both cases we can effectively model the nonlinearity by adding a quartic potential to the mechanical potential energy (we show this later). This model is called a Duffing nonlinearity, named after Georg Duffing (1861-1944)~\cite{georgduffingErzwungeneSchwingungenBei1918,kovacicDuffingEquationNonlinear2011}.

\begin{figure}[ht]
    \centering
    \includegraphics[width=0.99\linewidth]{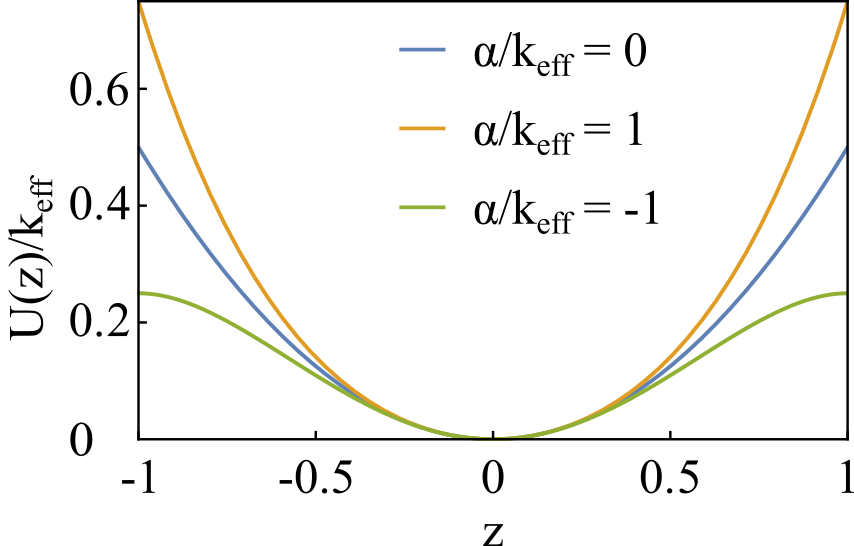}
    \caption{Potential energy $U(z)$ from Eq.~\eqref{eqn:Phononic-basics-duffing-model-potential}, normalised to the spring constant $k_\mathrm{eff}$.}
    \label{fig:Phononic-basics-potentials-quadratic+quartic}
\end{figure}

With Duffing nonlinearity the mechanical potential energy $U(z)$ reads:
\begin{equation}
    \label{eqn:Phononic-basics-duffing-model-potential}
    U(z)=\underbrace{\frac{1}{2}k_\mathrm{eff}z^2}_\mathrm{linear}+\underbrace{\frac{1}{4}\alpha_\mathrm{eff}z^4}_\mathrm{nonlinear},
\end{equation}
where $\alpha_\mathrm{eff}$ is the Duffing coefficient. Fig.~\ref{fig:Phononic-basics-potentials-quadratic+quartic} plots this model potential. The case $\alpha_\mathrm{eff}>0$ corresponds to a stiffening nonlinearity, which increases the wall steepness. The case $\alpha_\mathrm{eff}<0$ corresponds to a softening nonlinearity, which decreases the wall steepness.

We can rearrange Eq.~\eqref{eqn:Phononic-basics-duffing-model-potential} as follows:
\begin{equation}
    U(z)=\frac{1}{2}k_\mathrm{eff}\left(1+\frac{\alpha_\mathrm{eff}}{2k_\mathrm{eff}}z^2\right)z^2.
\end{equation}
This implies the potential energy resembles that of a linear oscillator with modified spring constant $k'=k_\mathrm{eff}+\alpha_\mathrm{eff}z^2/2$. For small nonlinearities where $\alpha_\mathrm{eff}z^2/2\ll k_\mathrm{eff}$, the system has a modified resonance frequency $\Omega'$ of:
\begin{align}
    \Omega' &= \sqrt{\frac{k'}{m_\mathrm{eff}}}=\sqrt{\frac{k_\mathrm{eff}}{m_\mathrm{eff}}}\sqrt{\frac{k'}{k_\mathrm{eff}}} \nonumber\\
    &= \Omega_0\sqrt{1+\frac{\alpha_\mathrm{eff}}{2k_\mathrm{eff}}z^2} \nonumber\\
    &\approx \Omega_0\left(1+\frac{\alpha_\mathrm{eff}z^2}{4k_\mathrm{eff}}\right).
    \label{eqn:Phononic-basics-duffing-frequency-shift}
\end{align}
With Duffing nonlinearity the resonance frequency depends on the amplitude of oscillation. For a stiffening nonlinearity ($\alpha_\mathrm{eff}>0$) the membrane will increase in resonance frequency at larger amplitudes. For a softening nonlinearity ($\alpha_\mathrm{eff}<0$) the membrane will decrease in resonance frequency at larger amplitudes.

The shifting resonance frequency of a Duffing oscillator has widespread implications. It limits the dynamic range of a frequency-shift based nanomechanical sensor~\cite{postmaDynamicRangeNanotube2005}, and it complicates the problem of coupling multiple nonlinear oscillators~\cite{jinCascadingNanomechanicalResonator2023}. It also introduces new functionalities, such as amplitude bistability which we will look at shortly in Section~\ref{sec:Phononic-basics-duffing-bistability}.

To incorporate Duffing nonlinearity into the lumped element model, we note from Eq.~\eqref{eqn:Phononic-basics-duffing-model-potential} that the restoring force $F_\mathrm{res}$ is altered from Hooke's law to $F_\mathrm{res}=-\upd U/\upd z=-k_\mathrm{eff}z-\alpha_\mathrm{eff}z^3$. Substituting the additional term into the lumped element dynamical equation (Eq.~\eqref{eqn:Phononic-basics-lumped-element-with-damping}) with a sinusoidal drive, we obtain the Duffing equation:
\begin{equation}
    \label{eqn:Phononic-basics-Duffing-equation}
    m_\mathrm{eff}\frac{\upd^2z}{\upd t^2}+\Gamma_\mathrm{eff}\frac{\upd z}{\upd t}+k_\mathrm{eff}z+\alpha_\mathrm{eff}z^3=F_0\cos(\Omega_dt).
\end{equation}

Analysing the Duffing oscillator is a complex but standard textbook procedure~\cite{nayfehNonlinearOscillations2004,lifschitzReviewsNonlinearDynamics2008,schmidFundamentalsNanomechanicalResonators2016}. Like most nonlinear equations there is no closed form analytic solution; however we can figure out the key behaviour with approximate methods. Here we will use a harmonic balance analysis~\cite{brennanJumpupJumpdownFrequencies2008a}. 

\begin{figure}[ht]
    \centering
    \vspace{5mm}
    \includegraphics[width=\columnwidth]{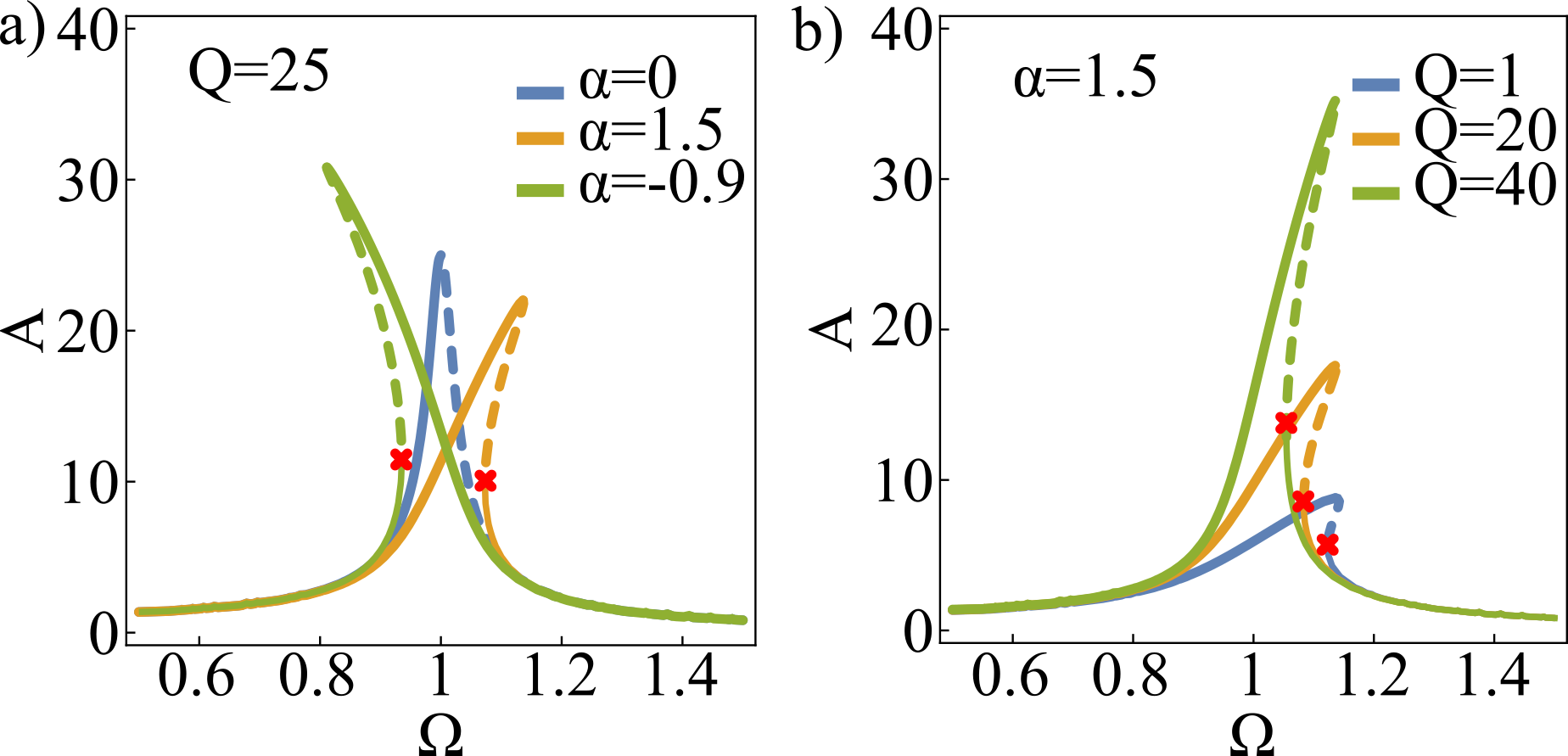}
    \caption[Frequency response of a Duffing oscillator]{Duffing frequency response from Eq.~\eqref{eqn:Phononic-basics-duffing-dimless-response}, (a) with varying Duffing coefficients $\alpha_\mathrm{eff}$, and (b) with varying quality factors $Q$. Solid lines: physical solutions. Dashed lines: unstable solutions. Red crosses: responses at the jump-up transition frequencies (see Subsection~\eqref{sec:Phononic-basics-duffing-bistability}).}
    \label{fig:Phononic-basics-duffing-freq-response}
\end{figure}

We first non-dimensionalise Eq.~\eqref{eqn:Phononic-basics-Duffing-equation} with the substitutions: $\tilde{F}_0=F_0/k_\mathrm{eff}$, $\tilde{t}=\Omega_0t$ (where $\Omega_0=\sqrt{k_\mathrm{eff}/m_\mathrm{eff}}$), $\tilde{z}=z(t)/\tilde{F}_0$, $\kappa=\alpha_\mathrm{eff} Z_0^2/k_\mathrm{eff}$, and $\tilde{\Omega}=\Omega_d/\Omega_0$. These substitutions transform the equation into:
\begin{equation}
    \frac{\mathrm{d}^2\tilde{z}}{\mathrm{d}\tilde{t}^2}+\frac{1}{Q}\frac{\mathrm{d}\tilde{z}}{\mathrm{d}\tilde{t}}+\tilde{z}+\kappa \tilde{z}^3=\cos(\tilde{\Omega}\tilde{t})
\end{equation}
Assuming that $Q=\Omega_0/\gamma\gg1$ and that $\kappa \tilde{z}^2\ll1$ such that the nonlinear restoring force is much weaker than the linear restoring force (we will justify these assumptions shortly), we can solve this equation by looking for a solution of the form $\tilde{z}(\tilde{t})=A\cos(\tilde{\Omega} \tilde{t}+\varphi)$, where $A$ is a real number describing the amplitude, $\varphi$ is a phaseshift, and we neglect fast rotating terms proportional to $\cos(3\Omega\tau)$. This yields an equation relating $A$, the drive frequency $\tilde{\Omega}$, and the drive force (through the normalisation of $\tilde{z}(\tilde{t})$)~\cite{brennanJumpupJumpdownFrequencies2008a}:
\begin{equation}
    \label{eqn:Phononic-basics-duffing-dimless-response}
    \left((1-\tilde{\Omega}^2)A+\frac{3}{4}\kappa A^3\right)^2+\frac{\tilde{\Omega}^2A^2}{Q^2}=1.
\end{equation}
Equation~\eqref{eqn:Phononic-basics-duffing-dimless-response} quantifies the response of the resonator to the applied driving force. Figure~\ref{fig:Phononic-basics-duffing-freq-response}(a) illustrates this response. As predicted in Eq.~\eqref{eqn:Phononic-basics-duffing-frequency-shift}, we see the resonance frequency shifts depending on the amplitude of vibration, with the direction of the shift corresponding to the sign of $\alpha_\mathrm{eff}$. The peak response increases for softening nonlinearities ($\alpha_\mathrm{eff}<0$), because to first order a softening nonlinearity reduces the effective spring constant of the oscillator (even as the force $F_0$ remains constant).

Another important effect is that at large magnitudes of $\alpha_\mathrm{eff}$, $A(\Omega_d)$ becomes multivalued. Subsection~\ref{sec:Phononic-basics-duffing-bistability} below discusses this in more detail.

\subsection{Justification of assumptions}
The harmonic balance analysis assumed that the quality factor is large and that the Duffing nonlinearity is relatively weaker than the linear terms. The quality factor assumption can be justified knowing that for materials such as silicon nitride (which have low intrinsic dissipation, low clamping losses, and a high stress such that dissipation dilution further increases the quality factor~\cite{sementilliNanomechanicalDissipationStrain2022,engelsenUltrahighqualityfactorMicroNanomechanical2024}) $Q\gg1$ is easily achievable for most resonator geometries. Even at room temperature and atmospheric pressure, small resonators of comparable size to the atmospheric mean free path length can achieve significant quality factors up to $Q\simeq10^2-10^3$~\cite{dantanMembraneSandwichSqueeze2020,liUltrasensitiveNEMSbasedCantilevers2007,leTemperatureRelativeHumidity2021}.

To justify that the nonlinearity is weak compared to the linear behaviour, consider the nonlinear version of the wave equation (derived in Section~\ref{sec:Phononic-basics-mechanical-nonlinearity}):
\begin{widetext}
\begin{equation}
    \label{eqn:Phononic-basics-nonlinear-wave-equatioin}
        \rho \frac{\partial^2u}{\partial t^2}=\left(\frac{\partial^2u}{\partial x^2}+\frac{\partial^2u}{\partial y^2}\right)\left(\underbrace{\sigma}_\mathrm{linear}+\underbrace{\frac{Y}{2}\left(\left(\frac{\partial u}{\partial x}\right)^2+\left(\frac{\partial u}{\partial y}\right)^2\right)}_\mathrm{nonlinear}\right).
\end{equation}
\end{widetext}
Here $u(t,x,y)$ is the displacement as a function of time and spatial coordinates, and $\sigma$, $\rho$, and $Y$ are respectively the tensile stress, density, and Young's modulus of the membrane.

The ratio of nonlinear to linear forces is $Y/2\sigma$ multiplied by the squared spatial derivatives. To quantitatively examine this ratio, we use the example of a nonlinear membrane resonator that was used as a logic gate in Ref.~\cite{romeroAcousticallyDrivenSinglefrequency2024}. The resonator was a square membrane operating in its fundamental mode. Neglecting motion at the boundary (see Section~\ref{sec:Phononic-basics-boundary-conditions}) the displacement can be written as:
\begin{equation}
    \label{eqn:Phononic-basics-justification-of-assumptions-logic-gate-displacement}
    u(x,y)=Z_0\sin\left(\frac{\pi x}{W}\right)\sin\left(\frac{\pi y}{W}\right).
\end{equation}
In this expression we have ignored the time dependence, which is not directly relevant to the geometric nonlinearity. The resonator had side length $W=80\,\mathrm{\upmu m}$ and maximum amplitude of $Z_0=19\,\mathrm{nm}$. The measured stress was $\sigma\simeq0.67\,\mathrm{GPa}$, and the Young's modulus can be taken to be around $250\,\mathrm{GPa}$~\cite{boeOnchipTestingLaboratory2009,khanYoungsModulusSilicon2004}.

We can use the displacement in Eq.~\eqref{eqn:Phononic-basics-justification-of-assumptions-logic-gate-displacement} to estimate the maximum value of the ratio of nonlinear to linear restoring forces in Eq.~\eqref{eqn:Phononic-basics-nonlinear-wave-equatioin}:
\begin{align}
    \frac{\text{nonlinear forces}}{\text{linear forces}}&\leq\frac{Y}{2\sigma}\left[\max\left(\left(\frac{\partial u}{\partial x}\right)^2+\left(\frac{\partial u}{\partial y}\right)^2\right)\right] \nonumber\\
    &\leq\frac{Y}{2\sigma}\times2\left(\frac{\pi Z_0}{W}\right)^2\nonumber\\
    &\simeq2\times10^{-4}.
\end{align}
Hence the nonlinear forces in the wave equation are much smaller than the linear forces, justifying the harmonic balance treatment used earlier.

It is worth noting that although the nonlinear force is smaller compared to the linear force, the nonlinear behaviour can still dominate the dynamics, so long as the nonlinear frequency shift is large compared to the damping rate $\gamma$ (i.e., for a sufficiently high $Q$). For example in this case, the mechanical logic gate had a quality factor of $Q=28,000$ and utilised the nonlinearity to create an amplitude bistability (explained below).


\subsection{Bistability}
\label{sec:Phononic-basics-duffing-bistability}
We saw in Fig.~\ref{fig:Phononic-basics-duffing-freq-response} that at large enough values of $\alpha_\mathrm{eff}$ the original Lorentzian response curve tilts over and becomes multivalued, accepting three distinct amplitude solutions. The middle amplitude is not experimentally realisable because it corresponds to a saddle point of the dynamical system~\cite{nayfehNonlinearOscillations2004}. However, the other two amplitudes are stable solutions. When driven at a frequency at which bistability occurs, the resonator can respond at either of those two different amplitudes of motion. Which amplitude it responds at depends on the system history; in other words it exhibits hysteresis~\cite{nayfehNonlinearOscillations2004,schmidFundamentalsNanomechanicalResonators2016,brennanJumpupJumpdownFrequencies2008a}.

To understand how this works in practice, imagine holding the drive amplitude constant and let $\alpha_\mathrm{eff}>0$, so we follow the orange curve in Fig.~\ref{fig:Phononic-basics-duffing-freq-response}. As $\Omega_d$ is swept from $0$ to $+\infty$, the response amplitude will slowly increase until it achieves its peak value, then sudden `jump down' to the lower branch. If we perform the opposite sweep from $+\infty$ to $0$, the drive will smoothly move along the lower branch until the point marked by the red cross, then `jump up' to the higher branch. On the green curve where $\alpha_\mathrm{eff}<0$, the same jumps will occur but during the opposite sweeps.

\begin{figure}[ht]
    \centering
    \includegraphics[width=0.99\linewidth]{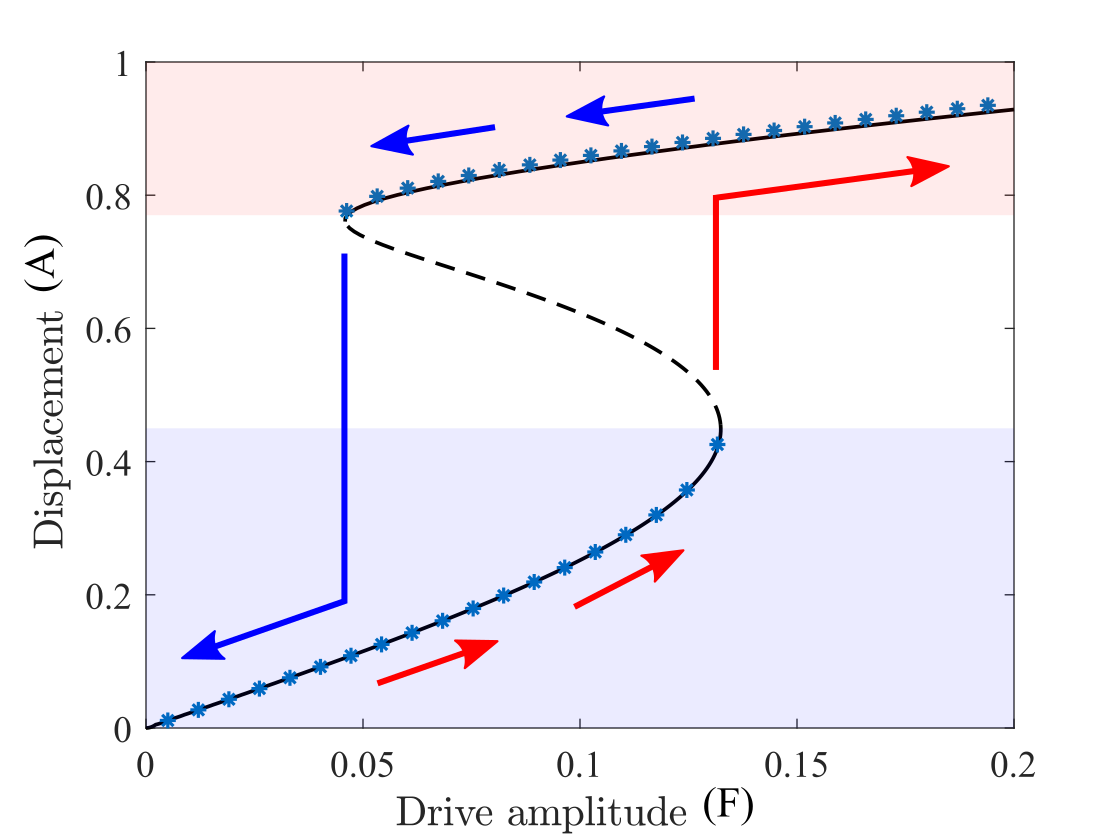}
    \caption[Bistability of a Duffing oscillator]{Bistability of a Duffing oscillator. Black line: analytic solution from solving Eq.~\eqref{eqn:Phononic-basics-duffing-dimless-response}, showing stable solutions (solid) and also unstable solutions (dashed). Markers: solutions obtained by numerically solving Eq.~\eqref{eqn:Phononic-basics-Duffing-equation}. Red arrows: increasing drive amplitude with `jump up' transition. Blue arrows: decreasing amplitude sweep with `jump down' transition. The drive force $F$ and displacement $A$ are in arbitrary units.}
    \label{fig:Phononic-basics-bistability}
\end{figure}

We can also observe bistability with a fixed drive frequency while varying the drive amplitude. The amplitude response from Eq.~\eqref{eqn:Phononic-basics-duffing-dimless-response} with a fixed frequency is illustrated in Fig.~\ref{fig:Phononic-basics-bistability}. The response demonstrates an `S' shape, with a finite range of input amplitudes that can yield
either a high or low response.

Just as in the frequency-sweep situation, the amplitude response switches between the high and low responses at the edges of the bistable region. For example, beginning at the very left of Fig.~\ref{fig:Phononic-basics-bistability}, as the drive smoothly increases so does the response, until at a drive of approximately 0.13 a.u. the response jumps up to the higher solution. Conversely, an oscillator initially sitting on the high response curve will smoothly follow that curve until the drive amplitude drops below approximately 0.05 a.u., at which point its oscillation amplitude will suddenly drop to the lower solution.

Bistability provides a natural way to represent a logical bit, so it has been used to create mechanical logic and memory devices~\cite{yaoLogicmemoryDeviceMechanical2014,guerraNoiseAssistedReprogrammableNanomechanical2010,hafizEfficientActivationNanomechanical2019, tadokoroHighlySensitiveImplementation2021,romeroAcousticallyDrivenSinglefrequency2024}. Bistability has also been leveraged to make sensors, typically by positioning the system at the edge of a jump-up or jump-down transition such that small changes in load can create significant responses~\cite{khaterBinaryMEMSGas2014,benjaminDesignImplementationBistable2018}.


\subsection{Critical amplitude}
Having seen in Fig.~\ref{fig:Phononic-basics-duffing-freq-response} and Fig.~\ref{fig:Phononic-basics-bistability} that nonlinear behaviour gradually appears as the vibration amplitude increases, it is useful to define a `critical' amplitude $Z_{0,\mathrm{crit}}$ that marks where nonlinearity becomes significant. Different definitions of `critical' amplitude are used in the literature: for example, the minimum amplitude of oscillation at which bistability appears~\cite{febboCriticalForcingAmplitude2013,brennanJumpupJumpdownFrequencies2008a}, or the amplitude at which the linear behaviour breaks down and nonlinear frequency shift can no longer be ignored~\cite{postmaDynamicRangeNanotube2005,juillardAnalysisResonantSensors2018,nayfehNonlinearOscillations2004}.

Here we define critical amplitude in the latter sense, as where the amplitude-dependent frequency shift described in Eq.~\eqref{eqn:Phononic-basics-duffing-frequency-shift} is of the same magnitude as the resonator linewidth $\gamma=\Omega_0/Q$, that is:
\begin{equation}
    \frac{\Omega_0}{Q}=\frac{\Omega_0\alpha_\mathrm{eff}Z_{0,\mathrm{crit}}^2}{4k_\mathrm{eff}}.
\end{equation}
This rearranges for the critical amplitude as:
\begin{equation}
\label{eqn:Phononic-basics-critical-amplitude}
    Z_{0,\mathrm{crit}}=\sqrt{\frac{4k_\mathrm{eff}}{\alpha_\mathrm{eff}Q}}.
\end{equation}

When $\sqrt{\alpha_\mathrm{eff}/k_\mathrm{eff}}$, the ratio of nonlinear to linear spring constants, increases, the critical amplitude decreases. It also decreases when $Q$ is larger, because narrower resonance peaks require smaller frequency shifts to reach bistability (see Fig.~\ref{fig:Phononic-basics-duffing-freq-response}(b)).

\subsection{Geometric Duffing nonlinearity}
\label{sec:Phononic-basics-mechanical-nonlinearity}

As stated at the beginning of Section~\ref{sec:Phononic-basics-duffing}, there are two key sources of Duffing nonlinearity for suspended membrane devices: elongation stress in the material during vibration, and electrostatic forces. In this section we explain the origin of elongation stresses and derive an expression for the corresponding Duffing coefficient.

At small scales gravity is a far weaker force than the internal stresses of a membrane. For example, consider a silicon nitride membrane with density $\rho=3200\,\mathrm{kg\cdot m^{-3}}$ and tensile stress $\sigma=1\,\mathrm{GPa}$. In one dimension, when hanging under its own weight, it assumes a catenary (hyperbolic cosine) curve with minimum radius of curvature $\sigma/(g\rho)\approx32\,\mathrm{km}$~\cite{ganderCatenaryCurve2004}. This is vastly greater than the membrane's horizontal scale, so we can safely say the membrane is flat when at rest~\cite{gardeniersLPCVDSiliconrichSilicon1996,beliaevOpticalStructuralComposition2022}. Therefore, because the edges of the membrane are fixed, any curvature of the membrane (such as from vibration) must require some elongating strain in the material. This is the root cause of what is called the geometric Duffing nonlinearity~\cite{schmidFundamentalsNanomechanicalResonators2016}.

\begin{figure}[ht]
    \centering
    \includegraphics[width=0.99\linewidth]{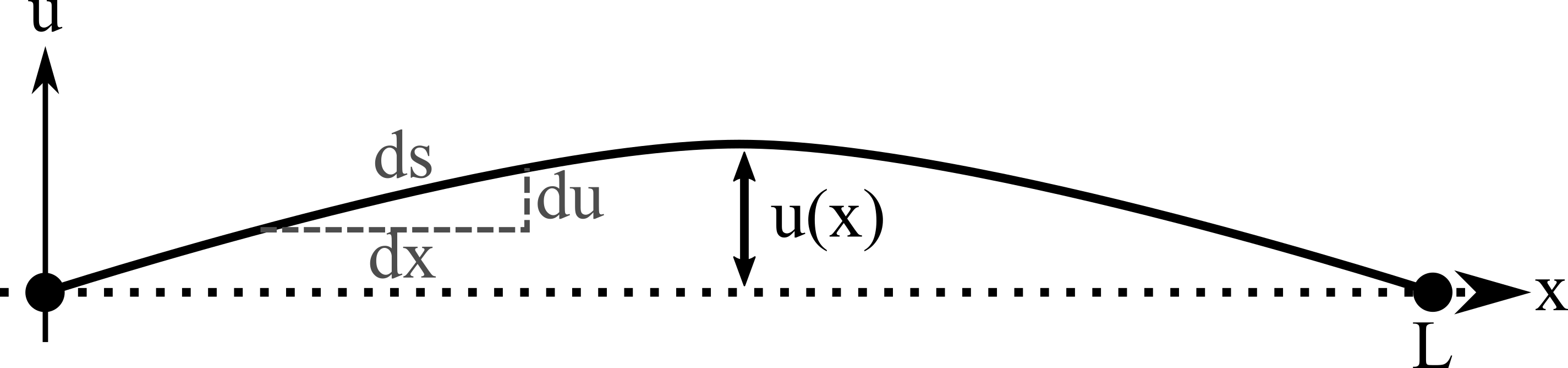}
    \caption[Appearance of nonlinear elongation in a one-dimensional string of length $L$]{Appearance of nonlinear elongation in a one-dimensional string of length $L$. The length element $\mathrm{d}s$ is greater than the unstretched distance $\mathrm{d}x$ because of the out-of-plane deflection $du$.}
    \label{fig:Phononic-basics-stretching-duffing-1d}
\end{figure}

To see how this works in one dimension, consider a string of length $L$ with amplitude $u(x)$ and fixed endpoints $u(0)=u(L)=0$, as illustrated in Fig.~\ref{fig:Phononic-basics-stretching-duffing-1d}. In the limit of small deflection (i.e. $\max(u)/L\ll1$), The string is horizontal at rest, and when curved experiences elongation, causing longitudinal strain $\varepsilon$, with:
\begin{align}
    \varepsilon(x)&=\frac{\mathrm{d}s-\mathrm{d}x}{\mathrm{d}x}\nonumber\\
    &=\frac{\sqrt{\dx^2+\mathrm{d}u^2}}{\dx}-1\nonumber\\
    &=\sqrt{1+\left(\frac{\mathrm{d}u}{\dx}\right)^2}-1\nonumber\\
    &\approx\frac{1}{2}\left(\frac{\upd u}{\dx}\right)^2.
    \label{eqn:Phononic-basics-strain-gradient-1D}
\end{align}
This strain produces a longitudinal stress $\sigma'$ in additional to the intrinsic stress. Because typical transverse deflections are nanometre scale and device dimensions are micron scale, we can safely assume a linear stress-strain relationship $\sigma'=Y\varepsilon(x)$, where $Y$ is the Young's modulus. In this case the energy per unit length $\delta E_\mathrm{strain}(x)$ associated with this additional stress is $\delta E_\mathrm{strain}(x)=\frac{1}{2}A\sigma'\varepsilon(x)$, where $A$ is the cross sectional area of the string.  Utilising Eq.~\eqref{eqn:Phononic-basics-strain-gradient-1D}, the total strain energy is:
\begin{align}
    E_\mathrm{strain}&=\frac{1}{2}AY\int_0^L\varepsilon(x)^2\,\dx \nonumber\\
    &=\frac{1}{8}AY\int_0^L\left(\frac{\mathrm{d}u}{\dx}\right)^4\,\dx.
\end{align}
Writing $u(x)=Z_0\psi(x)$ as the product of the maximum displacement and a modeshape $\psi(x)$ where $\max(|\psi(x)|)=1$ shows us the strain energy is proportional to the fourth power of $Z_0$:
\begin{align}
    E_\text{strain}&=\frac{1}{4}\alpha_\mathrm{eff} Z_0^4 \nonumber\\
    \text{where}\;\alpha_\mathrm{eff}&=\frac{1}{2}AY\int_0^L\left(\frac{\mathrm{d}\psi(x)}{\dx}\right)^4\,\dx.
    \label{eqn:Phononic-basics-duffing-coeff-derivation}
\end{align}
We saw earlier that this quartic potential corresponds to a Duffing nonlinearity. Equation~\eqref{eqn:Phononic-basics-duffing-coeff-derivation} provides an analytic estimate for $\alpha_\mathrm{eff}$ that can be substituted into the lumped element equation (Eq.~\eqref{eqn:Phononic-basics-Duffing-equation}). It can be extended to two dimensional membranes as:
\begin{equation}
    \label{eqn:Phononic-basics-duffing-coeff-derivation-2D}
    \alpha_\mathrm{eff}=\frac{1}{2}Yh\iint\left[\left(\frac{\upd \psi}{\dx}\right)^4+\left(\frac{\upd\psi}{\dy}\right)^4\right]\,\dx\dy,
\end{equation}
where $h$ is the thickness of the membrane.


Equation~\eqref{eqn:Phononic-basics-duffing-coeff-derivation-2D} allow us to design resonators with maximal or minimal nonlinear behaviour, as has been of interest in the literature previously~\cite{douStructuralOptimizationNonlinear2015,liTailoringNonlinearResponse2017}. Increasing the nonlinearity can be useful, for example to reduce the input energy required to reach the critical amplitude and access nonlinear behaviours~\cite{romeroAcousticallyDrivenSinglefrequency2024}. Decreasing the nonlinearity can also be useful, for example to allow greater drive amplitudes while remaining in the linear regime, as is optimal for frequency shift-based mass sensing~\cite{ekinciUltimateLimitsInertial2004}. 

\begin{figure}
    \centering
    \includegraphics[width=0.99\linewidth]{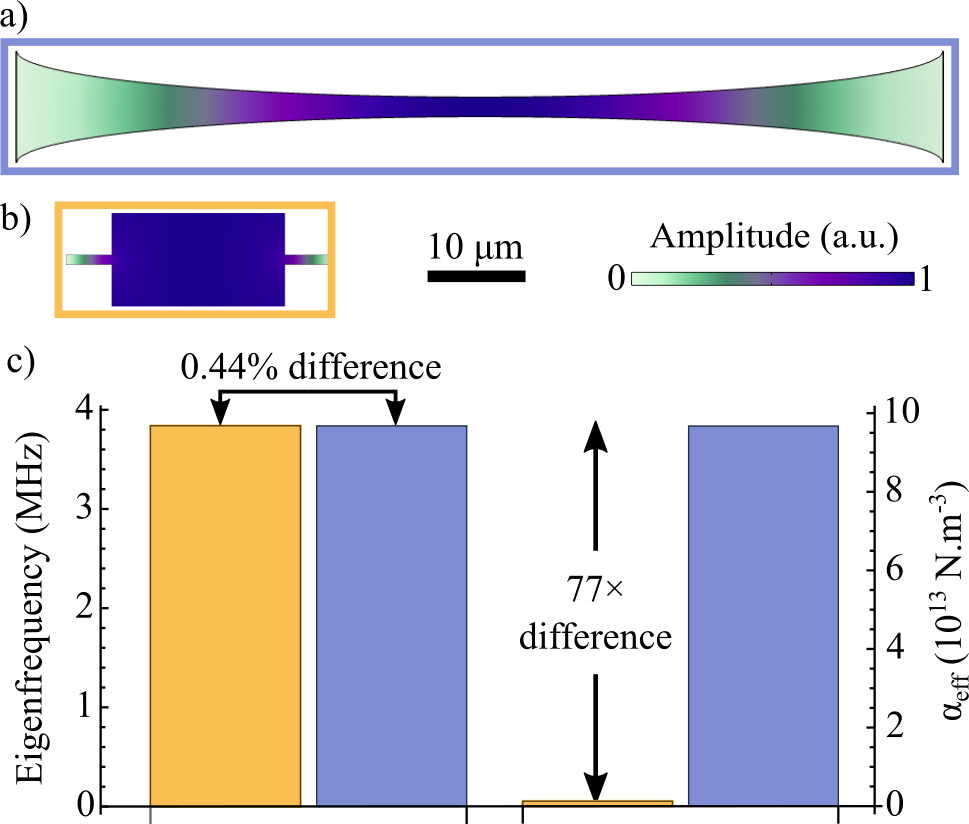}
    \caption{(a) Fundamental modeshape of a concave 2D beam. (b) Fundamental modeshape of a tethered rectangular resonator. Both resonators are drawn to the same scale. (c) Left: Eigenfrequencies from finite element modelling for the beam (blue) and trampoline (yellow) . Right: Duffing coefficient $\alpha_\mathrm{eff}$, calculated from the finite element model using Eq.~\eqref{eqn:Phononic-basics-duffing-coeff-derivation-2D}. Parameters: tensile stress $\sigma=1\,\mathrm{GPa}$, Young's modulus $Y=250\,\mathrm{GPa}$, thickness $h=100\,\mathrm{nm}$.}
    \label{fig:Phononic-basics-duffing-optimisaton-example}
\end{figure}

As an example, consider Fig.~\ref{fig:Phononic-basics-duffing-optimisaton-example}, where we use finite element software (COMSOL Multiphysics~\cite{COMSOLMultiphysics2020}) to model a pair of silicon nitride resonators (a), one a concave two-dimensional beam, and the other a doubly-tethered trampoline (b), with both resonators drawn to the same scale. Both resonators effectively share the same eigenfrequency around $3.8\,\mathrm{MHz}$. However, their effective Duffing coefficients are almost two orders of magnitude ($77\times$) different. This is because, as seen in Fig.~\ref{fig:Phononic-basics-duffing-optimisaton-example}(b), the design of the trampoline concentrates the flexural strain to the tethers. The tethers are clamped to the substrate, so motion there necessarily requires a large localised deformation of the material, leading to geometric nonlinearity. The concave beam (a) produces the opposite effect, creating a more even distribution of strain throughout the resonator. This is reminiscent to the technique of soft clamping~\cite{tsaturyanUltracoherentNanomechanicalResonators2017}, which minimises material curvature in order to improve quality factors. 

\subsection{Electrostatic actuation}
\label{sec:Phononic-basics-electrostatic-actuation}

In addition to geometric elongation, a second common source of Duffing nonlinearity in membrane phononics is electrostatic forces. Here we explain the principle of electrostatic actuation, derive the corresponding Duffing coefficient, and explain other nuances such as linear resonance frequency shifts and the existence of quadratic nonlinearity.

Electrostatic actuation is one of the most common methods for generating motion in nanomechanical devices~\cite{schmidFundamentalsNanomechanicalResonators2016,bekkerInjectionLockingElectrooptomechanical2017,hatanakaElectromechanicalMembraneResonator2012}. In membrane phononic circuits it is an attractive actuation method because it can be highly localised and achieve highly nonlinear amplitudes of motion. Additionally, as we will discuss, it can be used to perform non-actuation functions such as frequency tuning~\cite{jinEngineeringErrorCorrecting2024,chaElectricalTuningElastic2018,meiFrequencyTuningGraphene2018}.

\begin{figure*}[ht]
    \centering
    \includegraphics[width=0.8\linewidth]{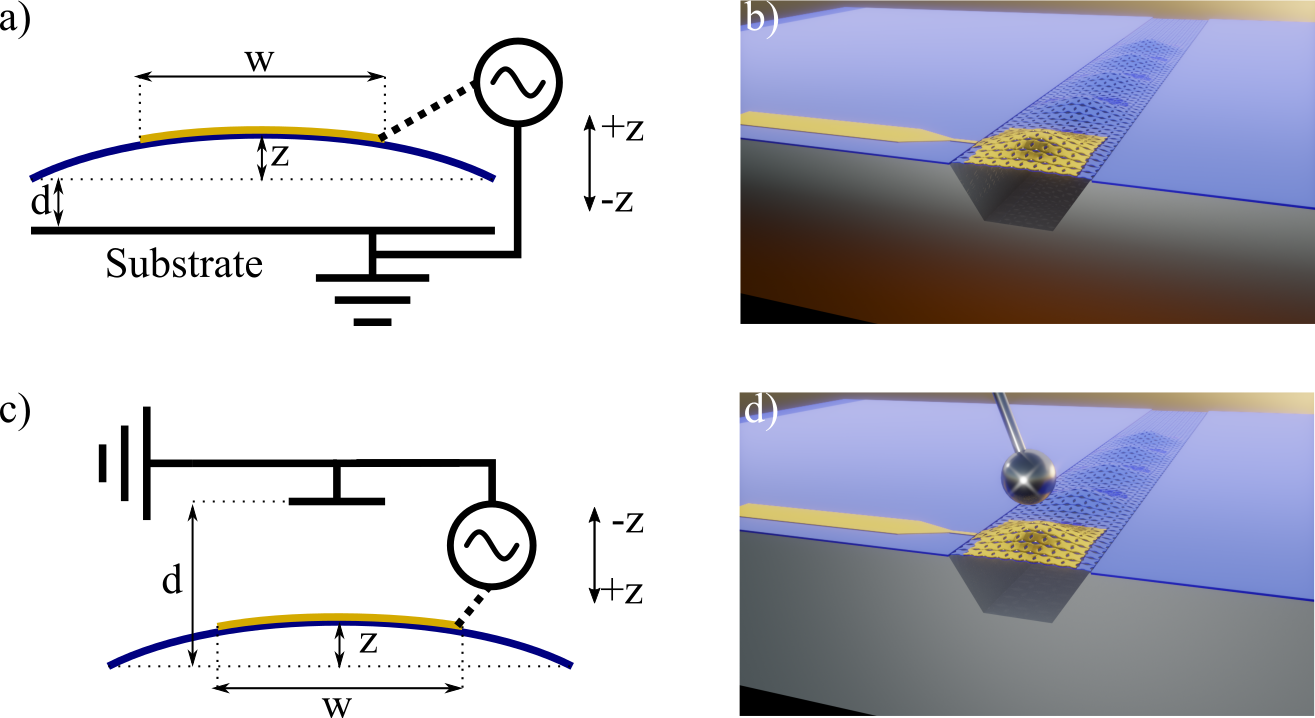}
    \caption[Electrostatic actuation mechanisms]{(a) Electrostatic actuation mechanism for a membrane where a voltage is applied to the membrane and the substrate is grounded. (b) 3D graphics illustration of how that scheme can be physically implemented. An electrode (gold) is fabricated, for example via metal lift-off, onto a suspended membrane (blue). The substrate (grey) is grounded (brown shading). (c) Alternative actuation scheme where a suspended electrode is used as the ground. Because the electrode is above the membrane, here upwards motion is defined as negative displacement $z$. (d) 3D graphics illustration of how this scheme can be physically implemented. A suspended ball electrode (silver) is placed above an electrode (gold) that sits on the suspended waveguide (blue).}
    \label{fig:Phononic-basics-electrostatic-actuation}
\end{figure*}

Membranes can be electrostatically actuated by depositing an electrode on the membrane surface, then creating a potential difference between that electrode and another fixed electrode. For example, one can fabricate gold wiring on top of the membrane via a metal liftoff process, then apply a voltage to the wires while grounding the electrically conductive substrate~\cite{romeroAcousticallyDrivenSinglefrequency2024}. Alternatively, the substrate could be replaced with a suspended electrode above the membrane~\cite{romeroPropagationImagingMechanical2019}. These two setups are illustrated in Fig.~\ref{fig:Phononic-basics-electrostatic-actuation}(a) and (b) respectively.

Consider the more common situation where the substrate is used as the fixed electrode~\cite{chaElectricalTuningElastic2018,romeroAcousticallyDrivenSinglefrequency2024}. The second electrode is patterned onto the membrane and has area $A$ and voltage $V$. Because the etch depth $d$ ($\approx 500 \,\mathrm{nm}$) is typically much less than the electrode size ($\approx 10\mathrm{\upmu m}$), and considering the extremely large ($\approx10^4)$ megahertz-range dielectric permittivity of doped silicon~\cite{romeroAcousticallyDrivenSinglefrequency2024}, the electric field is well-localised to the vertical column of depth $d$ beneath the electrode, as in an ideal parallel plate capacitor. Indeed finite element simulations show that for typical values the effective capacitance is only half a percentage point off the value for a parallel plate capacitor~\cite{romeroAcousticallyDrivenSinglefrequency2024}. 

In the parallel plate model the capacitive energy $E_\mathrm{cap}$ is~\cite{schmidFundamentalsNanomechanicalResonators2016}:
\begin{equation}
    \label{eqn:Phononic-basics-capacitive-energy}
    E_\mathrm{cap}\equiv\frac{1}{2}CV^2=\frac{1}{2}\frac{\varepsilon_0\varepsilon_rA}{d+z}V^2,
\end{equation}
where $C$ is the system capacitance, $z$ is the membrane displacement and the constants $\varepsilon_0$ and $\varepsilon_r$ are respectively the electric permittivity of free space and relative permittivity of the medium within the capacitor gap (generally air or vacuum where $\varepsilon_r\simeq1$). Because the potential energy depends on displacement  the membrane experiences a capacitive force:
\begin{equation}
    \label{eqn:Phononic-basics-capacitive-force}
    F_\mathrm{cap}(z)=\frac{-\mathrm{d}E_\mathrm{cap}}{\mathrm{d}z}=-\frac{1}{2}\frac{\varepsilon_0\varepsilon_rA}{(d+z)^2}V^2.
\end{equation}

We can understand the dynamical effect of this force by performing a Taylor expansion it about $z/d$:

\begin{align}
\label{eqn:Phononic-basics-capacitive-taylor-expansion}
F_\mathrm{cap}(z)=-\frac{1}{2}\frac{\varepsilon_0\varepsilon_r AV^2}{d^2}\bigg(&1-2\frac{z}{d}  +3\left(\frac{z}{d}\right)^2 -4\left(\frac{z}{d}\right)^3\\
 &+\mathcal{O}\left(\left(\frac{z}{d}\right)^4\right)\bigg).
\end{align}

We will analyse this electrostatic force using the lumped element model, considering each power of $(z/d)$ in turn.

The constant force is:
\begin{equation}
    F_{\mathrm{cap},0}=-\frac{1}{2}\frac{\varepsilon_0\varepsilon_rAV^2}{d^2}.
\end{equation}
This force produces a to a static displacement of the membrane towards the substrate, which we can find by considering the Duffing equation (Eq.~\eqref{eqn:Phononic-basics-Duffing-equation}) for a constant displacement $z(t)=Z_0$:
\begin{equation}
    \label{eqn:Phononic-basics-electrostatic-steadystate-forces}
    k_\mathrm{eff}Z_0+\alpha_\mathrm{eff}Z_0^3=F_{\mathrm{cap},0}.
\end{equation}
Here $k_\mathrm{eff}$ is the effective spring constant determined by the material and geometry using Eq.~\eqref{eqn:Phononic-basics-k_eff}.
Assuming small deflections such that $k_\mathrm{eff}\gg\alpha_\mathrm{eff}Z_0^2$, the new equilibrium displacement will be shifted by $\Delta Z_0\simeq F_{\mathrm{cap},0}/k_\mathrm{eff}$. For a linear resonator this displacement is all that occurs and the dynamical behaviour is unchanged. However, if we also factor in the Duffing nonlinearity from Section~\ref{sec:Phononic-basics-mechanical-nonlinearity} we expect the longitudinal strain from the displacement to modify the effective spring constant. By linearising Eq.~\eqref{eqn:Phononic-basics-electrostatic-steadystate-forces} around the displacement of $\Delta Z_0\simeq F_{\mathrm{cap},0}/k_\mathrm{eff}$, the change in spring constant can be obtained as:
\begin{equation}
    \Delta k_\mathrm{Duff}=3\alpha_\mathrm{eff}\left(\frac{F_{\mathrm{cap},0}}{k_\mathrm{eff}}\right)^2.
\end{equation}

\begin{figure}[ht]
    \centering
    \includegraphics[width=0.99\linewidth]{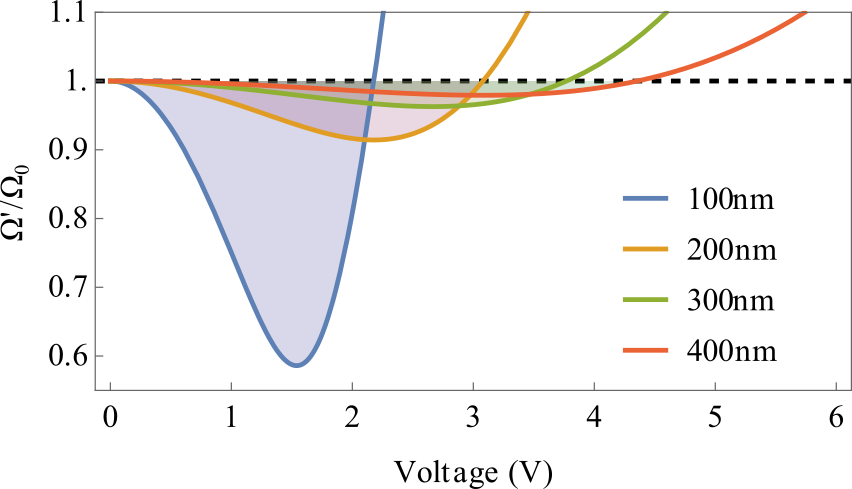}
    \caption[Resonance frequency shifts due to electrostatic actuation]{Solid lines: relative frequency shifts calculated from Eq.~\eqref{eqn:Phononic-basics-es-spring-correction}. Line colour corresponds to capacitive separation distance $d$. Shaded areas indicate where the net frequency shift is negative. Dashed line: $\Omega'/\Omega_0=1$. This example considers a square membrane of side length 50~{\textmu}m, electrode area $15^2$~{\textmu}m$^2$, Young's modulus 200~GPa, tensile stress $\sigma=1\,\mathrm{GPa}$, thickness 60~nm, and resonance frequency 1.5~MHz.}
    \label{fig:Phononic-basics-freq-shift-etch-depth}
\end{figure}

Now we turn to the linear term in Eq.~\eqref{eqn:Phononic-basics-capacitive-taylor-expansion}. This directly provides another correction to the spring constant:
\begin{equation}
    \Delta k_\mathrm{cap} = -\frac{\varepsilon_0\varepsilon_rV^2A}{d^3}. 
\end{equation}
In this case the term is negative (softening) so it acts to decrease the resonance frequency. A DC voltage can be used for this reason to decrease the eigenfrequency of a nanomechanical oscillator---this is the well known process of electrostatic softening~\cite{kozinskyTuningNonlinearityDynamic2006a,mahboobPhononLasingElectromechanical2013}. Comparing the spring constant with and without the electrostatic force, we can see that the resonance frequency shifts from $\Omega_0=\sqrt{k_\mathrm{eff}/m_\mathrm{eff}}$ to:
\begin{equation}
    \label{eqn:Phononic-basics-es-spring-correction}
    \Omega'=\sqrt{\frac{k_\mathrm{eff}+\Delta k_\mathrm{Duff}+\Delta k_\mathrm{cap}}{m_\mathrm{eff}}}.
\end{equation}
Whether the change in frequency is positive or negative depends on the applied voltage and the etch depth. $\Delta k_\mathrm{Duff}$ is positive and scales with $V^4/d^{4}$. $\Delta k_\mathrm{cap}$ is negative and scales with $V^2/d^{3}$. The different scaling laws imply the frequency shift can be positive or negative depending on $V$ and $d$, as illustrated in Fig.~\ref{fig:Phononic-basics-freq-shift-etch-depth}. It can be seen that the stiffening correction dominates for large voltages and the softening term dominates for small voltages (especially for smaller etch depths). Setting $\Delta k_\mathrm{Duff}=\Delta k_\mathrm{cap}$, we find that the transition occurs at $V=2k_\mathrm{eff}\sqrt{d/3\varepsilon_0\varepsilon_rA\alpha_\mathrm{eff}}$.

Equation~\eqref{eqn:Phononic-basics-es-spring-correction} implies that DC voltages can be used to tune the resonance frequency of the membrane. This can be used, for example, to compensate for fabrication imperfections among identical oscillators in applications like error correction~\cite{jinNanomechanicalErrorCorrection2025}.

\begin{figure*}
    \centering
    \includegraphics[width=0.99\linewidth]{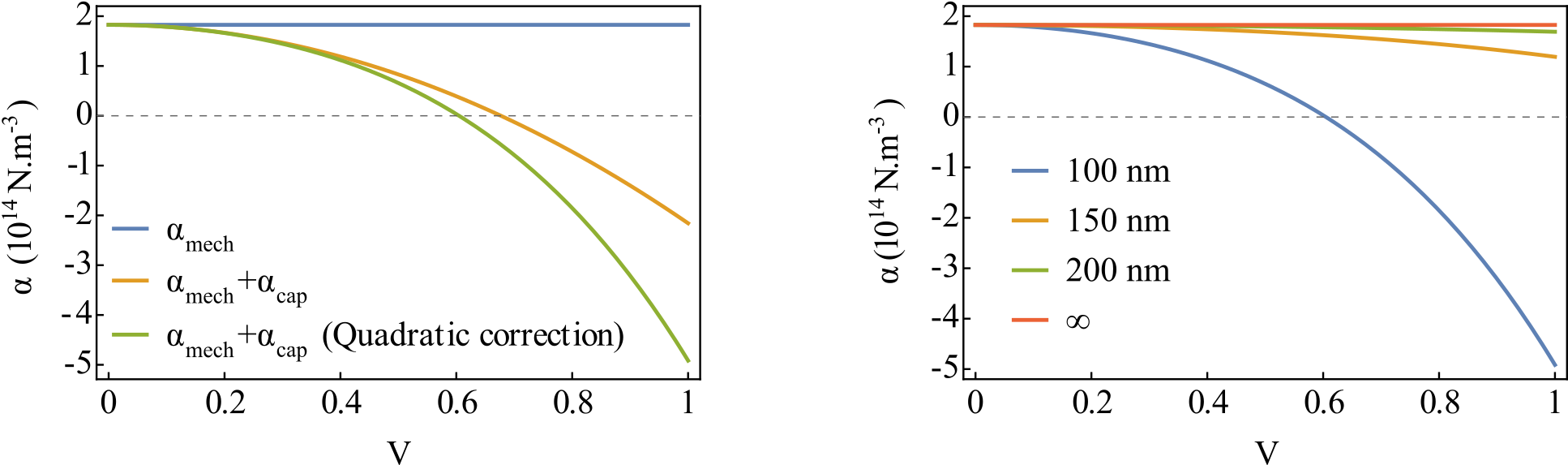}
    \caption[Change in Duffing nonlinearity under electrostatic actuation]{(a) Total Duffing coefficient $\alpha=\alpha_\mathrm{mech}+\alpha_\mathrm{cap}$, with and without the electrostatic term, and showing the effect of the quadratic contribution. Here  $d=100\,\mathrm{nm}$. (b) Total Duffing coefficient $\alpha=\alpha_\mathrm{mech}+\alpha_\mathrm{cap}$ at various etch depths. This example considers a square silicon nitride membrane of side length 50~{\textmu}m, gold electrode area $(15~\mathrm{\mu m})^2$, Young's modulus 200~GPa, stress $\sigma=1\,\mathrm{GPa}$, thickness 60~nm, and resonant frequency 1.5~MHz.}
    \label{fig:Phononic-basics-duffingcoeff-electrostatics}
\end{figure*}

Now we consider the quadratic and cubic terms from the electrostatic force~\eqref{eqn:Phononic-basics-capacitive-taylor-expansion}:
\begin{equation}
    \beta_\mathrm{cap}=\frac{3\varepsilon_0\varepsilon_rAV^2}{2d^4},\;\;\text{and}\;\alpha_\mathrm{cap}=-\frac{4}{3}\frac{\beta_\mathrm{cap}}{d}.
\end{equation}
These two terms dominate over higher order nonlinearities in the relevant scenarios where $z/d\ll1$. To first order, we can ignore $\beta_\mathrm{cap}$ because it corresponds to a cubic potential that only shifts the equilibrium position of the resonator and not the resonator frequency. To second order, we can lump it in with $\alpha_\mathrm{cap}$ as~\cite{nayfehNonlinearOscillations2004,sfendlaExtremeQuantumNonlinearity2021}:
\begin{equation}
    \alpha_\mathrm{cap}\bigg|_{\text{quadratic correction}}= \alpha_\mathrm{cap}-\frac{9}{10}\frac{\beta_\mathrm{cap}^2}{k_\mathrm{eff}}.
\end{equation}
Figure \ref{fig:Phononic-basics-duffingcoeff-electrostatics}(a) shows that this correction does not change the qualitative behaviour but does increase the magnitude of the electrostatic nonlinearity.

The electrostatic Duffing term $\alpha_\mathrm{cap}$ is negative and so produces softening behaviour. Additionally, it scales very sensitively to the fifth power of the capacitor separation distance. As Fig.~\ref{fig:Phononic-basics-duffingcoeff-electrostatics}(b) shows, increasing the etch depths greatly suppresses the electrostatic nonlinearity, leaving  the mechanical nonlinearity (here derived using two-dimensional version of Eq.~\eqref{eqn:Phononic-basics-duffing-coeff-derivation}) as the dominant term.

The tunability (via etch depth, electrode area, and applied voltage) of the electrostatic nonlinearity, combined with the geometric nonlinearity (which we have seen is sensitive to actuation amplitude and device geometry), can together produce a rich set of dynamical behaviours. For instance, the electrostatic nonlinearity can be tuned to effective nullify the geometric Duffing term, allowing higher order nonlinear terms to be observed and probed~\cite{kozinskyTuningNonlinearityDynamic2006a,samantaTuningGeometricNonlinearity2018,elshurafaNonlinearDynamicsSpring2011}. Alternatively, suppressing the nonlinearity allows operation at higher amplitudes and therefore more precise sensing~\cite{ekinciUltimateLimitsInertial2004, postmaDynamicRangeNanotube2005,sansaOptomechanicalMassSpectrometry2020}.

\subsection{DC bias for oscillatory motion}
\label{sec:Phononic-basics-electrostatic-actuation-DC-bias}
When using electrostatic actuation to drive oscillatory motion, it is usually helpful to add a DC bias to the alternating drive voltage. There are two reasons which both stem from the fact that the capacitive force in Eq.~\eqref{eqn:Phononic-basics-capacitive-force} is proportional to $V^2$.

Let the applied voltage be $V=V_\mathrm{DC}+V_\mathrm{AC}\cos(\Omega_d t)$. If no bias is used and $V_\mathrm{DC}=0$, then the force on the membrane will be proportional to $V_\mathrm{AC}^2\cos^2(\Omega_dt)=V_\mathrm{AC}^2(1+\cos(2\Omega_d t))/2$. The membrane will be driven at twice the input frequency, which is often not desirable in experiments.

If a DC bias is used, however, then the force on the membrane will be proportional to:
\begin{equation}
    V^2 = V_\mathrm{DC}^2+2V_\mathrm{DC}V_\mathrm{AC}\cos(\Omega_d t)+V_\mathrm{AC}^2\cos^2(\Omega_dt).
\end{equation}
The first term provides a constant force that changes the membrane equilibrium position and the effective spring constant as described in Eq.~\eqref{eqn:Phononic-basics-es-spring-correction}. The middle term provides the actuation at the desired input frequency. In experiments it is often the case that $V_\mathrm{DC}\gg V_\mathrm{AC}$, in which case the actuation force has been amplified beyond what could be achieved with an alternating potential alone. The last term can often be ignored, both because $V_\mathrm{DC}\gg V_\mathrm{AC}$ and because it is not a force at the frequency of interest.

\subsection{High-voltage limits}
\label{sec:Phononic-basics-electrostatic-actuation-high-voltage-limits}

The discussion in Section~\ref{sec:Phononic-basics-electrostatic-actuation} is valid only for voltages up to a certain limit. At high voltages two kinds of sudden and self-reinforcing behaviour can occur, which invalidate the lumped element model and can even destroy the membrane entirely.

\subsubsection{Pull-in}

The first behaviour is pull-in, which occurs when the electrostatic actuation forces become stronger than the mechanical restoring forces. Consider the situation of a membrane above a grounded substrate as illustrated in Fig.~\ref{fig:Phononic-basics-electrostatic-actuation}(a). Neglecting nonlinearity, the net force is the balance of tension pulling upwards and capacitive force pulling downwards:
\begin{equation}
\label{eqn:Phononic-basics-pullin-net-force}
    F_\Sigma= -\underbrace{ k_\mathrm{eff}z}_\mathrm{tension} -\underbrace{\frac{1}{2}\frac{\varepsilon_0\varepsilon_rAV^2}{(d+z)^2}}_\mathrm{electrostatic}.
\end{equation}
The tension force scales linearly with displacement, but the electrostatic force, as seen from its Taylor expansion~\eqref{eqn:Phononic-basics-capacitive-taylor-expansion}, has higher order terms that become significant when $z\sim -d$. Therefore there exists a displacement $z_\mathrm{pull}<0$ below which the electrostatic force increases faster than the upward tension restoring force as the membrane moves towards the substrate. A positive feedback loop forms causing runaway motion until the membrane collapses onto the substrate. This is called pull-in~\cite{zhangElectrostaticPullinInstability2014}, and the voltage at which it occurs is called the pull-in voltage. 

The pull-in voltage can be found analytically from Eq.~\eqref{eqn:Phononic-basics-pullin-net-force}~\cite{kaajakariNonlinearLimitsSingleCrystal2004}. Taking the derivative with respect to $z$ and evaluating at the equilibrium point gives:
\begin{equation}
    \frac{\partial F_\Sigma}{\partial z}\bigg|_{F_\Sigma=0}=\frac{2k_\mathrm{eff}z}{d+z}-k_\mathrm{eff}.
\end{equation}
The system is unstable when $\frac{\partial F}{\partial z}\geq0$, corresponding to downwards force increasing with downwards displacement. This inequality is satisfied when $z\leq-d/3$. 

\begin{figure}
    \centering
    \includegraphics[width=0.99\linewidth]{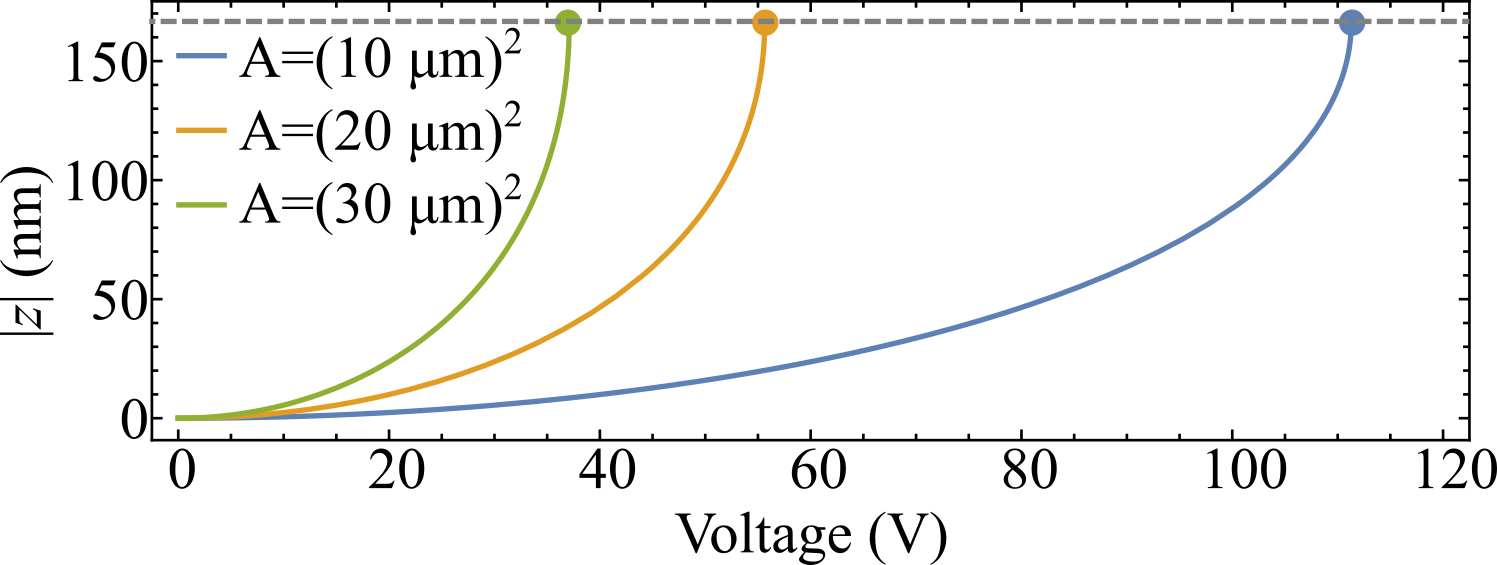}
    \caption[Equilibrium displacement magnitude $|z|$ and pull-in voltage from electrostatic actuation]{Equilibrium displacement magnitude $|z|$ calculated from Eq.~\eqref{eqn:Phononic-basics-pullin-net-force} for different electrode areas. Blue line: electrode area of $A=10\times10\,\mathrm{\mu m}^2$. Orange: $A=20\times20\,\mathrm{\mu m}^2$. Green: $A=30\times30\,\mathrm{\mu m}^2$. Circles mark the pull-in instability locations. The etch depth is $d=500\,\mathrm{nm}$. The effective spring constant was calculated analytically from Eq.~\eqref{eqn:Phononic-basics-k_eff} assuming a $30\,\mathrm{\mu}m$ silicon nitride square membrane vibrating in the fundamental mode with $\sigma=1\,\mathrm{GPa}$ and $\rho=3200\,\mathrm{kg\cdot m^{-3}}$.}
    \label{fig:Phononic-basics-pullin-example-plot}
\end{figure}

We can find the voltage that produces this unstable amplitude by plugging $z=-d/3$ into Eq.~\eqref{eqn:Phononic-basics-pullin-net-force}. It yields:
\begin{equation}
    \label{eqn:Phononic-basics-pullin-voltage}
    V_\mathrm{pull}=\sqrt{\frac{8}{27}\frac{kd^3}{\varepsilon_0\varepsilon_rA}}.
\end{equation}
Membranes with larger actuation electrodes and smaller capacitive gaps are more susceptible to pull-in. Examples of pull-in voltages for a typical silicon nitride membrane are given in Fig.~\ref{fig:Phononic-basics-pullin-example-plot}.

\subsubsection{Capacitor breakdown}
The second catastrophic failure that can occur at high voltages is the formation of a current between the membrane and fixed electrode. This can be due to Townsend discharge, where the electric field accelerates the free electrons present in the gas surrounding the device~\cite{sladeElectricalBreakdownAtmospheric2002}. This discharge current lowers the capacitance and hence the actuation efficiency of the electrostatic setup. At higher voltages the free electrons begin to ionise gas molecules during their journey to the positive terminal. Above a critical voltage each electron on average produces more than one extra seed electron (`avalanche ionisation'), producing a self-sustaining current~\cite{xiaoFundamentalTheoryTownsend2016}. This is called dielectric breakdown. The current essentially short circuits the parallel-plate capacitor, causing a total actuation failure and also possibly damage to the device from heating.

Current may also form via field emission of electrons from the metal electrode. This can cause a cascading failure mode where the field emission current produces ohmic heating, which then produces a thermionic current, quickly leading to  breakdown~\cite{fuElectricalBreakdownMacro2020}. Large areas of a micron-scale electrode may vaporise due to the highly concentrated ohmic heating~\cite{bakerHighBandwidthOnchip2016}. The field emission failure mode remains relevant at high vacuum applications where Townsend discharge is mitigated.


\section{Evanescent fields}
\label{sec:Coupling-evanescent-fields}

So far in this paper we have separately introduced membrane waveguides and resonators, along with the continuum and lumped-element models. We now shift focus towards how these circuit elements can be coupled. Coupling is an essential ingredient in phononic circuits; we will show that one of the outstanding aspects of suspended membranes is that the coupling can be engineered very flexibly and precisely. We will build towards this conclusion in the rest of the paper and finish by outlining several examples. Firstly we introduce evanescent acoustic fields, which underpin the mechanism by which acoustic phonons are coupled.

Evanescent waves are oscillations that do not propagate but instead exponentially decay with distance. They are highly important in photonics, for coupling waveguides, optical fibers, and on-chip photonic structures~\cite{yeReviewSiliconPhotonics2013,karabchevskyOnchipNanophotonicsFuture2020}; in optical sensing, using the evanescent field distributed along a fiber~\cite{luDistributedOpticalFiber2019} or concentrated using surface plasmon resonances~\cite{mayerLocalizedSurfacePlasmon2011}; and in other applications such as super-resolution imaging~\cite{vanhulstEvanescentfieldOpticalMicroscope1991,willetsSuperResolutionImagingPlasmonics2017} and optical tweezers~\cite{bouloumisFarFieldNearFieldMicro2020,zhaoOpticalFiberTweezers2020}.

Evanescent waves are also very useful for membrane phononic circuits, where they allow single mode waveguides and highly efficient coupling of heterogeneous membrane geometries~\cite{,fuPhononicIntegratedCircuitry2019,mauranyapinTunnelingTransverseAcoustic2021,hatanakaValleyPseudospinPolarized2024}. A simple use case for these functions is the coupling of a resonator with input and output single mode waveguides, allowing mechanical logic to be performed without electromechanical actuation~\cite{romeroAcousticallyDrivenSinglefrequency2024}. In this tutorial we will go further and detail several more phononic devices that can be achieved using evanescent couplers (see Section~\ref{sec:Coupling-examples-of-engineered-coupling}).

\begin{figure}[ht]
    \centering
    \includegraphics[width=\columnwidth]{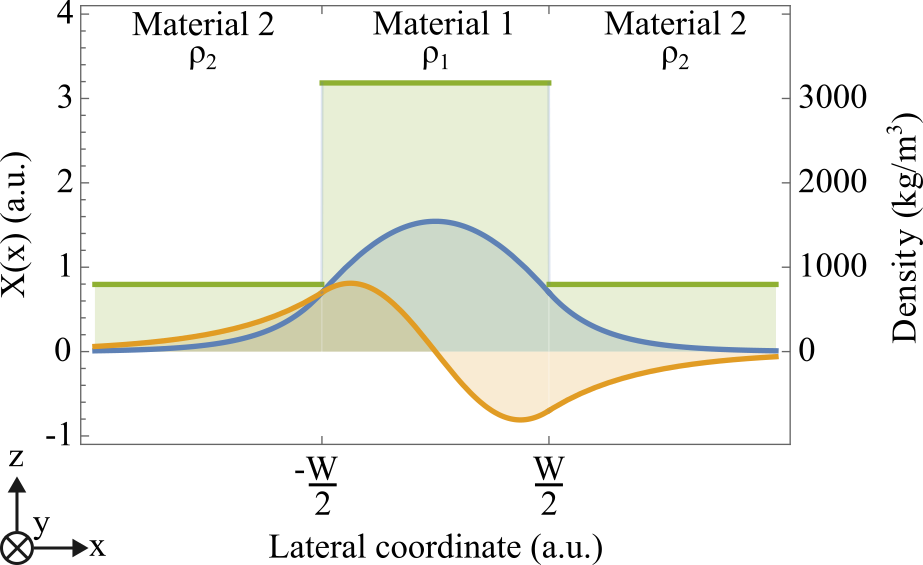}
    \caption[Waveguide modeshapes with evanescent field components]{Waveguide modeshapes with evanescent field components, from solving~\eqref{eqn:Coupling-evanescent-profile-cases}. Blue: first mode. Orange: second mode. Green: density profile of the membrane. This is for a waveguide of width $W=50\,\mathrm{\upmu m}$ made of silicon nitride with $1\,\mathrm{GPa}$ stress and density $\rho=3184\,\mathrm{kg\cdot m}^{-3}$ in the waveguide and $\rho/4$ outside the waveguide.}
    \label{fig:Coupling-densityprofile-and-modeshapes}
\end{figure}

To see how evanescent waves arise in membrane phononics, consider a waveguide of width $W$ defined by having a density $\rho_1$ greater than the density $\rho_2$ of the surrounding material, as illustrated in Fig.~\ref{fig:Coupling-densityprofile-and-modeshapes}. This is not the same mechanism of operation as membrane waveguides defined by the large impedance mismatch between the membrane and substrate (described earlier), but it has the same principle and makes for a simpler example.

As always, the basic equation of motion for the membrane is the wave equation, which we study here neglecting loss and nonlinearities:
\begin{equation}
    \label{eqn:Coupling-wave-eqn}
    \frac{1}{c^2}\frac{\partial^2u(t,x,y)}{\partial t^2}=\nabla^2u(t,x,y).
\end{equation}
The waveguide has a large extent (effectively infinite in the $y$ coordinate), so as seen in Section~\ref{sec:Phononic-basics-Dispersion} we can invoke translational invariance and look for a solution of the form $u(t,x,y)=Z_0e^{i(\Omega t-k_yy)}\psi(x)$. Expecting a transverse mode with evanescent components, we can use the following ansatz for $\psi(x)$:
\begin{equation}
    \label{eqn:Coupling-evanescent-profile-cases}
    \psi(x)=\begin{cases}
        \cos(k_xx) & |x|<\frac{W}{2}\\
        e^{-\kappa (|x|-\frac{W}{2})}&\text{elsewhere}.
    \end{cases}
\end{equation}
Here $\kappa$ is the exponential decay rate.

The smoothness of the membrane requires that both $\psi(x)$ and $\psi'(x)$ are continuous at the boundary where $|x|=W/2$. This provides the equation:
\begin{equation}
    \label{eqn:Coupling-evanescentprofile-Haus1}
    \tan\left(\frac{k_xW}{2}\right)=\frac{\kappa}{k_x}.
\end{equation}
Additionally, the wave equation applied inside and outside the waveguide region provides two more equations:
\begin{align}
    \label{eqn:Coupling-evanescentprofile-Haus2}
    \frac{\rho_1\Omega^2}{\sigma}&=k_y^2+k_x^2,\\
    \label{eqn:Coupling-evanescentprofile-Haus3}
    \frac{\rho_2\Omega^2}{\sigma}&=k_y^2-\kappa^2.
\end{align}
Combining Eqs.~\eqref{eqn:Coupling-evanescentprofile-Haus1},~\eqref{eqn:Coupling-evanescentprofile-Haus2}, and~\eqref{eqn:Coupling-evanescentprofile-Haus3} eliminates $k_y$ and $\kappa$, yielding a transcendental equation for $k_x$:
\begin{equation}
    \label{eqn:Coupling-evanescentprofile-Haus4}
    \tan(k_xd)=\sqrt{\frac{\Omega^2(\rho_1-\rho_2)}{\sigma k_x^2}-1}.
\end{equation}

The $\tan$ function restricts $k_x$ to discrete solutions. These solutions correspond to the discrete eigenmodes that arise from the clamped boundaries dispersion relationship (Eq.~\eqref{eqn:Phononic-basics-dispersion-clamp}). They are not exactly the same, because here we have allowed a nonzero evanescent field, while earlier we imposed the more stringent condition that $\psi(0)=\psi(W)=0$ at the boundaries.

\begin{figure}[ht]
    \centering
    \includegraphics[width=0.99\linewidth]{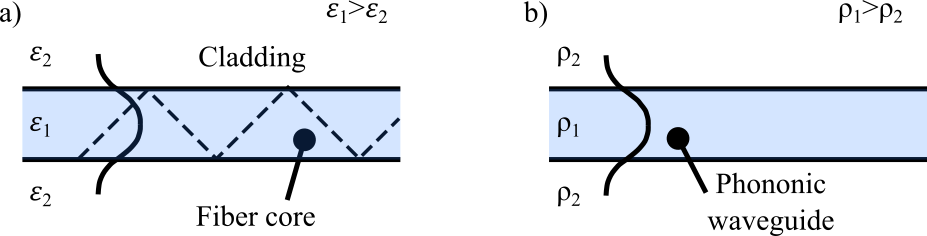}
    \caption[Equivalence between confinement of optical waves with dielectric index contrast and confinement of acoustic waves with density contrast]{Equivalence between confinement of optical waves with dielectric index contrast and confinement of acoustic waves with density contrast. (a) In a dielectric slab or optical fiber, the guided mode is confined into the core, which has high refractive index than cladding material. The mode evanescently extends into the cladding, but no energy is transferred outside, such that is can analogously be formulated in a ray optics description in terms of total internal reflection (dashed line). (b) A mathematically equivalent example is the phononic waveguide studied in this section, where density differences create a guided mode.}
    \label{fig:Coupling-dielectric-slab-evanescence-equivalence}
\end{figure}

Equation~\eqref{eqn:Coupling-evanescent-profile-cases} is mathematically equivalent to the case of the transverse electric mode of light in a dielectric waveguide, if we make the substitution $\rho\mapsto\varepsilon$~\cite{hausWavesFieldsOptoelectronics1984}; that is, if we replace density differences with electric permittivity differences. This equivalence is illustrated in Fig.~\ref{fig:Coupling-dielectric-slab-evanescence-equivalence}. Because refractive index scales with $\varepsilon^2$, the density contrast of $\rho_1/\rho_2=4$ illustrated in Fig.~\ref{fig:Coupling-densityprofile-and-modeshapes} is equivalent to a refractive index contrast of $n_2/n_1=2$, such as is found for silicon nitride photonic waveguides surrounded by air~\cite{stutiusSiliconNitrideFilms1977,bakerOpticalInstabilitySelfpulsing2012}.





\section{Coupled mode theory}
\label{sec:Coupling-coupled-mode-theory}

Evanescent acoustic fields facilitate the coupling of different, spatially separated phononic circuit components. To quantitatively model how this coupling arises, we can treat the coupling as a perturbation on the uncoupled behaviour. This strategy is called coupled mode theory (CMT)~\cite{hausCoupledmodeTheory1991,manolatouCouplingModesAnalysis1999,yarivCoupledmodeTheoryGuidedwave1973,rosencherOptoelectronicsEmmanuelRosencher2002}. Originally developed for vacuum tubes and transmission lines~\cite{pierceCouplingModesPropagation1954a,schelkunoffConversionMaxwellEquations1955}, in optics CMT has been an extremely powerful tool for the understanding and design of on-chip and fiber-based devices, such as fiber Bragg reflectors~\cite{hausCoupledmodeTheory1991,mccallApplicationCoupledMode2000}, grating and waveguide couplers~\cite{snyderCoupledModeTheoryOptical1972,hardyCoupledModeTheory1985}, tapered waveguiding structures~\cite{dongInvestigationEvanescentCoupling2012} and add/drop filters~\cite{manolatouCouplingModesAnalysis1999}. Here we translate it to membrane phononics.

CMT is generally applicable to situations involving waves at interfaces. It derives only from conservation of energy and the approximation of `small coupling', which assumes that to first order the presence of the coupling (i.e. perturbative) element does not modify the eigenmodes of the system~\cite{hausWavesFieldsOptoelectronics1984}. A key property of CMT is that the change in amplitude of one of the coupled elements is proportional to the amplitude of motion of the other coupled element~\cite{hausCoupledmodeTheory1991}. The constant of proportionality is called the coupling rate. For example if $a$ is the amplitude in the eigenmode of interest, which is coupled to $N$ other modes, we can write:
\begin{equation}
    \label{eqn:Coupling-coupled-mode-theory-proportionality}
    \frac{\upd a}{\dt}=\{\text{non-coupling terms}\}+\sum_{i=1}^N\gamma_ia_i.
\end{equation}
Here $\gamma_i$ and $a_i$ are respectively the coupling rate to and amplitude of the $i^\mathrm{th}$ external eigenmode. We usually normalise the amplitudes so that $|a|^2$ equals the energy in the mode of interest and $|a_i|^2$ equals the energy in the $i^\mathrm{th}$ eigenmode.

Equation~\eqref{eqn:Coupling-coupled-mode-theory-proportionality} implies that the coupling rates $\{\gamma_i\}$ determine the flow of energy through a coupled system. This is a key insight of CMT, and will further explored in Section~\ref{sec:Coupling-input-output}, where we demonstrate how choosing particular coupling rates can produce specific behaviour such as impedance matching, amplitude enhancement, and signal filtering.

\begin{figure}
    \centering
    \includegraphics[width=0.99\linewidth]{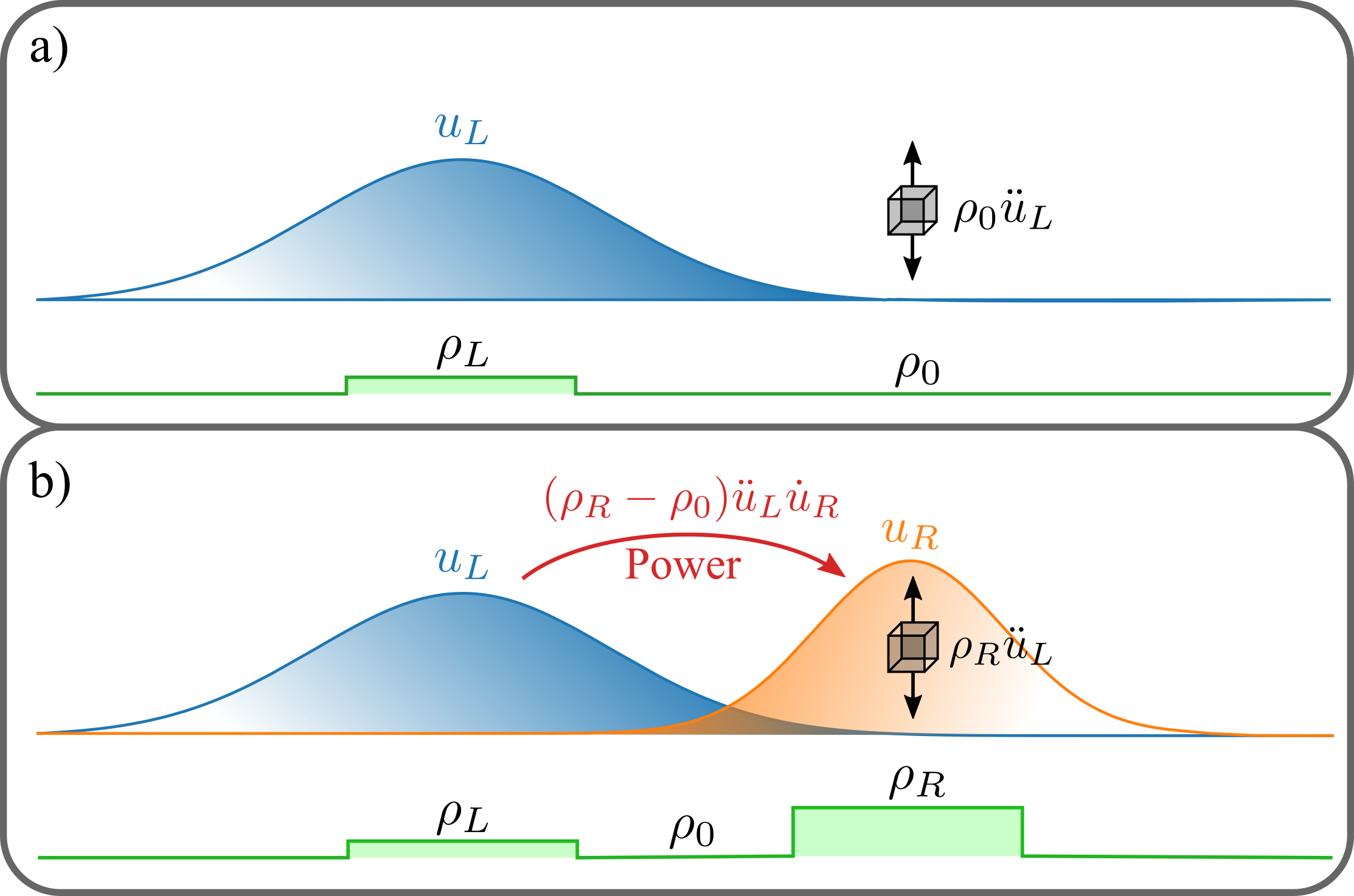}
    \caption[Origin of work and power transfer in phononic coupled mode theory]{Origin of work and power transfer in phononic coupled mode theory. (a) We consider an eigenmode (blue, above left) with time-varying displacement $u_L$, localised to a region of increased density $\rho_L>\rho_0$ (green, below left). Without external coupling, a volume element in the region of lower density (where the field is evanescent) experiences a force per unit volume of $\rho_0\ddot{u}_L$. The net work associated with this force is zero, with the force simply shifting energy between potential and kinetic forms. (b) We perturbatively introduce a second eigenmode (orange, right), with its own time-varying displacement $u_R$, localised to a region of increased density $\rho_R$. The force on a volume element in this region, due to the left (blue) eigenmode, is $\rho_R\ddot{u}_L$. Consequently, the net flow of power from the left mode to the right mode (red arrow) associated with this volume element is $(\rho_R-\rho_0)\ddot{u}_L\dot{u}_R$. Depending on the relative phase of the oscillations, this power flow may be positive or negative.}
    \label{fig:Coupling-coupled-mode-work-diagram}
\end{figure}

The origin of coupling in phononic CMT is illustrated in Figure~\ref{fig:Coupling-coupled-mode-work-diagram}. In Fig.~\ref{fig:Coupling-coupled-mode-work-diagram}(a) we consider an eigenmode (blue shading) with a time-varying displacement $u_L$, localised to a region of increased density $\rho_L>\rho_0$, where $\rho_0$ is the background density. In the surrounding material with density $\rho_0$ the acoustic field from the mode is evanescent. However, the nonzero amplitude still exerts a force per unit volume of $\rho_0\ddot{u}_L$. In the absence of coupling, the net work associated with this force is zero: it simply converts energy between kinetic and potential forms as the membrane vibrates. In Fig.~\ref{fig:Coupling-coupled-mode-work-diagram}(b) we introduce a second eigenmode (orange shading) with time-varying displacement $u_R$, localised to a region of increased density $\rho_R>\rho_0$. Under the perturbative approach of the coupled-modes formalism, the introduction of the $u_R$ mode does not affect the $u_L$ mode. Therefore, we can say that the force on a volume element in the $u_R$ mode, due to the vibration of the $u_L$ mode, is $\rho_R\ddot{u}_L$. The associated instantaneous power per unit volume (proportional to the force times velocity) from the $u_L$ mode to the $u_R$ mode, $P_{LR}$ (red arrow) is:
\begin{equation}
    \label{eqn:Coupling-CMT-power-per-unit-volume}
    P_\mathrm{LR}=(\rho_R-\rho_0)\ddot{u}_L\dot{u}_R.
\end{equation}
The subtraction here is required because the force $\rho_0\ddot{u}_L$ was already present before the perturbation.

\subsection{Analytic expression for coupling rates}
\label{sec:Coupling-two-waveguides-example}

We will use Eq.~\eqref{eqn:Coupling-CMT-power-per-unit-volume} to derive an analytic expression for the coupling rates between evanescently coupled phononic resonators and waveguides. We restrict our attention to evanescent coupling because CMT, as a perturbative treatment, is only valid when the coupling is relatively weak (specifically, when it does not change the shape of the modes).

\begin{figure}[ht]
    \centering
    \includegraphics[width=0.99\linewidth]{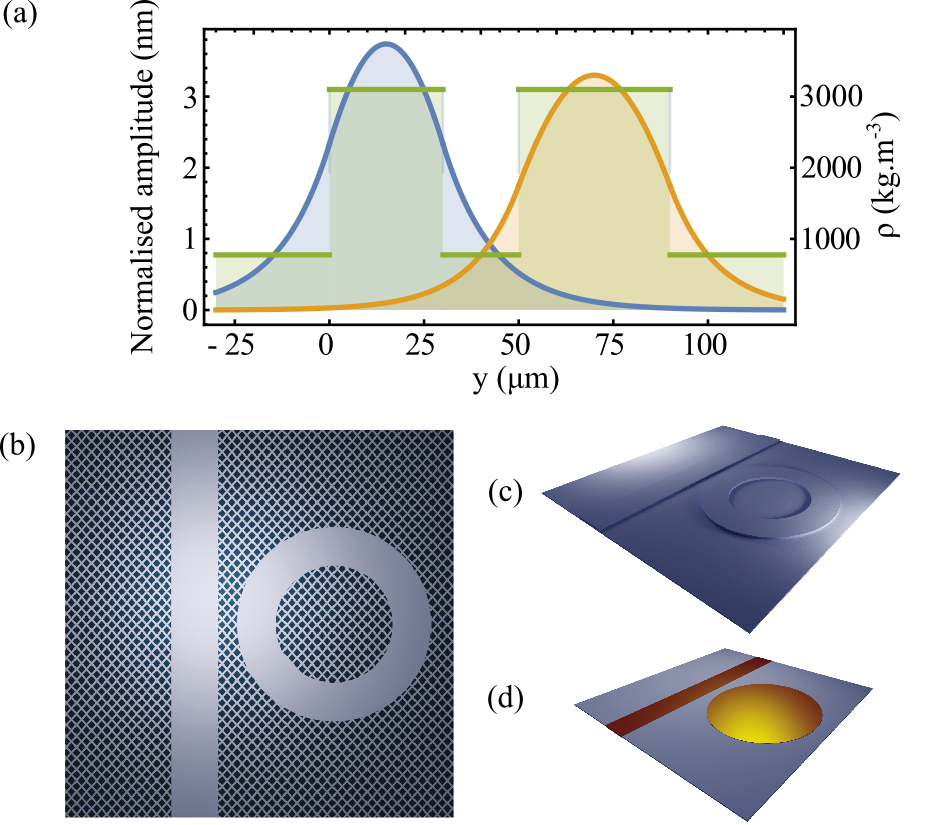}
    \caption[Evanescent coupling via engineered density differences]{Evanescent coupling via engineered density differences. (a) Two guided acoustic modes with overlapping evanescent fields. The left and right waveguides have respective widths $W_L=30$~{\textmu}m and $W_R=40$~{\textmu}m; waveguide separation: 20~{\textmu}m; $\sigma=1$~GPa; $\rho_L=\rho_R=3100~\mathrm{kg\cdot m}^{-3}$; $\rho_0=0.25\,\rho_L$ (green trace); $\Omega/2\pi=10\,\mathrm{MHz}$. Left (blue) and right (orange) mode amplitudes are normalized to correspond to an identical power flow of $P_0=1$~W.m$^{-1}$. (b) High tensile stress membrane patterned with a subwavelength mesh~\cite{romeroPropagationImagingMechanical2019} outside the membrane guiding area. (c) Ridge acoustic waveguide and resonator formed within a high tensile stress membrane. (d) Localised higher density regions can also be achieved through material deposition.}
    \label{fig:Coupling-modal-overlap-density-modification}
\end{figure}

Our example system of interest is two parallel, evanescently coupled waveguides, as illustrated in Fig.~\ref{fig:Coupling-modal-overlap-density-modification}(a). The two waveguides are defined by local increases in volumetric density (green shading): $\rho_L$ on the left and $\rho_R$ on the right, both being greater than the background density $\rho_0$. Superimposed on the waveguide densities we have plotted the fundamental eigenmodes of each waveguide (blue and orange curves), calculated from Eq.~\eqref{eqn:Coupling-evanescentprofile-Haus4} (see caption for parameter values). The eigenmodes both have evanescent components which will produce the coupling. The next subfigures provides examples for how this density modification could be fabricated: by reducing the density with a patterned mesh outside the waveguiding region (Fig.~\ref{fig:Coupling-modal-overlap-density-modification}(b)); by increasing density with a ridge acoustic waveguide set within the mesh (Fig.~\ref{fig:Coupling-modal-overlap-density-modification}(c)); and by increasing the density by depositing material such as metal (Fig.~\ref{fig:Coupling-modal-overlap-density-modification}(d)).


Assume there is one drive frequency $\Omega$ and that power is flowing in the same direction in both waveguides. Under those assumptions we can write the displacement in the left waveguide as:
\begin{equation}
    \label{eqn:Coupling-parallel-waveguides-1}
    u_L(t,x,y)=a_L(y)\tilde{u}_L(x)\cos(\Omega t-k_Ly),
\end{equation}
and in the right waveguide as:
\begin{equation}
    \label{eqn:Coupling-parallel-waveguides-2}
    u_R(t,x,y)=a_R(y)\tilde{u}_R(x)\cos(\Omega t-k_Ry+\Delta\varphi).
\end{equation}
Here $a_L$ and $a_R$ the amplitudes of displacement in the left and right waveguides, normalised so that $|a_L|^2$ and $|a_R|^2$ respectively equal the power per unit thickness flowing through the left waveguide and right waveguides. $k_L$ and $k_R$ are the longitudinal wavenumbers of the propagating waves in each waveguide (they may not be the same because the waveguide geometries could differ). The phaseshift $\Delta\varphi$ describes the phase difference between the waves in each waveguide. The functions $\tilde{u}_L(x)$ and $\tilde{u}_R(x)$ are normalised modeshapes found by dividing the displacement by the power travelling through the relevant waveguide. For example the normalised modeshape in the left waveguide is:
\begin{equation}
    \label{eqn:Coupling-parallel-waveguides-2.1}
    \tilde{u}_L(x)=\frac{\psi_L(x)}{\sqrt{v_{g,L}\times\frac{1}{2}\Omega^2\int_{-\infty}^\infty\rho(x)\psi_L(x)^2\,\dx}},
\end{equation}
where $\psi_L(x)$ is the dimensionless function with $\max(|\psi_L(x)|)=1$ that describes the eigenmode from Eq.~\eqref{eqn:Coupling-evanescent-profile-cases}.

A volume element in the right waveguide experiences a net power $P_{LR}$ due to the motion in the left waveguide according to Eq.~\eqref{eqn:Coupling-CMT-power-per-unit-volume}. Substituting $u_L$ and $u_R$ with Eqs.~\eqref{eqn:Coupling-parallel-waveguides-1} and~\eqref{eqn:Coupling-parallel-waveguides-2}, then evaluating the time derivatives will pull out a factor of $\Omega^3\cos(\Omega t)\sin(\Omega t+\Delta\varphi)$, which on average equals $\frac{1}{2}\Omega^3\sin(\Delta\varphi)$. Therefore the time averaged power per unit volume flowing from the left waveguide to the right is:
\begin{equation}
    \label{eqn:Coupling-parallel-waveguides-5}
    \langle P_{LR}\rangle=\frac{\Omega^3}{2}\sin(\Delta\varphi)(\rho_R-\rho_0)\,a_L(y)a_R(y)\,\tilde{u}_L(x)\tilde{u}_R(x).
\end{equation}
The $\sin(\Delta\varphi)$ factor is significant as it determines the direction of power flow. The greatest power flow occurs when $|\Delta\varphi|=\pi/2$ and no power flow occurs when $|\Delta\varphi|=0\;\text{or}\;\pi$. This resembles the spectral response of a driven lumped-element oscillator (see Fig.~\ref{fig:Phononic-basics-spectral-response} from Section~\ref{sec:Phononic-basics-lumped-element-model}); in that case the greatest response occurs when the drive is $\pi/2$ ahead of the resonator in phase. Here the left waveguide does work on the right waveguide when $\Delta\varphi$ is in the first or second quadratures, corresponding to the waves in the left waveguide appearing `ahead' of the waves in the right waveguide. Vice versa holds when power flows in the opposite direction. We see this borne out later in finite difference time domain simulations shown in Figure~\ref{fig:Coupling-FDTD-parallel-waveguides}(a).

If we integrate Eq.~\eqref{eqn:Coupling-parallel-waveguides-6} over the transverse ($x$) direction, we get the instantaneous power transferred from the left to right waveguides as the waves propagate down the waveguides. In our scenario, coupling is the only reason for waveguide power to fluctuate, so this power transfer equals the rate of change of $|a_R(y)|^2$. Written mathematically, we have:
\begin{align}
    \label{eqn:Coupling-parallel-waveguides-6}
    \frac{\upd |a_R(y)|^2}{\dy}=\frac{\Omega^3}{2}\sin(\Delta\varphi)(\rho_R-\rho_0) \nonumber\\ \times a_L(y)a_R(y)
    \int_R\tilde{u}_L(x)\tilde{u}_R(x)\,\dx.
\end{align}
The integral here is over the right waveguide (i.e. where in Fig.~\ref{fig:Coupling-modal-overlap-density-modification}(a) the density has increased from $\rho_0$ to $\rho_R$).

Now we will pause at Eq.~\eqref{eqn:Coupling-parallel-waveguides-6} and derive the same quantity using CMT. The coupled modes description of the same two waveguides is a pair of coupled differential equations~\cite{hausCoupledmodeTheory1991}:
\begin{align}
    \frac{\upd a_L(y)}{\dy}&=-ik_La_L(y)+\gamma_{RL}a_R(y) \nonumber\\
    \frac{\upd a_R(y)}{\dy}&=-ik_Ra_R(y)+\gamma_{LR}a_L(y).    
    \label{eqn:Coupling-parallel-waveguides-7}
\end{align}
The coupling rates are written such that $\gamma_{AB}$ corresponds to the coupling from waveguide $A$ into waveguide $B$.

The total power per unit thickness flowing in both waveguides is $|a_L(y)|^2+|a_R(y)|^2$. Conservation of energy demands that it does not change with $y$, that is:
\begin{equation}
    \label{eqn:Coupling-parallel-waveguides-8}
    \frac{\upd |a_L(y)|^2}{\dy}+\frac{\upd |a_R(y)^2|}{\dy}=0.
\end{equation}
Substituting Eq.~\eqref{eqn:Coupling-parallel-waveguides-7} into Eq.~\eqref{eqn:Coupling-parallel-waveguides-8} and simplifying will yield that $\gamma_{RL}=-\gamma_{LR}^*$, which also means that $|\gamma_{LR}|^2=|\gamma_{RL}|^2=|\gamma|^2$.

From Eq.~\eqref{eqn:Coupling-parallel-waveguides-7} we can derive the rate of change of $|a_R(y)|^2$ (we omit arguments of $y$ for clarity):
\begin{equation}
    \label{eqn:Coupling-parallel-waveguides-8.1}
    \frac{\upd |a_R|^2}{\dy}=a_R^*\frac{\upd a_R}{\dy}+a_R\frac{\upd a_R^*}{\dy}=\gamma_{LR}a_La_R^*+\gamma_{LR}^*a_L^*a_R.
\end{equation}

Comparing Eq.~\eqref{eqn:Coupling-parallel-waveguides-8.1} with Eq.~\eqref{eqn:Coupling-parallel-waveguides-6} suggests that the coupling coefficient $\gamma_{LR}$ is:
\begin{equation}
    \label{eqn:Coupling-parallel-waveguides-8.2}
    \gamma_{LR}=\frac{1}{4}\Omega^3(\rho_R-\rho_0)\int_R\tilde{u}_L(x)\tilde{u}_R(x)\,\dx.
\end{equation}
This can be heuristically justified by taking the real part of Eq.~\eqref{eqn:Coupling-parallel-waveguides-8.1}, or by comparing to similar calculations in the literature derived in the case of evanescent coupling of electromagnetic waves~\cite{manolatouCouplingModesAnalysis1999,hausCoupledmodeTheory1991}. In the next section we will also quantitatively justify this expression with a numerical simulation. The symmetric equation for coupling into the left waveguide is:
\begin{equation}
    \label{eqn:Coupling-parallel-waveguides-8.3}
    \gamma_{RL}=\frac{1}{4}\Omega^3(\rho_L-\rho_0)\int_L\tilde{u}_L(x)\tilde{u}_R(x)\,\dx,
\end{equation}
where the integral is over the left waveguide.

A key insight of Eqs.~\eqref{eqn:Coupling-parallel-waveguides-8.2} and~\eqref{eqn:Coupling-parallel-waveguides-8.3} is that the coupling rate is proportional to the \emph{overlap integral} of the normalised modeshapes. This immediately provides an intuitive explanation for the coupling of modes. We can see that the modeshapes must be spatially proximate to achieve coupling. Additionally, because the overlap integral is a signed integral, modes with differing even and odd symmetry will couple with each other minimally. We will use this second principle later (see Section~\ref{sec:Coupling-mode-division-multiplexer}) to design a mode demultiplexing power splitter.

In Section~\ref{sec:Coupling-resonator-waveguide-analytic-coupling} we will extend Eq.~\eqref{eqn:Coupling-parallel-waveguides-8.3} to the case of a resonator coupled to a waveguide. For now we will stay with our parallel waveguides and numerically test our expression for the coupling rates with a simulation.

We note that this CMT approach is quite general and can be applied as well to model the coupling of Rayleigh-like or Love-like acoustic modes in on-chip high acoustic index contrast slab waveguides~\cite{fuPhononicIntegratedCircuitry2019,fengGigahertzPhononicIntegrated2023,bicerGalliumNitridePhononic2022,xuHighfrequencyTravelingwavePhononic2022,mayorGigahertzPhononicIntegrated2021}. The specific case of coupling of flexural modes of acoustic membranes detailed here provides simple expressions (Eqs.~\eqref{eqn:Coupling-parallel-waveguides-8.2} and~\eqref{eqn:Coupling-parallel-waveguides-8.3}) reminiscent of the results obtained in the optical domain~\cite{hausCoupledmodeTheory1991,manolatouCouplingModesAnalysis1999,rosencherOptoelectronicsEmmanuelRosencher2002}, as the displacement is purely one dimensional (out of plane) and does not involve any material anisotropy~\cite{mayorGigahertzPhononicIntegrated2021}

\subsection{Numerical test}
\label{sec:Coupling-FDTD-parallel-waveguides}

The coupling rate determines the distance over which energy completes an oscillation between the coupled waveguides, as well as what fraction of energy can be transferred (this ratio is less than unity for dissimilar waveguides). Therefore we can check Eq.~\eqref{eqn:Coupling-parallel-waveguides-8.2} and Eq.~\eqref{eqn:Coupling-parallel-waveguides-8.3} by analytically deriving that transfer wavelength and transfer ratio in terms of the coupling rate, and comparing the expected values with observations from a numerical simulation.

\begin{figure*}[ht!]
    \centering
    \includegraphics[width=\textwidth]{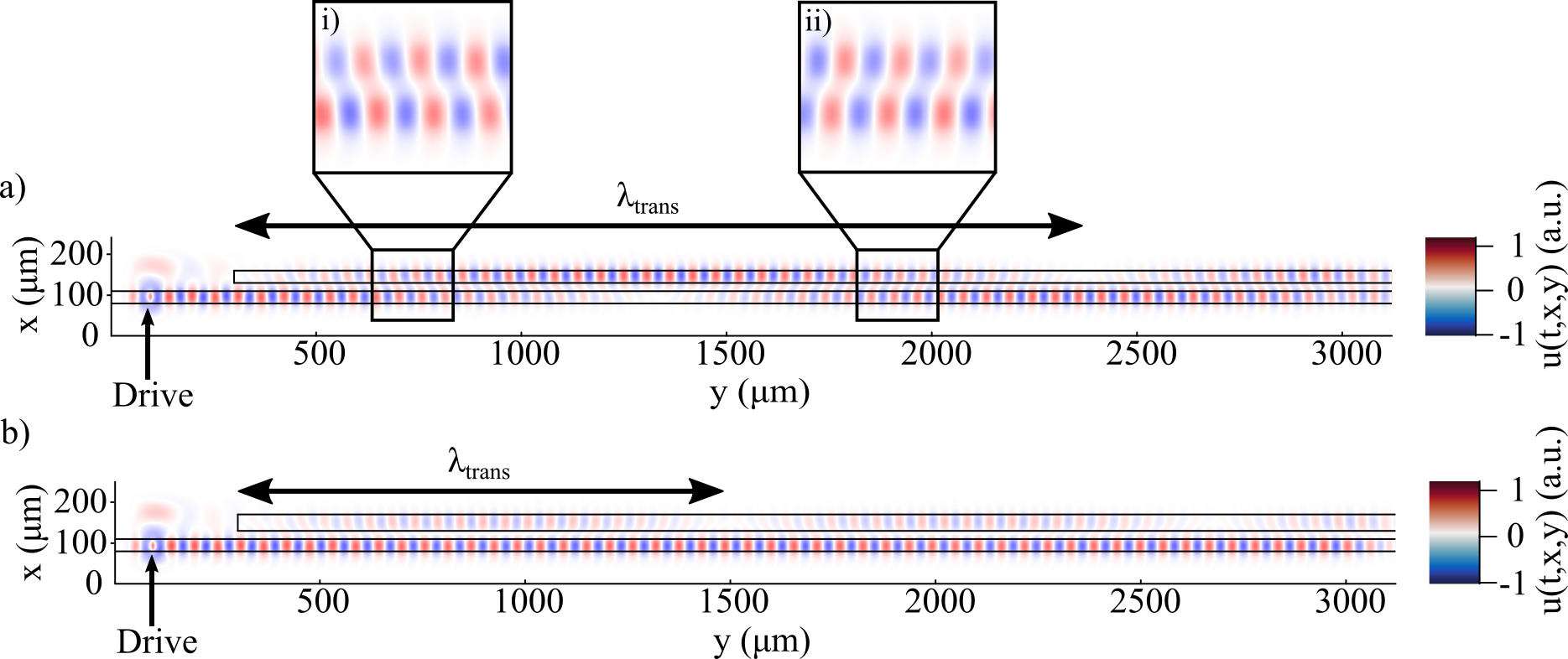}
    \caption[Finite difference time domain (FDTD) simulation of evanescent coupling between two parallel waveguides]{Finite difference time domain (FDTD) simulation of evanescent coupling between two parallel waveguides. Waves are launched at the left side of the simulation at frequency $\Omega/2\pi=12\,\mathrm{MHz}$. The heatmap plots the displacement in arbitrary units, after sufficient time such that waves have propagated across the length of the waveguide. (a) Coupling between two identical waveguides. The waveguides have width $30\,\mathrm{\upmu m}$ and separation $20\,\mathrm{\upmu m}$. Insets (i) and (ii) highlight the phase lag predicted by Eq.~\eqref{eqn:Coupling-parallel-waveguides-5}, which changes sign with the direction of power transfer. (b) Coupling in a nearly identical simulation where the width of the non-drive waveguide is increased to $40\,\mathrm{\upmu m}$. Parameters: stress $\sigma$ = 1 GPa; density in both waveguides is $\rho=3100\,\mathrm{kg\cdot m^{-3}}$; and density outside the waveguides is $\rho/4$. }
    \label{fig:Coupling-FDTD-parallel-waveguides}
\end{figure*}

We begin by returning to Eq.~\eqref{eqn:Coupling-parallel-waveguides-7}. If we assume the slowly-varying amplitudes change with $y$ as $e^{ik y}$ for some $k$, then the equations can be rearranged as:
\begin{equation}    
    \label{eqn:Coupling-parallel-waveguides-9}
    \begin{bmatrix}
        -ik_L+ik & \gamma_{RL}\\\gamma_{LR}&-ik_R+ik
    \end{bmatrix}\begin{bmatrix}
        a_L(y)\\a_R(y)
    \end{bmatrix}=\begin{bmatrix}
        0\\0
    \end{bmatrix}.
\end{equation}
For nontrivial solutions we require the square matrix to have zero determinant. This produces a quadratic equation in $k$, with solutions:
\begin{equation}    
    \label{eqn:Coupling-parallel-waveguides-10}
    k=-\frac{(k_L+k_R)}{2}\pm k_c,
\end{equation}
where
\begin{equation}    
    \label{eqn:Coupling-parallel-waveguides-11}
    k_c=\sqrt{\left(\frac{\Delta k}{2}\right)^2+|\gamma|^2},
\end{equation}
and we have defined $\Delta k=k_L-k_R$ and recall that $|\gamma|^2=|\gamma_{LR}|^2=|\gamma_{RL}|^2$ from energy conservation.

The two solutions for $k$ in Eq.~\eqref{eqn:Coupling-parallel-waveguides-10} follow the frequency anticrossing typical of coupled systems. When the waveguides are symmetric and $k_L=k_R$, the difference between the solutions of $k$ are $2k_c=2|\gamma|^2$. Conversely when the coupling is zero, the two solutions are just $k_L$ and $k_R$.

Given initial conditions $a_L(0)$ and $a_R(0)$ and our assumptions of codirectional and positive group velocities, Eq.~\eqref{eqn:Coupling-parallel-waveguides-7} can be solved for the two amplitudes~\cite{hausWavesFieldsOptoelectronics1984} to give:
\begin{widetext}
\begin{align}
    a_L(y)&=\left[a_L(0)\left(\cos(k_c y)-i\left(\frac{\Delta k}{2k_c}\right)\sin(k_c y)\right)+\frac{\gamma_{RL}}{k_c}a_R(0)\sin(k_c y)\right] \nonumber\\
    a_R(y)&=\left[a_R(0)\left(\cos(k_cy)+i\left(\frac{\Delta k}{2k_c}\right)\sin(k_c y)\right)+\frac{\gamma_{LR}}{k_c}a_L(0)\sin(k_c(y)\right].    
    \label{eqn:Coupling-parallel-waveguides-12}
\end{align}
\end{widetext}

These equations are mathematically equivalent to those that govern optical couplers and switches~\cite{hausWavesFieldsOptoelectronics1984,rosencherOptoelectronicsEmmanuelRosencher2002}. The acoustic power oscillates between the waveguides with a complete cycle (i.e. energy transferring back and forth) occurring over a wavelength of $\lambda_\mathrm{trans}=2\pi/k_c$. In terms of coupling rates this is equal to:
\begin{equation}    
    \label{eqn:Coupling-parallel-waveguides-13}
    \lambda_\mathrm{trans}=\frac{2\pi}{\sqrt{\left(\frac{\Delta k}{2}\right)^2+|\gamma|^2}}.
\end{equation}
If we assume that initially all the power is in one waveguide, say $a_R(0)=0$, then the maximum amount of power is in waveguide $R$ when $\cos(k_cy)=0$ and $\sin(k_cy)=\pm1$. This occurs for example at $y=\pi/2k_c$, implying the maximum fraction of power that can be transferred is:
\begin{equation}    
    \label{eqn:Coupling-parallel-waveguides-14}
    \frac{|a_R(\pi/2k_c)|^2}{|a_R(\pi/2k_c)|^2+|a_L(\pi/2k_c)|^2}=\frac{|\gamma|^2}{|\gamma|^2+\left(\frac{\Delta k}{2}\right)^2}.
\end{equation}

To test our theory we perform a finite difference time domain (FDTD) simulation of a system of two parallel, evanescently coupled waveguides, with parameters relevant to suspended silicon nitride membranes. The results of the simulation are shown in Table~\ref{tab:Coupling-FDTD-parallel-waveguides} and Fig.~\ref{fig:Coupling-FDTD-parallel-waveguides}. We initialise the system with zero amplitude everywhere and an oscillatory drive in only waveguide. By evolving the system until a steady state is reached we can measure the transfer wavelength and calculate the power transfer ratio, comparing the results with Equations~\eqref{eqn:Coupling-parallel-waveguides-13} and~\eqref{eqn:Coupling-parallel-waveguides-14}.  We use a custom FDTD code that is described in detail later in Section~\ref{sec:Coupling-FDTD}. 
 Finite element method simulations could also be used to test the theory~\cite{romerosanchezPhononicsEngineeringControl2019a}). 

\begin{table}
    \centering
    \begin{tabular}{m{0.25\columnwidth}||m{0.15\columnwidth}|m{0.15\columnwidth}||m{0.15\columnwidth}|m{0.15\columnwidth}}
         & \multicolumn{2}{c}{$\lambda_\mathrm{trans}$ ($\mathrm{\upmu m})$} & \multicolumn{2}{c}{Transfer ratio}\\
        Waveguide widths ($\mathrm{\upmu m}$) & Coupled modes & FDTD  & Coupled modes & FDTD\\
        \hline
        30/30 & 1788 & 1988 & 100\% & 90\% \\
        30/40 & 1086 & 1140 & 13\% & 16\% 
    \end{tabular}
    \caption[Comparison of results from analytic CMT and FDTD simulation for two evanescently coupled parallel waveguides]{Comparison of results from analytic CMT and FDTD simulation for two evanescently coupled parallel waveguides. The coupled modes expressions for $\lambda_\mathrm{trans}$ and the transfer ratio are taken from Equations~\eqref{eqn:Coupling-parallel-waveguides-13} and~\eqref{eqn:Coupling-parallel-waveguides-14}.}
    \label{tab:Coupling-FDTD-parallel-waveguides}
\end{table}

The results of the simulation and our coupled modes expression are summarised in Table~\ref{tab:Coupling-FDTD-parallel-waveguides}. The relative error between the analytic theory and simulation is generally $\lesssim 10\%$, and we can observe the qualitative behaviour predicted by the theory. Specifically, we see oscillation of power between the two waveguides. The two insets highlight the phase lag between the wavefronts predicted by Eq.~\eqref{eqn:Coupling-parallel-waveguides-5}. We can see for example in inset (i) that the bottom waveguide appears `ahead' of the top waveguide at the position in space where there is a net power flow from it to the top waveguide. In contrast, inset (ii) shows how the relative phase between the waveguides is switched further down in the direction of travel, at a position where the net power flux is now from the top waveguide to the bottom waveguide.

We ascribe differences between the theory and simulation to possible changes in the eigenmodes (due to the coupling) in violation of the small-coupling approximation (the separation distance between the waveguides is only $20\,\mathrm{\upmu m}$, less than the waveguide widths themselves), and to numerical error inherent in the finite difference method arising due to the coarseness of the grid. Another source of error particular to the simulation is the imperfect nature of the reflectionless boundaries, which are meant to emulate a waveguide of infinite spatial extent (also known as a perfectly matched layer). This is likely why the transfer ratio observed for identical waveguides is less than unity.

\section{Longitudinal evanescent coupling}
\label{sec:Coupling-evanescent-coupling}

\begin{figure}[ht]
    \centering
    \includegraphics[width=0.99\linewidth]{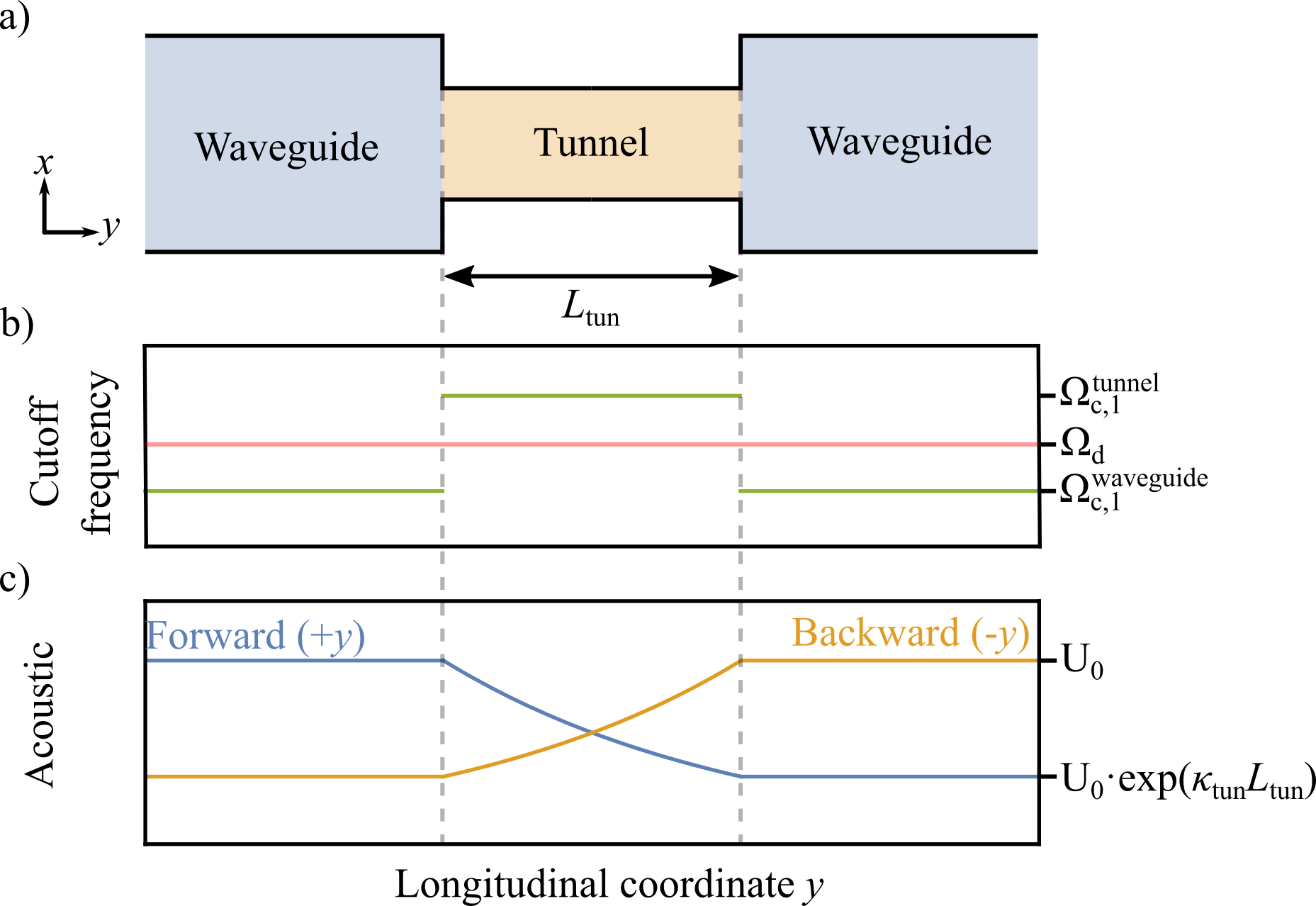}
    \caption[Operating principle of evanescent tunnel couplers]{Operating principle of evanescent tunnel couplers. (a) Diagram of an evanescent tunnel of length $L_\mathrm{tun}$ being used to couple two waveguides. (b) Illustration of the desired arrangement of cutoff frequencies, with the drive frequency sandwiched between the upper and lower cutoffs. (c) Illustration of the acoustic amplitudes for forward (blue) and reverse (orange) direction of travel through the tunnel. The amplitude decays exponentially in the tunnel with rate of decay $\kappa_\mathrm{tun}$. This value was derived in Section~\ref{sec:Coupling-evanescent-fields} as $\kappa_\mathrm{tun}=\sqrt{k_y^2-k_\mathrm{tun}^2}$, where $k_\mathrm{tun}$ is the maximum wavenumber that can be supported in the tunnel.}
    \label{fig:Coupling-evanescent-tunnel-diagram}
\end{figure}

The coupling of parallel waveguides discussed in Section~\ref{sec:Coupling-two-waveguides-example} relies on the lateral evanescent field extending into the lower density region. This method is appropriate for waveguides defined by acoustic impedance mismatches such as created by lift-off deposition of metals or variation of release hole size. However, there is an even more compact way to couple using evanescent fields. Evanescent fields can be created in the direction of wave propagation by abruptly narrowing the waveguide, so that the cutoff frequency rises above the operating frequency. This is shown in Fig.~\ref{fig:Coupling-evanescent-tunnel-diagram}. Evanescent waves in this `tunnel' region will exponentially decrease in amplitude as they propagate. By placing the tunnel between two waveguides, the corresponding evanescent fields can be made to overlap, producing coupling.

This is the idea behind evanescent tunnel couplers~\cite{mauranyapinTunnelingTransverseAcoustic2021}. These have the advantage over the side-by-side coupling examined in Section~\ref{sec:Coupling-two-waveguides-example} of being more compact, and obviating the need to fabricate differences in the membrane. In common with the side-by-side scheme, using evanescent waves for the coupling avoids resonances that can occur when using using propagating modes~\cite{fangOpticalTransductionRouting2016}.

Evanescent tunnel couplers are simply defined by width and length. The width determines the cutoff frequency $\Omega_{c,n}$ from Section~\ref{sec:Phononic-basics-Dispersion}:
\begin{equation}
    \Omega_{c,n}=\sqrt{\frac{\sigma}{\rho}}\frac{n\pi}{W_\mathrm{tun}}.
\end{equation}
Here $n$ is the guide wave mode number. 

\begin{figure}[ht]
    \centering
    \includegraphics[width=0.99\linewidth]{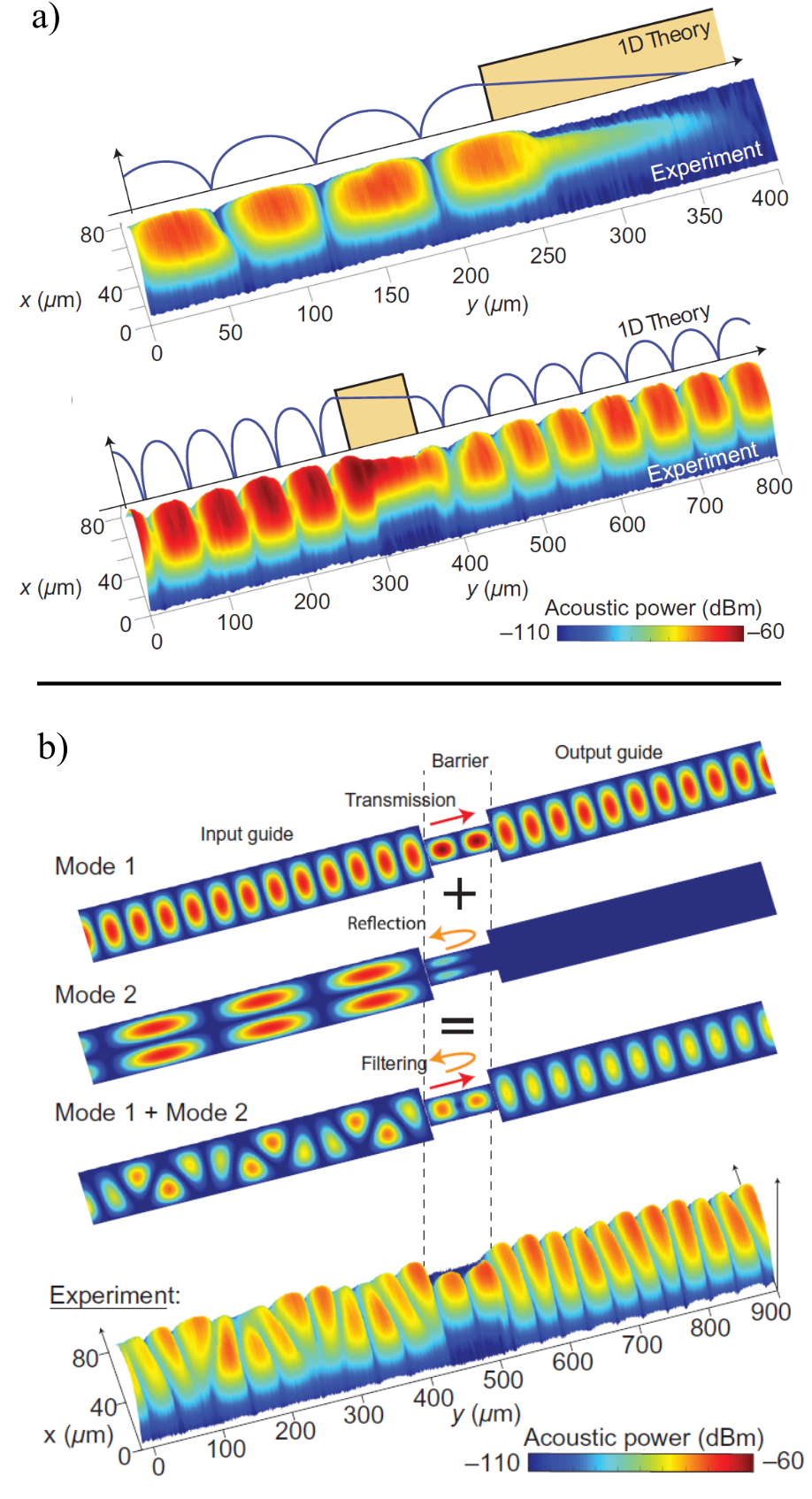}
    \caption[Experimental demonstration of evanescent tunnel couplers]{(a) Measurements of evanescent decay through a tunnel barrier. Top: an infinite tunnel barrier. Bottom: a finite tunnel barrier. (b) Mode filtering using a tunnel barrier. Top: theoretical prediction. Bottom: experimental results. Reproduced with permission from~\cite{mauranyapinTunnelingTransverseAcoustic2021}.}
    \label{fig:Coupling-mauranyapin2019-figs}
\end{figure}

An evanescent wave that propagates a distance $L$ will decrease in amplitude by $\exp(\kappa_\mathrm{tun L})$, where $\kappa_\mathrm{tun}$ is the exponential decay rate. This is shown experimentally in Fig.~\ref{fig:Coupling-mauranyapin2019-figs}(a). $\kappa_\mathrm{tun}$ can be calculated by rearranging Eq.~\eqref{eqn:Phononic-basics-dispersion-clamp}, the dispersion relationship for waves in a clamped-edges waveguide:
\begin{equation}
    \label{eqn:Coupling-tunnel-decay-rate}
    \kappa_\mathrm{tun}=\sqrt{\left(\frac{\pi}{W_\mathrm{tun}}\right)^2-\Omega_d^2\frac{\rho}{\sigma}}.
\end{equation}
Here $W_\mathrm{tun}$ is the width of the tunnel, $\Omega_d$ is the frequency at which waves are being driven, and $\sigma$ and $\rho$ are the tensile stress and the density of the membrane. 

Because different wave modes cut off at different frequencies, tunnels can be used as modal filters. This is shown in Fig.~\ref{fig:Coupling-mauranyapin2019-figs}(b) for the case of the first and second modes~\cite{mauranyapinTunnelingTransverseAcoustic2021}. In this experiment the tunnel was sized such that $\Omega_{c,1}^{\mathrm{waveguide}}<\Omega_{c,2}^\mathrm{waveguide}<\Omega_d$ in the waveguide, but $\Omega_{c,1}^\mathrm{tunnel}<\Omega_d<\Omega_{c,2}^\mathrm{tunnel}$ in the tunnel. Waves in the fundamental mode of the waveguide then propagated through the tunnel while higher order modes were exponentially filtered out.

A significant advantage of tunnel couplers is that they are directly compatible with highly-acoustically confining, highly-impedance mismatched suspended membranes. In contrast, coupling schemes like the parallel waveguides example in Section~\ref{sec:Coupling-two-waveguides-example} rely on the lateral evanescent field and therefore must have a comparatively low impedance contrast. To achieve such a low impedance mismatch with suspended membranes would require a more complex fabrication process and quite possibly perform worse---for example, waveguides could be defined by depositing metal but this may introduce surface losses~\cite{schmidFundamentalsNanomechanicalResonators2016}. In other words, evanescent tunnel couplers are the natural solution for creating coupling in suspended membrane devices. They can also be used to create resonators with high (and tailorable) quality factors, as we will now discuss.

\section{Input-output formalism}
\label{sec:Coupling-input-output}
\begin{figure}[ht]
    \centering
    \includegraphics[width=0.99\linewidth]{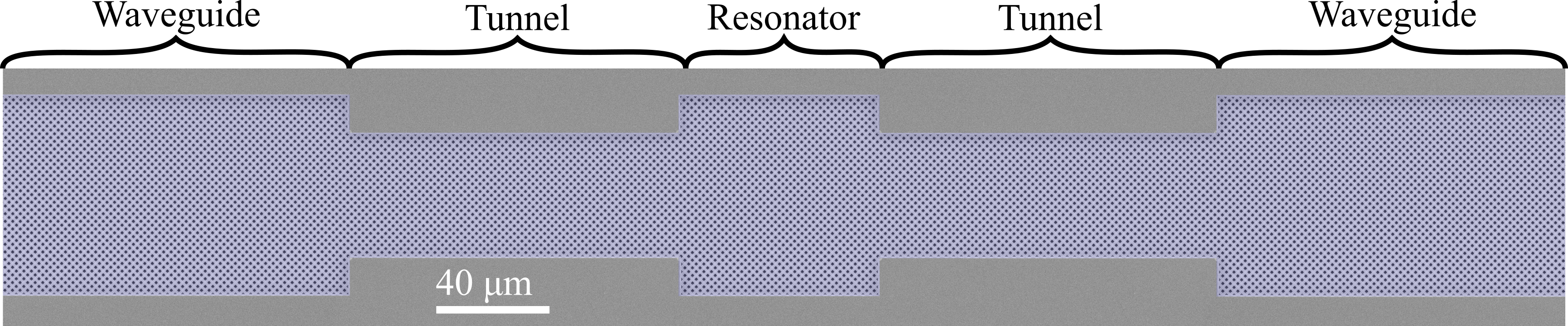}
    \caption[False-colour scanning electron micrograph of a phononic cavity coupled to two waveguides by evanescent tunnels]{False-colour scanning electron micrograph of a phononic cavity coupled to two waveguides by evanescent tunnels. Solid grey: unreleased silicon nitride. Blue: released membrane. This cavity was used to perform single frequency acoustic logic~\cite{romeroAcousticallyDrivenSinglefrequency2024}. By extending the tunnel lengths the evanescent coupling between the resonator mode and waveguide modes decreases, leading to cavity quality factors as high as 275,000~\cite{romeroAcousticallyDrivenSinglefrequency2024}.}
    \label{fig:Coupling-mechanical-logic-gate-SEM}
\end{figure}

Tunnel barriers allow distinct membrane resonators to be integrated into a phononic circuit. The basic recipe for this is to place two (or more) couplers on either side of a section of wider membrane, as demonstrated in Fig.~\ref{fig:Coupling-mechanical-logic-gate-SEM}. This system is the acoustic analog to a  Fabry-P\'erot cavity~\cite{romeroAcousticallyDrivenSinglefrequency2024}, which has a variety of applications including sensing, band pass filtering~\cite{venghausWavelengthFiltersFibre2014}, amplitude enhancement for readout~\cite{huangRoomtemperatureQuantumOptomechanics2024} and nonlinear enhancement~\cite{bischofbergerTheoreticalExperimentalStudy1979}.

The phononic version of a Fabry-P\'erot cavity presents a range of applications, for example an entirely acoustically coupled mechanical logic device~\cite{romeroAcousticallyDrivenSinglefrequency2024}, or signal processing functions (explored later in Section~\ref{sec:Coupling-examples-of-engineered-coupling}). Designing the cavity for these applications requires a way to model the flow of energy into and out of the resonator. In this section, we do this by translating to phononics the input-output formalism, which was originally developed optoelectronics and used in quantum optics for describing the behaviour of optical cavities~\cite{hausWavesFieldsOptoelectronics1984,yarivCoupledmodeTheoryGuidedwave1973, gardinerInputOutputDamped1985,wallsQuantumOptics2008,clerkIntroductionQuantumNoise2010}.

The input-output formalism, like CMT, is a useful model that makes generally acceptable approximations. Here we apply it to the case of a phononic resonator interacting with external fields. The model assumes that the couplings between those fields and the cavity are linear and constant with respect to frequency, and makes the rotating wave approximation~\cite{gardinerInputOutputDamped1985}. The constant coupling approximation is good when dealing with narrow band scenarios such as high quality factor cavities. The rotating wave approximation discards fast-rotating terms, which is justifiable when the external fields are oscillating near to the cavity resonance frequency and the coupling is weak.

\begin{figure}[ht]
    \centering
    \includegraphics[width=0.99\linewidth]{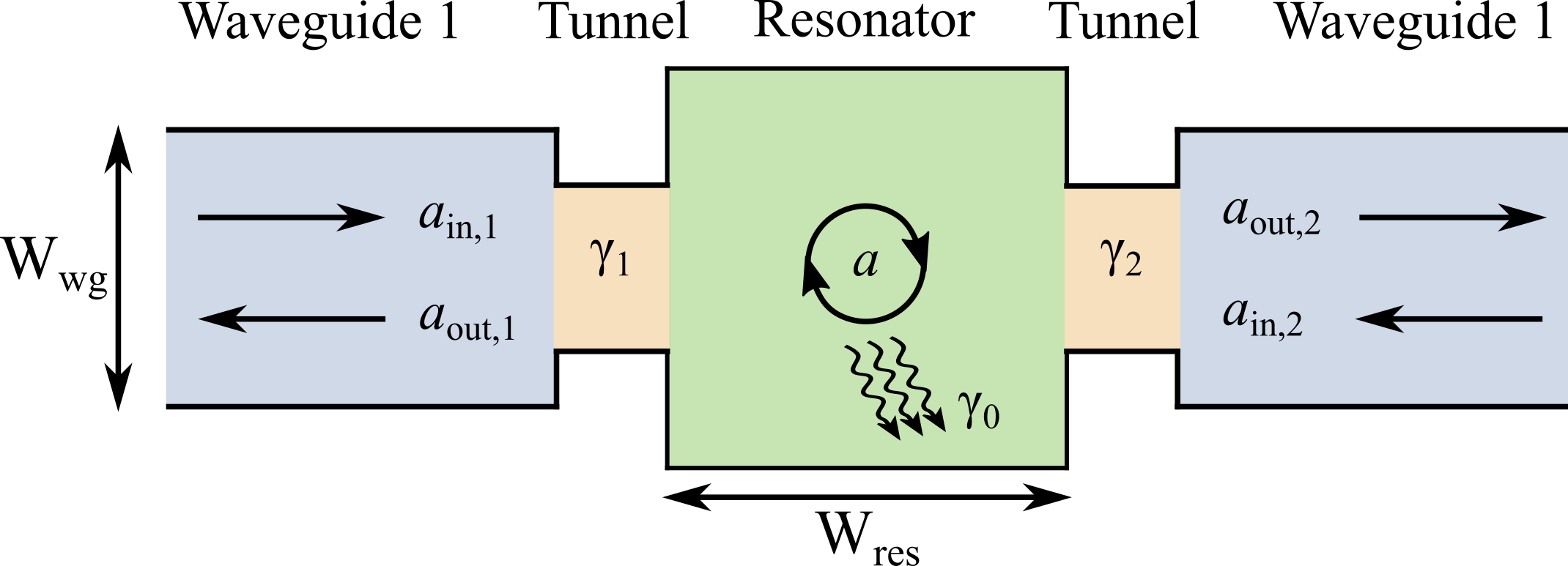}
    \caption[The input-output formalism for a membrane resonator]{The input-output formalism for a membrane resonator (green shading), connected to two waveguides (blue shading) by evanescent tunnels (yellow shading). Resonator has amplitude of vibration $a$ and is coupled to two waveguides with respective coupling rates $\gamma_i$. It is also coupled to the environment with loss rate $\gamma_0$. Each waveguide has two amplitudes which correspond to waves travelling towards and away from the resonator.}
    \label{fig:Coupling-input-output-diagram}
\end{figure}

As an example, consider a membrane resonator coupled to two waveguides as shown in Fig.~\ref{fig:Coupling-input-output-diagram}. The system is being driven at frequency $\Omega_d$ and the resonator has an eigenfrequency $\Omega_0$. The corresponding detuning is $\Delta=\Omega_d-\Omega_0$. Define $a$ as the amplitude of the acoustic field in the resonator, and $a_{\mathrm{in},i}$ and $a_{\mathrm{out},i}$ as the respective amplitudes of incoming and outgoing acoustic fields in waveguide $i$. The amplitudes are normalised similarly to Section~\ref{sec:Coupling-two-waveguides-example}, such that $|a|^2$ is the energy in the resonator while $|a_{\mathrm{in},i}|^2$ and $|a_{\mathrm{out},i}|^2$ correspond to the incoming and outgoing powers in the waveguides. Under the input-output formalism approximations, and in the absence of thermal noise driving, the resonator amplitude will follow the equation~\cite{aspelmeyerCavityOptomechanics2014}:

\begin{equation}
    \label{eqn:Coupling-input-output-equation}
    \dot{a}=-\frac{\gamma}{2}a+i\Delta a+\sum_{i=1}^{N}\sqrt{\gamma_i}a_\textrm{in,i}.
\end{equation}

Here $\gamma_1$ and $\gamma_2$ (with dimension $T^{-1}$) are the coupling rates to waveguides $1$ and $2$ respectively, and $\gamma_0$ is the intrinsic or internal loss rate describing coupling to the environment (which could for example correspond to air damping or thermoelastic damping in the membrane). The total loss rate $\gamma$ is equal to the sum of the couplings to the waveguides and the environment: $\gamma=\gamma_0 +\gamma_1+\gamma_2$. The situation can be generalised to resonators coupled to $N$ waveguides by considering additional coupling rates $\gamma_3,\,\gamma_4,\,\ldots$ and so on. If the resonator is not being driven we can easily derive from Eq.~\eqref{eqn:Coupling-input-output-equation} that the rate of energy loss is $\frac{\upd|a|^2}{\dt}=-\gamma|a|^2$, consistent with our previous experience that $\gamma/2$ is the rate of amplitude decay.

For efficient power routing the coupling rates should be engineered such that $\gamma_0$ is small compared to the coupling rates $\gamma_1$ and $\gamma_2$; that is, the decay rate of the resonator should be dominated by its coupling to the waveguides. This can easily be achieved in practice, as the intrinsic Q factor of such membrane resonators can be well in excess of $10^5$~
\cite{thompsonStrongDispersiveCoupling2008}, corresponding to an intrinsic dissipation rate $\gamma_0$ in the Hertz range. Meanwhile the coupling rate to waveguides can be reliably and controllably be made several orders of magnitude larger (see Fig.~\ref{fig:Coupling-yQ_diffCalcMethods} for instance), such that $\gamma_\mathrm{tot}=\gamma_0+\gamma_1+\gamma_2\approx2\gamma_1$ (for symmetric waveguide couplers where $\gamma_1=\gamma_2$).

As illustrated in Fig.~\ref{fig:Coupling-input-output-diagram}, the acoustic field in each waveguide is a superposition of waves travelling towards and away from the resonator. Because we assume the waveguides have infinite spatial extent, any waves travelling towards the resonator ($a_\mathrm{{in,1}},\,a_\mathrm{{in,2}}$) must originate from external sources; in contrast waves travelling away can either come from the resonator or be reflected inputs. It can be shown by applying time reversal symmetry that the loss rate for each waveguide also defines its reflection coefficient~\cite{gardinerInputOutputDamped1985}, such that the outgoing amplitude in each waveguide is simply:
\begin{equation}
    \label{eqn:Coupling-input-output-reflection}
    a_{i,out}=a_{i,in}-\sqrt{\gamma_i}a.
\end{equation}

Equations~\eqref{eqn:Coupling-input-output-equation} and~\eqref{eqn:Coupling-input-output-reflection} can be used to find the coupling rates required for various functionalities. We will now show some examples.

\subsection{Example: Impedance matching}
\label{sec:Coupling-input-output-impedance-matching}

Consider a two-port resonator similar to Fig.~\ref{fig:Coupling-mechanical-logic-gate-SEM} with input in waveguide 1. No back-reflection demands $a_{1,\mathrm{out}}=0$, which by Eq.~\eqref{eqn:Coupling-input-output-reflection} implies $a_{1,\mathrm{in}}=\sqrt{\gamma_1}a$. This is the condition for perfect destructive interference with the outgoing field. Input only in waveguide 1 means $a_{2,\mathrm{in}}=0$. Substituting both these conditions into Eq.~\eqref{eqn:Coupling-input-output-equation} yields that $\gamma_1=\gamma_0+\gamma_2$, where $\gamma_0$ is the intrinsic loss of the resonator. This is the condition of impedance matching of input and output.

\subsection{Example: Intracavity amplitude enhancement}
A resonator can be used to locally increase the amplitude of motion along a waveguide. This can be useful for applications such as sensing, readout, or logic operations~\cite{schmidFundamentalsNanomechanicalResonators2016,sunOpticalRingResonators2011,hatanakaBroadbandReconfigurableLogic2017,romeroAcousticallyDrivenSinglefrequency2024}.

Consider a square resonator of width $W_\mathrm{res}$ bookended by two waveguides of width $W_\mathrm{wg}$, as illustrated in Fig.~\ref{fig:Coupling-input-output-diagram}. The resonator is impedance matched so that $\gamma_1=\gamma_2=\gamma/2$, were $\gamma_i$ is the coupling rate of the resonator to the $i^\mathrm{th}$ waveguide and $\gamma$ is the total decay rate of the resonator (we ignore intrinsic losses, i.e. $\gamma_0\approx0$).

Consider the case where there is a drive in one waveguide ($a_\mathrm{in}=a_\mathrm{in,1}$ and $a_\mathrm{in,2}=0$), and unidirectional propagation. The drive frequency $\Omega$ is in the single mode regime of the waveguides and equals the resonator fundamental eigenfrequency ($\Delta=0$). Plugging the assumptions into Eq.~\eqref{eqn:Coupling-input-output-reflection} we find:
\begin{equation}
    \label{eqn:Coupling-input-output-energyvspower-enhancement}
    \frac{|a|^2}{|a_\mathrm{in}|^2}=\frac{2}{\gamma}.
\end{equation}
This is equivalent to noting that the circulating power within the resonator is enhanced by a factor of $\mathcal{F}/(2\pi)$ where $\mathcal{F}$ is the resonator finesse~\cite{fowlesIntroductionModernOptics1989}.

Because both the waveguides and resonator are operating in the single mode regime (i.e. vibrating in their fundamental modes), we have simple expressions for the displacement in the waveguide and the resonator: $u_\mathrm{wg}(t,x,y)=Z_{0,\mathrm{wg}}e^{i(\Omega t-k_yy)}\sin(\pi x/W_\mathrm{wg})$, and $u_\mathrm{res}(t,x,y)=Z_{0,\mathrm{res}}e^{i(\Omega t+\Delta\varphi)}\sin(\pi x/W_\mathrm{res})\sin(\pi y/W_\mathrm{res}).$ Here $Z_\mathrm{0,\mathrm{wg}}$ and $Z_{0,\mathrm{res}}$ are the physical amplitudes (in dimensions of length) in the waveguide and resonator respectively. The phase shift $\Delta\varphi$ represents the relative phase lag between the drive waveguide and the resonator. 

We want to know the amplitude enhancement inside the resonator, that is $Z_{0,\mathrm{res}}/Z_\mathrm{0,\mathrm{wg}}$. We can substitute analytic expressions for the energy in the resonator and power flow in the waveguide built into the normalisation conditions for $a$ and $a_\mathrm{in}$: 
\begin{equation}
     \frac{|a|^2}{|a_\mathrm{in}|^2}=\frac{\frac{1}{2}\rho \iint \left|\frac{\partial u_\mathrm{res}(t,x,y)}{\partial t}\right|^2\,\dx\dy}{v_g\times\frac{1}{2}\rho\int\left|\frac{\partial u_\mathrm{wg}(t,x,y)}{\partial t}\right|^2\,\dx}=\frac{\frac{1}{2}\rho\Omega^2Z_{0,\mathrm{res}}^2\frac{1}{4}W_\mathrm{res}^2}{v_g\frac{1}{2}\rho\Omega^2Z_{0,\mathrm{wg}}^2\frac{1}{2}W_\mathrm{wg}}.
\end{equation}
Here $v_g$ is the group velocity in the waveguide from Eq.~\eqref{eqn:Phononic-basics-group-velocity}.

Combining the two equations reveals that the physical amplitude enhancement is inversely proportional to the square root of the total decay rate:
\begin{equation}
    \label{eqn:Coupling-intracavity-amplitude-enhancement}
    \frac{Z_{0,\mathrm{res}}}{Z_{0,\mathrm{wg}}}\simeq2\sqrt{\frac{v_gW_\mathrm{wg}}{W_\mathrm{res}^2\gamma}}.
\end{equation}
Substituting $\gamma=\Omega/Q$ we can see that the energy enhancement in the resonator scales $Q$-fold, and the amplitude enhancement scales with $\sqrt{Q}$. This can be a significant and useful enhancement; for example, consider a silicon nitride resonator used by our group for mechanical logic which had a measured $Q\simeq275,000$, corresponding to an amplitude enhancement factor of over $500$-fold~\cite{romeroAcousticallyDrivenSinglefrequency2024}. This enhancement allowed the critical amplitude for Duffing nonlinearity to be reached at lower drive powers, therefore reducing the power consumption of the mechanical logic gate.

Another insightful example is considering a resonator of the same width as the waveguide, $W_\mathrm{wg}=W_\mathrm{res}=W$. In this case Eq.~\eqref{eqn:Coupling-intracavity-amplitude-enhancement} transforms to:
\begin{equation}
    \label{eqn:Coupling-intracavity-amplitude-enhancement-2}
    \frac{Z_{0,\mathrm{res}}}{Z_{0,\mathrm{wg}}}\simeq2\sqrt{\frac{v_g}{W\gamma}}.
\end{equation}

In the limit where the frequency is well above cutoff, the group velocity in Eqs.~\eqref{eqn:Coupling-intracavity-amplitude-enhancement} and~\eqref{eqn:Coupling-intracavity-amplitude-enhancement-2} can be effectively approximated as $v_g=\sqrt{\sigma/\rho}$.

\subsection{Example: Filtering}
A phononic cavity can be placed into a waveguide to act as a notch pass filter. Consider a filter composed of a cavity evanescently coupled to two waveguides as in Fig.~\ref{fig:Coupling-input-output-diagram}. Assuming it operates without reflection in the steady state ($a_{1,\mathrm{out}}=a_{2,\mathrm{in}}=\dot{a}=0$), we can solve Equations~\eqref{eqn:Coupling-input-output-equation} and $\eqref{eqn:Coupling-input-output-reflection}$ to obtain $\gamma_1=\gamma_0+\gamma_2$. This the same impedance matching expression from Section~\ref{sec:Coupling-input-output-impedance-matching}. The total loss rate of the resonator is then $\gamma_\mathrm{total}=\gamma_0+\gamma_1+\gamma_2\simeq2\gamma_\mathrm{1}$, assuming the coupling rate is engineered to be much greater than the intrinsic damping rate. This total loss rate $\gamma$ equals the full width at half maximum (in angular units) of the filter.

This system presents a widely customisable filter. We can easily change the width of the notch by extending or shortening the evanescent tunnel lengths (i.e. changing $\gamma_1$), and change the location of the notch (i.e. the resonance frequency) by enlarging or shrinking the size of the resonator.

\section{Numerically finding coupling rates}
\label{sec:Coupling-numerically-finding-coupling-rates}

To design useful membrane devices one needs to know the relationship between device performance and device geometry. We have seen in Section~\ref{sec:Coupling-input-output} how the input-output formalism links device performance to coupling rates, and in Section~\ref{sec:Coupling-two-waveguides-example} how coupling rates can be calculated from modal overlap integrals. This technically completes the link between performance and geometry because the overlap integrals are defined by the device geometry. However, except for specific situations like the coupled waveguides in Section~\ref{sec:Coupling-two-waveguides-example}, computing the integrals is not analytically possible. To solve that problem, here we explain numerical methods of calculating coupling rates that can be used for arbitrary geometries.

Numerical methods are needed because because phononic membranes can be fabricated into a large variety of shapes that resist analytic treatment. As described in Section~\ref{eqn:Phononic-basics-dispersion-clamp}, the acoustic confinement provided by the impedance mismatch between the suspended membrane and the bulk substrate is very large (from the combination of contrasting speeds of sound and from the large difference in mass between the membrane and substrate~\cite{sementilliNanomechanicalDissipationStrain2022}) enabling low-loss
acoustic wave transmission as experimentally demonstrated
in our earlier work with silicon nitride membranes~\cite{romeroPropagationImagingMechanical2019}. This low loss---much lower than what is achievable for conventional semiconductor photonic high-index waveguides~\cite{thylenIntegratedPhotonics21st2014} or phononic high-index contrast circuits\cite{sarabalisGuidedAcousticOptical2016,mayorGigahertzPhononicIntegrated2021}---means that sharp bends and abrupt changes in geometry can be realised while still achieving extremely high quality factors. Indeed, a typical refractive index mismatch in silicon photonics is $n_\mathrm{Si}-n_\mathrm{vacuum}\simeq3.5-1=2.5$, corresponding to an impedance mismatch (for a plane electromagnetic wave in a nonconductive, nonmagnetic material) of $1/2.5=0.4$. In contrast, we saw in Section~\ref{sec:Phononic-basics-boundary-conditions} that silicon nitride membranes can have more pronounced impedance mismatches with the substrate of $Z_\mathrm{mem}/Z_\mathrm{sub}\simeq0.06$, even before considering thickness differences. The fabrication scope for membrane phononics is therefore more closely akin to what can be achieved in the optical realm with photonic crystal waveguides, where lossless $90$ degree bends are achievable~\cite{mekisHighTransmissionSharp1996}, or in the microwave regime with hollow conductive waveguides which impose that the parallel electric field components be zero at the boundaries~\cite{khanMicrowaveEngineeringConcepts2014}.

\begin{figure*}[ht]
    \centering
    \includegraphics[width=0.8\linewidth]{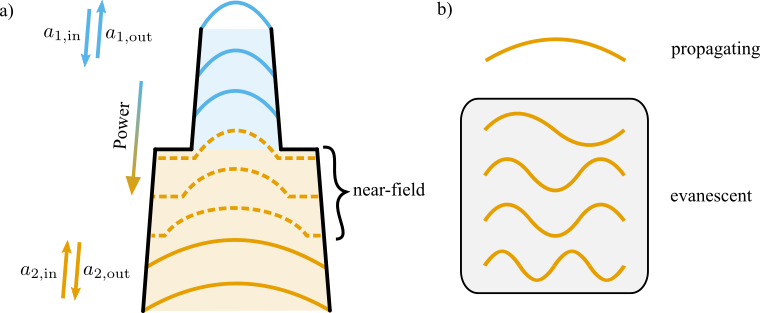}
    \caption[Evanescent mode contributions to coupling at waveguide junctions]{Evanescent mode contributions to coupling at waveguide junctions. (a) Diagram of single mode acoustic waves travelling through an abrupt change in waveguide width, from smaller (blue) to wider (orange). In the far-field region energy propagates in the form of fundamental eigenmodes (solid lines) and is fully defined by four propagating terms ($a_{1,\mathrm{in}},\,a_{1,\mathrm{out}},\,a_{2,\mathrm{in}}$ and $a_{2,\mathrm{out}}$), but near the junction the modeshape does not match any analytic eigenmode (dashed lines). (b) In that near-field region, energy transport is performed by both propagating and higher-order evanescent modes.}
    \label{fig:Coupling-linear-combination-of-modes}
\end{figure*}

In the parallel waveguides example from Section~\ref{sec:Coupling-two-waveguides-example},  symmetries and other simplifications allowed us to calculate the coupling rate analytically. However, this is not usually possible, even for simple geometries.

For example, consider two butt-coupled waveguides of different widths, as illustrated in Fig.~\ref{fig:Coupling-linear-combination-of-modes}(a). CMT tells us that the coupling between the waveguides will be proportional to the overlap integral of each waveguide's mode of oscillation. In the parallel waveguides example we could describe the perturbative field using a single travelling wave eigenmode. However, because in this case the waveguide width changes abruptly (and the eigenmodes are solutions for a fixed-width waveguide), no single eigenmode exactly represents the perturbation field. The perturbation can still be expanded into eigenmodes, but the expansion will feature an infinite series of higher-order modes as shown in Fig.~\ref{fig:Coupling-linear-combination-of-modes}(b). The higher-order modes will propagate evanescently and do not feature in the far-field transport, but in the near-field they do play a significant role. Therefore, an analytic expression for the coupling rate between the two far-field modes would actually require summing over infinitely many overlap integrals between higher order modes~\cite{muehleisenModalCouplingAcoustic2002}.

\begin{figure}[ht]
    \centering
    \includegraphics[width=0.99\linewidth]{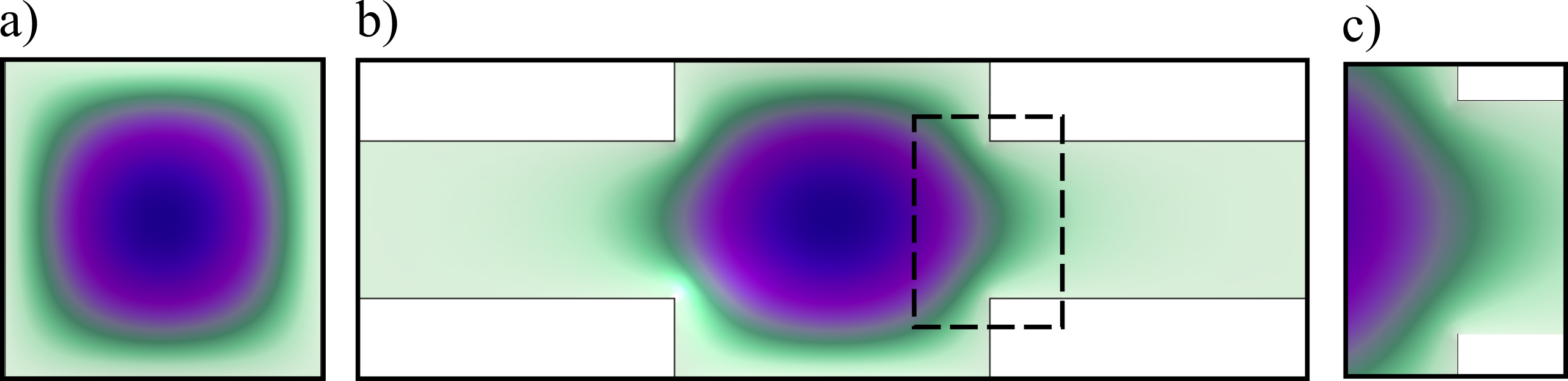}
    \caption[Change in resonator eigenmode from the addition of an evanescent coupler]{Change in resonator eigenmode from the addition of an evanescent coupler, calculated from finite element modelling with COMSOL Multiphysics~\cite{COMSOLMultiphysics2020}. (a) Fundamental mode of a 100~{\textmu}m square silicon nitride resonator (b) Fundamental mode of the same resonator attached to two 30~{\textmu}m wide couplers. Dashed lines highlight detail area. (c) Detail of the mode `leaking' into the coupler. In this simulation the membrane is 60~nm thick, has density $3200\,\mathrm{kg\cdot m^{-3}}$, external stress of 1~GPa, and Young's modulus of 230~GPa.}
    \label{fig:Coupling-modeshape-with-constriction}
\end{figure}

The same reasoning applies to other geometries. For example, in the coupling of a resonator to a waveguide as seen in Fig.~\ref{fig:Coupling-modeshape-with-constriction}, the resonator modeshape `leaks' into the tunnel region and kinks around the 90 degree corners. There is no good simple expression for this modeshape, so again an analytic approach would require an expansion over known eigenmodes.

The first numerical method we explain is simply to calculate the overlap integrals with numerically calculated modeshapes, avoiding the need to expand into another basis. We then explain two methods that avoid using overlap integrals entirely: finite difference time domain simulations, and finite element simulations with damping boundary conditions. Finally, we show how transfer matrix theory can be used to model acoustic wave propagation through a heterogeneous chain of phononic waveguides, tunnels, and resonators.

The following subsections explain each of these four calculation methods.



\subsection{FEM simulations and overlap integration}
\label{sec:Coupling-resonator-waveguide-analytic-coupling}

Our first approach is to extend our analytic expression for the coupling rates that we derived in Section~\ref{sec:Coupling-two-waveguides-example}, Equations~\eqref{eqn:Coupling-parallel-waveguides-8.2} and~\eqref{eqn:Coupling-parallel-waveguides-8.3}, to the situation of a resonator coupled to a waveguide via an in-line evanescent tunnel. Unlike Section~\ref{sec:Coupling-two-waveguides-example} where we considered varying densities and modeshapes extending evanescently beyond guiding regions, here we assume a constant density $\rho$ and clamped boundary conditions, consistent with confinement by the impedance mismatch between suspended and unsuspended material. In this context, the natural extension of Eqs.~\eqref{eqn:Coupling-parallel-waveguides-8.2} and~\eqref{eqn:Coupling-parallel-waveguides-8.3} is to define the coupling rate of the resonator to the waveguide as:
\begin{equation}
    \label{eqn:Coupling-FEM-overlap-integration-1}
    \sqrt{\gamma_\mathrm{wg}}=\frac{ \rho\Omega^3}{4}\iint\tilde{u}_\mathrm{res}\tilde{u}_\mathrm{wg}\,\dx\dy,
\end{equation}
where the integral is over the coupling region, which should correspond to the area inside the tunnel barrier (if it does not, i.e. the tunnel is short enough that non-negligible acoustic fields extend through it and out the other side, we are departing from the perturbative limit upon which CMT is based). The amplitudes $\tilde{u}_\mathrm{res}$ and $\tilde{u}_\mathrm{wg}$ are normalised against the energy in the resonator and the power in the waveguide. For example, the normalised amplitude in the resonator $\tilde{u}_\mathrm{res}(x,y)$ is:
\begin{equation}
    \label{eqn:Coupling-FEM-overlap-integration-2}
    \tilde{u}_\mathrm{res}(x,y)=\frac{\psi_\mathrm{res}(x,y)}{\sqrt{\frac{1}{2}\rho\Omega^2\iint\psi_\mathrm{res}(x,y)^2\,\dx\dy}},
\end{equation}
where $\psi_\mathrm{res}(x,y)$ is the dimensionless, normalised ($\max(|\psi_\mathrm{res}(x,y)|)=1$) modeshape of the resonator eigenmode, and the double integral is over the resonator area. 

To normalise the amplitude in the waveguide, we note that the power carried by the waveguide $\langle P_\mathrm{wg}\rangle$ is equal to the integrated linear energy density across its cross section, multiplied by the wave's group velocity:
\begin{equation}
    \label{eqn:Coupling-FEM-overlap-integration-2.5}
    \langle P_\mathrm{wg}\rangle=v_g\times\frac{1}{2}\rho \Omega^2\int_0^{W_\mathrm{wg}}u_\mathrm{wg}(x,y_0)^2\,\dx,
\end{equation}
where $u_\mathrm{wg}$ is the flexural displacement of the waveguide, $W_\mathrm{wg}$ is the waveguide width, and $y_0$ is some coordinate far enough away from the resonator that the coupling is negligible and the total power travelling through the waveguide can be evaluated with a transverse line integral. Given Eq.~\eqref{eqn:Coupling-FEM-overlap-integration-2.5} we can write the normalised waveguide amplitude $\tilde{u}_\mathrm{wg}=u_\mathrm{wg}/\langle P_\mathrm{wg}\rangle$ as:
\begin{equation}
    \label{eqn:Coupling-FEM-overlap-integration-3}
    \tilde{u}_\mathrm{wg}(x,y)=\frac{\psi_\mathrm{wg}(x,y)}{\sqrt{v_g\times\frac{1}{2}\rho\Omega^2\int_0^{W_\mathrm{wg}} \psi(x,y_0)^2\,\dx}},
\end{equation}
where $\psi_\mathrm{wg}$ is the dimensionless, normalised ($\max(|\psi_\mathrm{wg}(x,y)|)=1$) waveguide modeshape.

\begin{figure}[ht]
    \centering
    \includegraphics[width=0.99\linewidth]{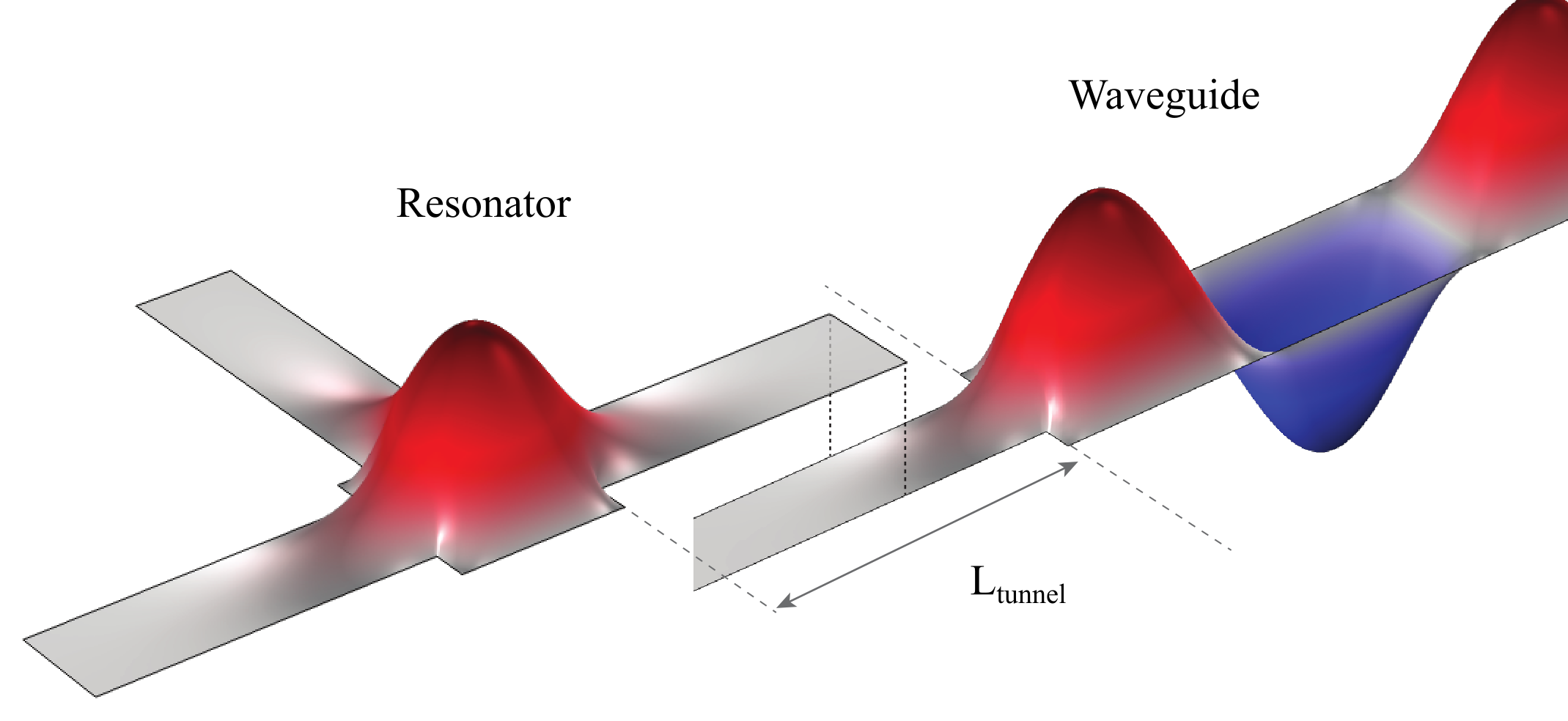}
    \caption[Principle of FEM-aided overlap integrals]{Principle of FEM-aided overlap integrals. The resonator modeshape and waveguide modeshapes (here calculated for a three-port power splitter) are first independently calculated as if there was an infinite evanescent tunnel. Then the two exported modeshapes are overlaid to create a particular tunnel length, and the overlap integral~\eqref{eqn:Coupling-FEM-overlap-integration-1} is calculated.}
    \label{fig:Coupling-overlap-integral-principle}
\end{figure}

In contrast to Section~\ref{sec:Coupling-two-waveguides-example}, in Eq.~\eqref{eqn:Coupling-FEM-overlap-integration-1} we do not subtract a reference density $\rho_0$ since the density is constant everywhere. Another point of contrast is that the waveguide mode is not univariate but now depends on both $x$ and $y$. This is because it breaks longitudinal symmetry where it evanescently extends into the tunnel coupler, as shown in Fig.~\ref{fig:Coupling-overlap-integral-principle}.

The modeshapes $\psi_\mathrm{res}$ and $\psi_\mathrm{wg}$ are difficult to calculate analytically so we simulate them using finite element modelling software (COMSOL Multiphysics). In keeping with the small-coupling approximation we simulate the resonator and waveguide modes separately, using the boundary condition of an infinitely long tunnel. We then export the modeshapes to a Mathematica script where we can represent different tunnel lengths as an offset to the $y$ coordinate. As illustrated in Fig.~\ref{fig:Coupling-overlap-integral-principle}, this allows us to calculate the coupling rate from Eq.~\eqref{eqn:Coupling-FEM-overlap-integration-1} for varying tunnel lengths.

\subsection{Complex eigenfrequencies}
\label{sec:Coupling-complex-eigenfrequency}
A perk of using FEM simulations is that one can avoid calculating overlap integrals entirely. This can be done by creating reflectionless boundaries on the simulation domain, for example by using a perfectly matched layer~\cite{berengerPerfectlyMatchedLayer1994} or using Rayleigh damping~\cite{liuFormulationRayleighDamping1995,hirschDirectionalEmissionOnchip2024}. This causes energy to be lost from the model, which corresponds to the system having a complex eigenfrequency $\Omega = \sqrt{\Omega_0^2-(\gamma/2)^2}+i(\gamma/2)$~\cite{schmidFundamentalsNanomechanicalResonators2016}. Here $\Omega_0=\sqrt{k_\mathrm{eff}/m_\mathrm{eff}}$ is the undamped eigenfrequency and $\gamma$ is the damping rate. The quality factor and damping force coefficient can be obtained from the complex eigenfrequency as~\cite{schmidFundamentalsNanomechanicalResonators2016,COMSOLMultiphysics2020}:
\begin{equation}
    \label{eqn:Coupling-Q-from-complex-eigenfrequency}
    Q=\frac{\Omega_0}{\gamma}=\frac{\Re(\Omega)}{2\Im(\Omega)}.
\end{equation}
where $\Re(\Omega)$ and $\Im(\Omega)$ respectively correspond to the real and imaginary parts of the complex eigenfrequency. 
Eq.~\eqref{eqn:Coupling-Q-from-complex-eigenfrequency} is very useful as it allows rapid prototyping by sweeping simulation parameters and immediately observing the coupling rates without having to perform any overlap integrals.

\begin{figure}[ht]
    \centering
    \includegraphics[width=0.99\linewidth]{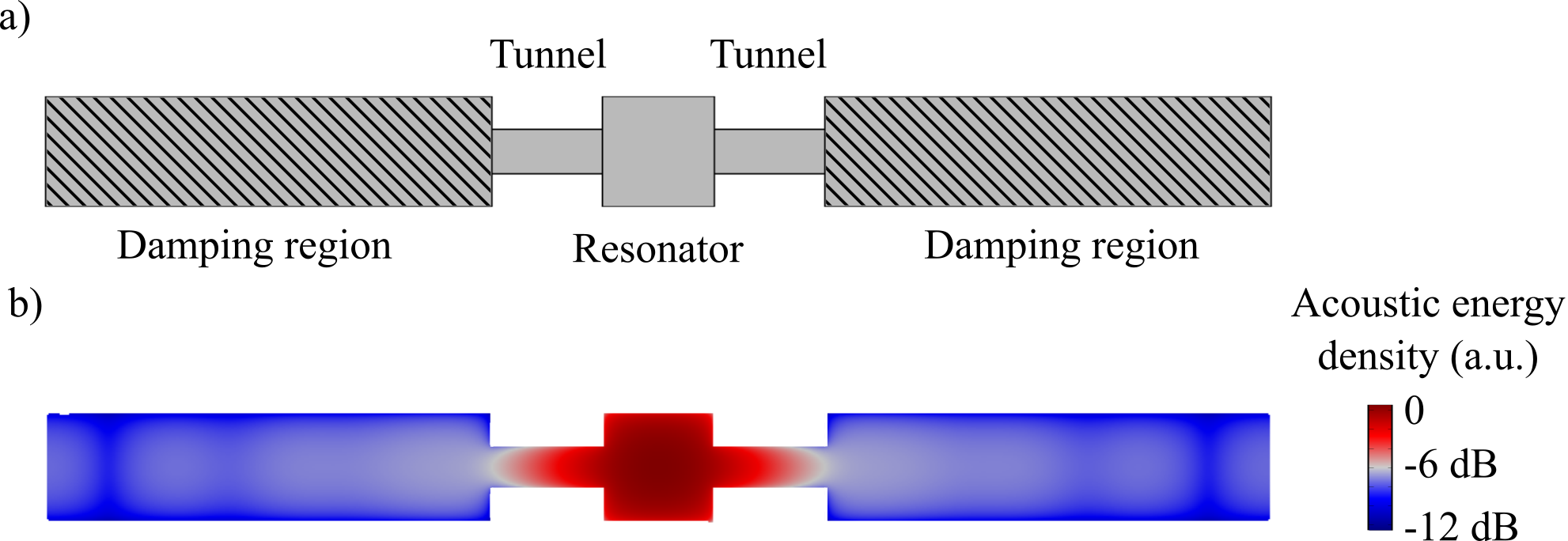}
    \caption[Finite element modelling of a damped waveguide coupled acoustic resonator]{Finite element modelling of a damped waveguide coupled acoustic resonator. (a) The model consists of a resonator out-coupled through two evanescent tunnels to two waveguides. Geometrical parameters such as the tunnels lengths and resonator size can be freely modified. Rayleigh damping is added to the model in the region of the out-coupling waveguides. (b) Distribution of kinetic energy density of the fundamental resonator mode, plotted on a logarithmic scale. One can observe exponential decay through the tunnels characteristic of evanescent waves, followed by damping of the excited travelling waves in the waveguides due to the Rayleigh damping.}
    \label{fig:Coupling-complex-eigenfrequency-diagram}
\end{figure}

Figure~\ref{fig:Coupling-complex-eigenfrequency-diagram} illustrates an example finite element simulation performed in COMSOL Multiphysics, where we include damping so that the resonator eigenfrequency is complex. 

\subsection{Time domain simulations}
\label{sec:Coupling-FDTD}
\begin{figure}[ht]
    \centering
    \includegraphics[width=0.99\linewidth]{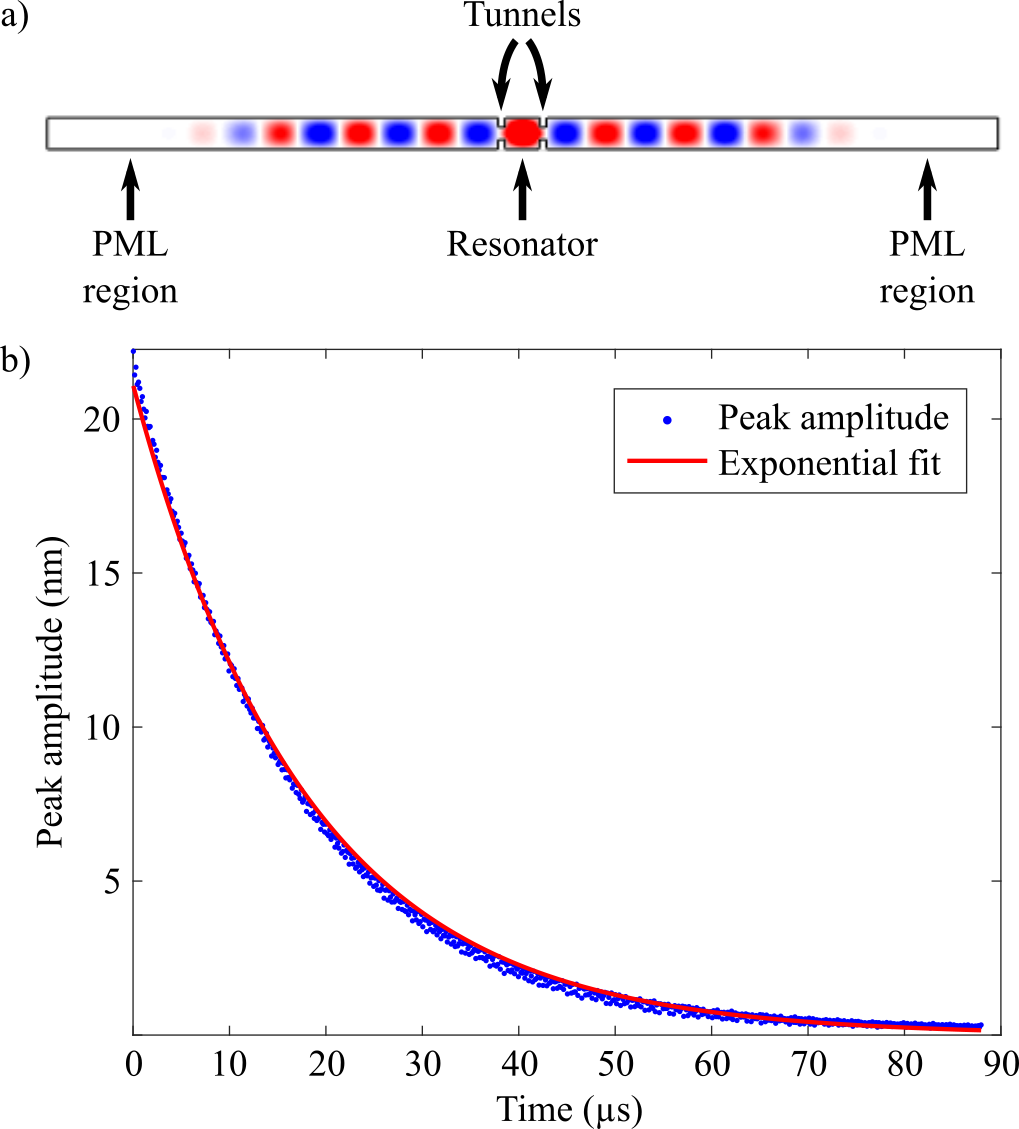}
    \caption[FDTD simulation of ringdown in a two-port resonator]{FDTD simulation of ringdown in a two-port resonator. (a) Surface plot of displacement at a simulation time step, showing decay of waves in the perfectly matched layer (PML). (b) Plot of the maximum displacements per cycle in the resonator. The maximum displacements follow an exponential fit as predicted by Eq.~\eqref{eqn:Coupling-exponential-decay-damping-rate}. Note that this simulation does not include Duffing nonlinearity so the inferred decay rate is independent of the starting amplitude. Simulation parameters: Square resonator has the same width as the waveguide of $50\,\mathrm{\upmu m}$, tunnel length $10\,\mathrm{\mu m}$ and width $20\,\mathrm{\upmu m}$, $\sigma=1\,\mathrm{GPa}$, $\rho=3100\,\mathrm{kg\cdot m^{-3}}$.}
    \label{fig:FDTD-decayrate-principle}
\end{figure}

Alternatively, instead of solving for the eigenfrequencies---a frequency-domain solution---one can simulate the decay of the resonator in the time domain, such as with a finite-difference time-domain (FDTD) code. This is the simulated version of a ringdown measurement~\cite{schmidFundamentalsNanomechanicalResonators2016,sementilliLowDissipationNanomechanicalDevices2025}. In this method the system is initialised with with nonzero displacement in the resonator and no external drive. Under these circumstances, we saw in Section~\ref{sec:Phononic-basics-lumped-element-model} that the maximum displacement per cycle of the damped, undriven resonator $Z(t)$ will exponentially decay at a rate proportional to the total coupling rate $\gamma$:
\begin{equation}
    \label{eqn:Coupling-exponential-decay-damping-rate}
    Z(t)\propto\exp\left(-\frac{\gamma}{2}t\right).
\end{equation}
An example simulation exhibiting this exponential decay in amplitude is shown in Fig.~\ref{fig:FDTD-decayrate-principle}.

We use a homemade FDTD code for our simulations. It solves the damped linear wave equation:
\begin{equation}
    \label{eqn:Coupling-linear-operator-wave-equation}
    \mathcal{L}u(t,x,y)=0,
\end{equation}
where $\mathcal{L}$ is the differential operator:
\begin{equation}
    \label{eqn:Coupling-linear-wave-operator-definition}
    \mathcal{L}(\cdot)=\rho\frac{\partial^2\cdot}{\partial t^2}+\rho\gamma\frac{\partial\cdot}{\partial t}-\sigma\nabla^2\cdot
\end{equation}
with $\rho$ the density and $\sigma$ the tensile stress as defined previously.

\begin{figure}[ht]
    \centering
    \includegraphics[width=0.98\linewidth]{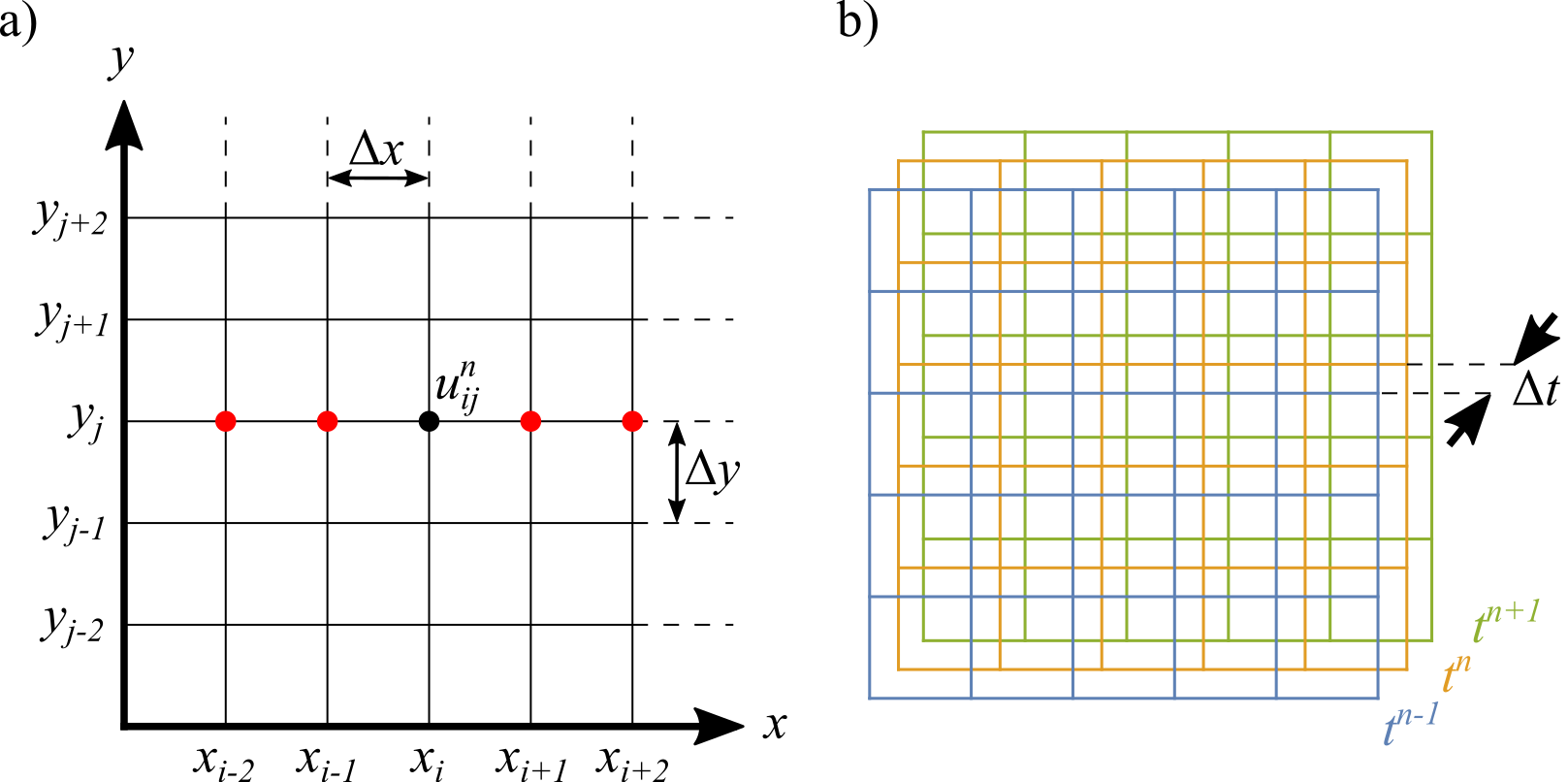}
    \caption[The finite difference time domain method]{The finite difference time domain (FDTD) method. (a) For each timestep the spatial dimensions $x$ and $y$ are discretised into a regular grid with respective grid spacings $\Delta x$ and $\Delta y$. For each combination of grid indices $i,j,n$ the solution is calculated and stored as a value $u^n_{ij}$. Derivatives are evaluated as linear combinations of stored solutions; for example, the red dots indicate the values that would be used to evaluate $\partial^2u^n_{ij}/\partial x^2$ $ $using a second-order, centred-differencing scheme (Eq.~\eqref{eqn:Coupling-FDTD-second-order-centred-difference-scheme}). (b) The time domain is also discretised with timestep $\Delta t$ in a mathematically equivalent way. The full domain over $t$, $x$ and $y$ can be pictured as layers of two-dimensional spatial matrices, each layer corresponding to a different timestep. Here we illustrate the $t^\mathrm{n-1}$, $t^\mathrm{n}$ and $t^\mathrm{n+1}$ matrices in blue, orange and green respectively.}
    \label{fig:Coupling-FDTD-code-explanation}
\end{figure}

The FDTD method converts $\mathcal{L}$ into a set of operations that can be performed on a computer by discretising the spatial and time domains into a three dimensional grid, as illustrated in Fig.~\ref{fig:Coupling-FDTD-code-explanation}. Instead of a continuous solution $u(t,x,y)$ we solve for a discretised solution $u_{ij}^n$. Our notation is that $u_{ij}^n$ means the value of the displacement at the $i^\mathrm{th}$ $x$-coordinate, $j^\mathrm{th}$ y-coordinate, and $n^\mathrm{th}$ time coordinate. Physical properties of the material can be replaced with discretised versions, for example density $\rho_{ij}\equiv\rho(x_{i},y_{j})$.

The analytic derivatives used in Eq.~\eqref{eqn:Coupling-linear-wave-operator-definition} are defined as:
\begin{equation}
    \label{eqn:Coupling-derivative-definition}
    \frac{\partial f}{\partial t}=\lim_{h\rightarrow0}\frac{f(t+h)-f(t)}{h}.
\end{equation}
To discretise Eq.~\eqref{eqn:Coupling-derivative-definition} we replace the infinitesimal step $h$ with the appropriate fixed step $\Delta x$, $\Delta y$ or $\Delta t$. There are a variety of ways to do this, each corresponding to a different finite difference scheme~\cite{levequeFiniteDifferenceMethods2007}. For example, directly substituting $h\mapsto\Delta t$ into Eq.~\eqref{eqn:Coupling-derivative-definition} produces the first order forward difference scheme:
\begin{equation}
    \frac{\partial f}{\partial t}=\frac{f(t+\Delta t)-f(t)}{\Delta t}+\mathcal{O}(\Delta t).
\end{equation}
In our code we use this scheme to represent the single time derivative in the damping term. For the other derivatives we use centred difference schemes, which generally have smaller truncation errors than forward and backward schemes~\cite{levequeFiniteDifferenceMethods2007}. We substitute the time derivative with a first order centred difference scheme:
\begin{equation}
    \frac{\partial^2u(t,x,y)}{\partial t^2}=\frac{1}{(\Delta t)^2}(u^{n+1}_{ij}-2u^n_{ij}+u^{n-1}_{ij})+\mathcal{O}((\Delta t)^2),
\end{equation}
and the spatial derivatives with a second order centred difference scheme~\cite{fornbergGenerationFiniteDifference1988}:
\begin{widetext}
\begin{align}
    \frac{\partial u(t,x,y)}{\partial x^2}&=\frac{1}{(\Delta x)^2}\left(\frac{-1}{12}u^n_{(i-2)j}+\frac{4}{3}u^n_{(i-1)j}-\frac{5}{2}u^n_{ij}+\frac{4}{3}u^n_{(i+1)j}-\frac{1}{12}u^n_{(i+2)j}\right) +\mathcal{O}((\Delta x)^4)\nonumber\\
    \frac{\partial u(t,x,y)}{\partial y^2}&=\frac{1}{(\Delta y)^2}\left(\frac{-1}{12}u^n_{i(j-2)}+\frac{4}{3}u^n_{i(j-1)}-\frac{5}{2}u^n_{ij}+\frac{4}{3}u^n_{i(j+1)}-\frac{1}{12}u^n_{i(j+2)}\right)+\mathcal{O}((\Delta y)^4).
    \label{eqn:Coupling-FDTD-second-order-centred-difference-scheme}
\end{align}

Substituting these schemes into Eq.~\eqref{eqn:Coupling-linear-operator-wave-equation} and rearranging gives a formula for iterating the solution to the next time step:

\begin{align}
    u^{n+1}_{ij}&=2u^n_{ij}-u^{n-1}_{ij}+\frac{(\Delta t)^2\sigma_{ij}}{12\rho_{ij}}\bigg\{\frac{1}{(\Delta x)^2}(16u^n_{(i+1)j}+16u^n_{(i-1)j}-u^n_{(i+2)j}-u^n_{(i-2)j}-30u^n_{ij}) \nonumber\\
    &+\frac{1}{(\Delta y)^2}(16u^n_{i(j+1)}+16u^n_{i(j-1)}-u^n_{i(j+2)}-u^n_{i(j-2)}-30u^n_{ij})\bigg\}+\gamma_{ij}(\Delta t)(u^n_{ij}-u^{n-1}_{ij}).
    \label{eqn:Coupling-FDTD-increment-formula}
\end{align}
\end{widetext}

For small enough fixed steps this expression accurately represents the action of the differential operator in Eq.~\eqref{eqn:Coupling-linear-wave-operator-definition}. The simulation iteratively solves Eq.~\eqref{eqn:Coupling-FDTD-increment-formula} until the desired number of time steps has been reached.

It should be noted that Eq.~\eqref{eqn:Coupling-FDTD-increment-formula} is not the only algorithm that can accurately time-evolve the system; other iteration equations derived from different choices of finite difference scheme can also be used.

We enforce reflectionless boundary conditions on the simulation by employing a effective perfectly-matched layer (PML)~\cite{berengerPerfectlyMatchedLayer1994}, highlighted in Fig.~\ref{fig:FDTD-decayrate-principle}(a). This is done by a gradual linear ramping up of the damping rate $\gamma$ over a distance of several wavelengths near the spatial borders. The effective PML layer emulates a spatially infinite medium by damping out all incident acoustic energy without back-reflections.

In addition to the ringdown simulations described above, we can also simulate driven systems by, at each timestep, adding an oscillatory term to the solution matrix before applying Eq.~\eqref{eqn:Coupling-FDTD-increment-formula}.

Time-domain simulations have the advantage over frequency-domain simulations in that the nonlinearity can be relatively straightforwardly added. For instance, to add the geometric nonlinearity from material elongation, the wave equation operator $\mathcal{L}$ can be replaced with a nonlinear version, $\mathcal{N}u=0$ where:
\begin{equation}
    \mathcal{N}(\cdot)=\mathcal{L}(\cdot)-\frac{Y}{2}(\nabla^2\cdot)\left(\left(\frac{\partial\,\cdot}{\partial x}\right)^2+\left(\frac{\partial\,\cdot}{\partial y}\right)^2\right).
    \label{eqn:Coupling-FDTD-wave-operator-nonlinear}
\end{equation}
Here $Y$ is the Young's modulus of the material. Equation~\eqref{eqn:Coupling-FDTD-wave-operator-nonlinear} is derived by adding an additional term to the tensile stress: $\sigma\mapsto\sigma+Y\varepsilon(x,y)$. Here $\varepsilon(x,y)$ is the longitudinal strain due to flexural motion, derived in Eq.~\eqref{eqn:Phononic-basics-strain-gradient-1D} in Section~\ref{sec:Phononic-basics-mechanical-nonlinearity}.

Equation~\eqref{eqn:Coupling-FDTD-wave-operator-nonlinear} can be converted into an iterative formula like Eq.~\eqref{eqn:Coupling-FDTD-increment-formula} by substituting the additional derivatives with well-chosen finite difference schemes.

Being able to straightforwardly include nonlinearities is a strength of time-domain simulations. This allows them to model intrinsically nonlinear behaviour like bistability~\cite{jinCascadingNanomechanicalResonator2023} and transient dynamics~\cite{jinEngineeringErrorCorrecting2024}. On the flip side, time-domain simulations are inefficient for modelling steady-state or long time scale behaviour, and run slower for larger simulation domains.

\subsection{Transfer Matrices}
\label{sec:Coupling-transfer-matrices}
For larger systems involving many changes in geometry, analysing the wave propagation using finite element or finite difference methods can be resource-intensive. In cases where only the steady-state solution is needed, we can greatly speed up the calculation by employing the transfer matrix method. The transfer matrix method was developed decades ago to study the propagation of optical waves through multilayered slabs of dielectric material~\cite{mackayTransferMatrixMethodElectromagnetics2020}, then later used for low frequency transmision lines, microwave waveguides~\cite{montgomeryPrinciplesMicrowaveCircuits1987} and acoustics~\cite{allardPropagationSoundPorous2009}. 

\begin{figure}[ht]
    \centering
    \includegraphics[width=0.99\linewidth]{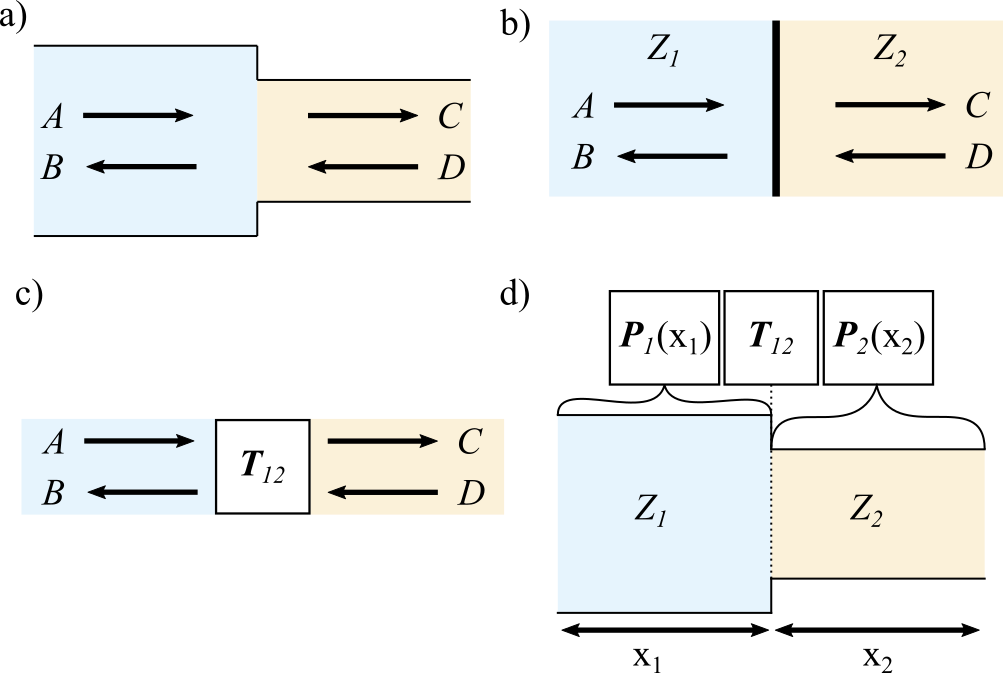}
    \caption[The transfer matrix representation of wave propagation]{The transfer matrix representation of wave propagation. (a) Diagram of a junction where the waveguide width suddenly changes. Arrows $A,B,C$ and $D$ represent incoming and and outgoing wave amplitudes. (b) Description of the same system using impedances $Z_1$ and $Z_2$ (c) Equivalent formulation in terms of a transfer matrix $\mathbf{T}_{12}$. (d) By combinging propagation and transfer matrices, one can represent the total effect of the two waveguides and the junction.}
    \label{fig:Coupling-transfer-matrix-diagram}
\end{figure}

The principle behind transfer matrices is illustrated in Fig.~\ref{fig:Coupling-transfer-matrix-diagram}, where we consider the common example of a sudden change in waveguide width (Fig.~\ref{fig:Coupling-transfer-matrix-diagram}(a)). At the junction some acoustic waves will be reflected and some will be transmitted, giving four amplitudes $A,B,C$ and $D$ to describe the situation (this assumes single-mode, single frequency operation). These amplitudes define the transfer matrix $\mathbf{T}$ corresponding to this junction:
\begin{equation}
    \label{eqn:Coupling-transfer-operation-definition}
    \begin{pmatrix}
        C\\D
    \end{pmatrix}=\mathbf{T}\begin{pmatrix}
        A\\B
    \end{pmatrix}.
\end{equation}
This is very similar to how scattering matrices in quantum mechanics are defined, except a scattering matrix $\mathbf{S}$ relates outgoing and incoming amplitudes~\cite{bachorGuideExperimentsQuantum2019}:
\begin{equation}
    \label{eqn:Coupling-scattering-matrix-definition}
    \begin{pmatrix}
        B\\C
    \end{pmatrix}=\mathbf{S}\begin{pmatrix}
        A\\D
    \end{pmatrix}.
\end{equation}

As depicted in Fig.~\ref{fig:Coupling-transfer-matrix-diagram}(b), we can describe the change in width as a change in acoustic impedance, $Z$. This is a one-dimensional treatment analogous to how transmission lines can be modelled~\cite{montgomeryPrinciplesMicrowaveCircuits1987}. Acoustic impedance $Z$ for sound waves propagating in one dimension, such as in membrane waveguides, can be written as~\cite{regtien9AcousticSensors2018,auldAcousticFieldsWaves1990} $Z=\rho v_\mathrm{ph}$, where $\rho$ is the volumetric density of the membrane and $v_\mathrm{ph}$ is the phase velocity of the waves. Substituting $v_\mathrm{ph}$ using the standard dispersion relationship (Eq.~\eqref{eqn:Phononic-basics-phase-velocity}), we can write the impedance as:
\begin{equation}
    \label{eqn:Coupling-impedance-calculation}
    Z(\Omega)=\rho\cdot\frac{\sqrt{\sigma/\rho}}{\sqrt{1-\left(\Omega_{c,n}/\Omega\right)^2}}.
\end{equation}
Here $\Omega$ is the frequency of the wave, and $\Omega_{c,n}$ is the cutoff frequency of the $n^\mathrm{th}$ waveguide mode (Eq.~\eqref{eqn:Phononic-basics-cutoff-frequency}). The impedance can be an imaginary number for drive frequencies below the cutoff frequency, in which case it indicates an evanescent field. 


Impedance is a useful quantity because it provides simple expressions for reflection and transmission coefficients. Consider a wave travelling from a medium with impedance $Z_1$ to $Z_2$. Assuming the junction is lossless, the junction boundary is perpendicular to the wave propagation direction, and the wave amplitude and transverse force are continuous at the boundary, the scattering matrix for this junction is~\cite{regtien9AcousticSensors2018}:
\begin{equation}
    \renewcommand\arraystretch{1.8}
    \mathbf{S}_{12}=\begin{pmatrix}
        \frac{Z_1-Z_2}{Z_1+Z_2}&\frac{2Z_1}{Z_1+Z_2}\\\frac{2Z_1}{Z_1+Z_2}&\frac{Z_1-Z_2}{Z_1+Z_2}
    \end{pmatrix}.
\end{equation}
By comparing Eq.~\eqref{eqn:Coupling-transfer-operation-definition} and Eq.~\eqref{eqn:Coupling-scattering-matrix-definition} it can be shown that the transfer matrix is:
\begin{equation}
    \label{eqn:transfer-matrix-impedance-definition}
    \mathbf{T}_{12}=\frac{1}{2Z_1}\begin{pmatrix}
        Z_1+Z_2&Z_1-Z_2\\Z_1-Z_2&Z_1+Z_2
    \end{pmatrix}.
\end{equation}

In addition to modelling waveguide junctions with transfer matrices, we can also model phase evolution in the waveguide with a propagation matrix $\mathbf{P}$:
\begin{equation}
    \mathbf{P}=\begin{pmatrix}
        e^{i kx}&0\\0&e^{-i kx}
    \end{pmatrix}.
\end{equation}
Here $x$ is the distance travelled and $k$ is the wavenumber, which depends on the waveguide width and frequency. In the case of evanescent propagation, $k$ is imaginary and the amplitude exponentially decays.

Figure~\ref{fig:Coupling-transfer-matrix-diagram}(d) illustrates how we can model a complete system involving multiple junctions and waveguides by using transfer and propagation matrices. Consider a chain of $N$ junctions in a row, describing propagation from a zeroth medium to an $N^\mathrm{th}$ medium. We can describe the wave amplitudes in the $i^\mathrm{th}$ medium as:
\begin{equation}
    \mathbf{U}_i=\begin{pmatrix}
        U_{if}\\U_{ib}
    \end{pmatrix},
\end{equation}
where $f$ and $b$ correspond to forward and backward propagating waves respectively. The total effect of the chain can be written as:
\begin{equation}
    \label{eqn:transfer-matrices-concatenated}
    \mathbf{U}_\mathrm{N}=\mathbf{T}_{(N-1)N}\mathbf{P}_{N-1}\mathbf{T}_{(N-2)(N-1)}\ldots\mathbf{P}_1\mathbf{T}_{01}\mathbf{U}_{0}. 
\end{equation}
By multiplying the matrices together one can evaluate the aggregate reflection and transmission through the system.

As an example, by setting $U_{Nb}=0$ and then multiplying both sides of Eq.~\eqref{eqn:transfer-matrices-concatenated} by the appropriate inverse matrices to solve for $\mathbf{U}_\mathrm{0}$, one can find the fraction of acoustic power that is reflected, $R$, as:
\begin{equation}
\label{eqn:Coupling-transfer-matrix-total-reflection}
    R=\frac{|U_{0b}|^2}{|U_{0f}|^2},
\end{equation}
and fraction that is transmitted total transmission coefficient, $T$, as~\cite{pfeiferAchievementsPerspectivesOptical2022}:
\begin{equation}
\label{eqn:Coupling-transfer-matrix-total-transmission}
    T=1-R.
\end{equation}

Reflection and transmission coefficients for a system of waveguides can be used find resonances in the system. Assuming no material losses are modelled, resonances will appear as peaks in the transmission spectrum (or equivalently, dips in the reflection spectrum). Resonance frequencies and linewidths factors can be measured directly from the spectrum. 

In situations where the width of the waveguide suddenly changes, the impedance calculation in Eq.~\eqref{eqn:Coupling-impedance-calculation} does not take into account significant near-field effects. A sudden reduction in width produces a junction effect analogous to the thin inductive aperture effect in microwave waveguides~\cite{montgomeryPrinciplesMicrowaveCircuits1987}. Fortunately, because this situation is physically analogous, we can correct for near-field effects using expressions developed in the microwave literature. Translated to membrane phononics, if the width suddenly and symmetrically changes at the boundary of waveguide 1 and waveguide 2, the corrected impedance $Z_2$ taking into account near-field effects is~\cite{montgomeryPrinciplesMicrowaveCircuits1987}:
\begin{equation}
    \label{eqn:Coupling-impedance-montgomery-correction}
    Z_2=\frac{Z_1}{n^2}\frac{i|\lambda_2|W_2}{\lambda_1W_1}, 
\end{equation}
where
\begin{equation}
    \label{eqn:Coupling-impedance-montgomery-correction-n2}
    n^2=\frac{4}{\pi}\frac{\cos\left(\frac{\pi}{2}\frac{W_2}{W_1}\right)}{1-\left(\frac{W_2}{W_1}\right)}.
\end{equation}
We use this correction in practice by calculating the impedance of the wider waveguide using Eq.~\eqref{eqn:Coupling-impedance-calculation}, and then the impedance of the narrower waveguide using Eq.~\eqref{eqn:Coupling-impedance-montgomery-correction}. In Eq.~\eqref{eqn:Coupling-impedance-montgomery-correction-n2} the variables $\lambda_1$ and $\lambda_2$ are the wavelengths of propagating waves in each waveguide, and $W_1$ and $W_2$ are the widths of each waveguide. We substitute for the wavelengths using our analytic dispersion relationship from Section~\ref{sec:Phononic-basics-Dispersion}.

\begin{figure}
    \centering
    \includegraphics[width=0.99\linewidth]{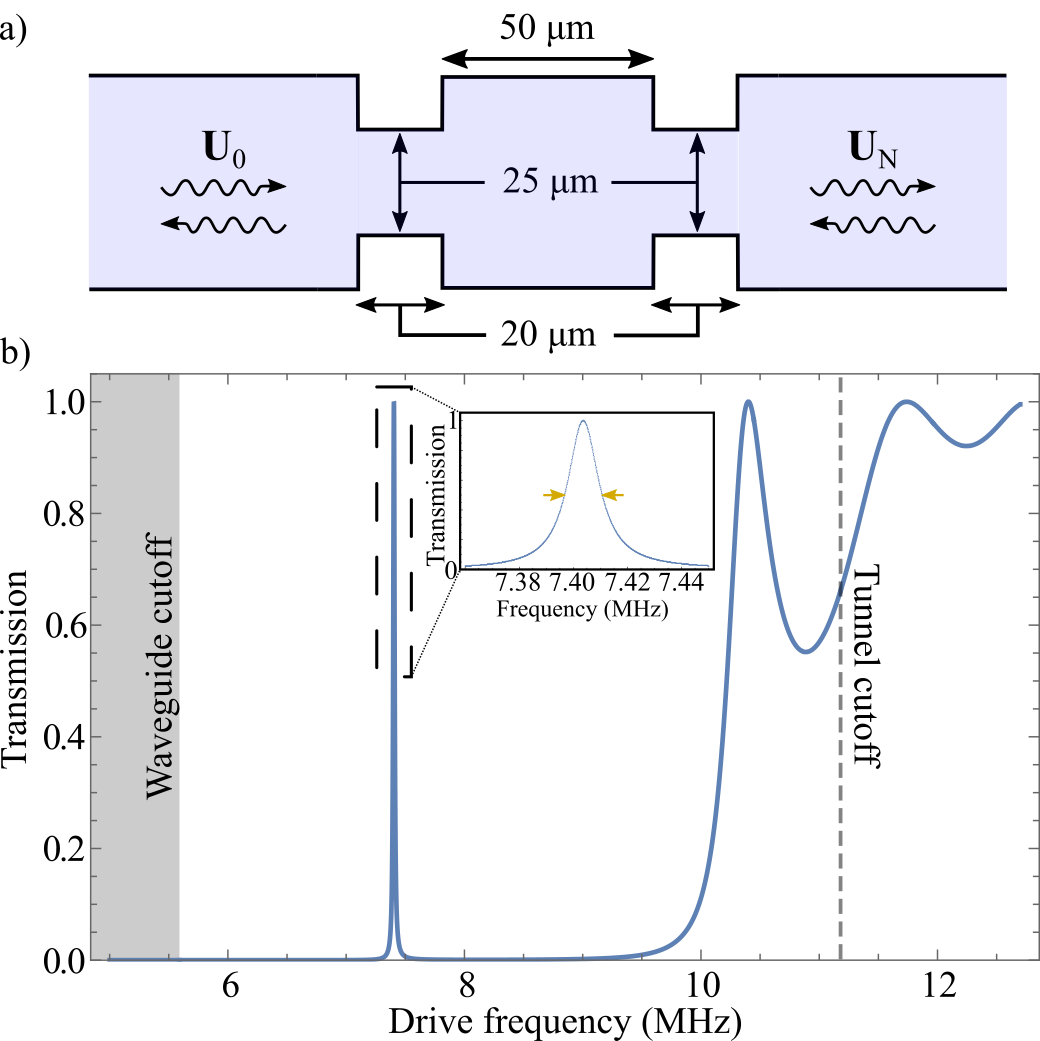}
    \caption[Transmission spectrum through a two-port resonator calculated using the transfer matrix method]{Wave propagation through a two-port resonator simulated using the transfer matrix method (Eq.~\eqref{eqn:Coupling-transfer-matrix-total-transmission}). (a) Diagram of the simulated resonator, coupled to input and output waveguides by evanescent tunnels. Dimensions: waveguide width = $50\,\mathrm{\upmu m}$, resonator is a square with side length = $50\,\mathrm{\upmu m}$, tunnel width = $25\,\mathrm{\upmu m}$, tunnel length = $20\,\mathrm{\upmu m}$. The membrane has volumetric density $\rho=3200\,\mathrm{kg\cdot m^{-3}}$ and tensile stress $\sigma=1\,\mathrm{GPa}$, corresponding to highly stressed silicon nitride. (b) Transmission spectrum, showing a resonance at approximately $7.4\,\mathrm{MHz}$. Inset: close-up of the resonance. Yellow arrows indicate the full width half maximum of the peak, approximately $14\,\mathrm{kHz}$, corresponding to a quality factor of $Q\simeq530$, determined by the coupling losses from the resonator to both waveguides. The left grey box and right dashed line respectively highlight the cutoff frequency $\Omega_{c,1}$ of the waveguides/resonator and the tunnel.}
    \label{fig:Coupling-transfer-matrix-resonance-example}
\end{figure}

To demonstrate using the transfer matrix method, in Figure~\ref{fig:Coupling-transfer-matrix-resonance-example} we model acoustic transmission through a two-port resonator. The transmission spectrum is done by iteratively calculating Eq.~\eqref{eqn:transfer-matrices-concatenated} over a range of frequencies, with the mode number $n=1$ corresponding to single-mode operation. We use the impedance correction from Eq.~\eqref{eqn:Coupling-impedance-montgomery-correction}. This simulation does not include complex terms in the propagation matrices, so the quality factor of the resonator is purely determined by the coupling rate through the evanescent tunnels. One can see that a transmission peak occurs at the resonance frequency of the phononic resonator, analogous to the transmission spectrum of an optical Fabry-P\'erot cavity~\cite{hungerFiberFabryPerot2010}.

\subsection{Comparison of numerical methods}
\label{sec:Coupling-comparison-of-numerical-methods}

We have now introduced four techniques that can be used to model evanescent coupling in membrane phononics: FEM simulations partnered with overlap integrals (Section~\ref{sec:Coupling-resonator-waveguide-analytic-coupling}), complex eigenfrequencies directly from FEM simulation (Section~\ref{sec:Coupling-complex-eigenfrequency}), FDTD simulations (Section~\ref{sec:Coupling-FDTD}) and the transfer matrix method (Section~\ref{sec:Coupling-transfer-matrices}).

\begin{figure*}[ht]
    \centering
    \includegraphics[width=0.99\linewidth]{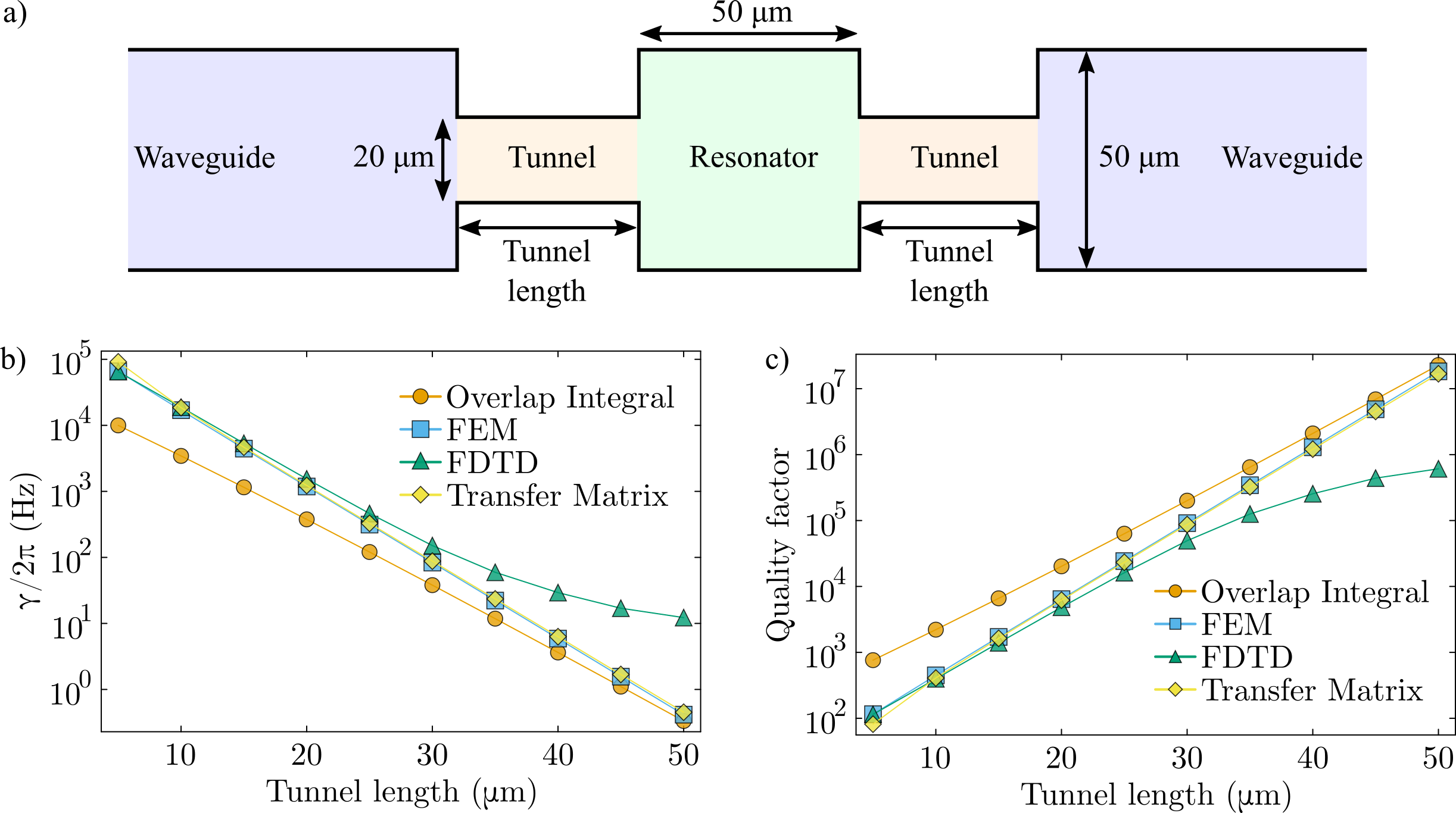}
    \caption[Comparison of model predictions of the coupling rate of a realistic two-port membrane resonator]{Comparison of model predictions of the coupling rate of a realistic two-port membrane resonator. (a) Illustration of the simulated device: a 50~{\textmu}m square silicon nitride resonator (green) coupled to two 50~{\textmu}m wide waveguides (blue) via 20~{\textmu}m wide evanescent tunnels (orange) of variable length. The next two subplots show the calculated total decay rate $\gamma_\mathrm{tot}$ (b) and quality factor (c) of the resonator. Orange: results from calculating the overlap integral (Eq.~\eqref{eqn:Coupling-FEM-overlap-integration-1}) using numerically calculated modeshapes (Section~\ref{sec:Coupling-resonator-waveguide-analytic-coupling}). Blue: results from using finite element modelling to find a complex eigenfrequency (Section~\ref{sec:Coupling-complex-eigenfrequency}). Green: results obtained using a custom finite difference time domain simulation (Section~\ref{sec:Coupling-FDTD}). Yellow: results from a transfer matrix model with the impedance correction (Eq.~\eqref{eqn:Coupling-impedance-montgomery-correction}) to account for varying lateral dimensions along the propagation direction. Note the yellow points are almost directly on top of the blue points. The coupling is calculated at the fundamental eigenfrequency of the resonator at approximately $7.62\,\mathrm{MHz}.$ Material parameters are stress $\sigma=1\,\mathrm{GPa}$ and density $\rho=3200\,\mathrm{kg\cdot m^{-3}}$.}
    \label{fig:Coupling-yQ_diffCalcMethods}
\end{figure*}

Figure~\ref{fig:Coupling-yQ_diffCalcMethods} compares these different methods for the situation of a 2-port square membrane connected by evanescent tunnels to straight waveguides on either side. This setup is the phononic version of the widely used optical Fabry-P\'erot cavity~\cite{romeroAcousticallyDrivenSinglefrequency2024}. In Section~\ref{sec:Coupling-examples-of-engineered-coupling} we also will show that variations on this layout correspond to a range of useful devices.

There is excellent qualitative agreement between the four calculation methods. As expected we see exponential attenuation of the wave amplitude within the tunnel coupler, where the drive frequency is below the cutoff frequency and the wavenumber is therefore complex~\cite{mauranyapinTunnelingTransverseAcoustic2021}. The exponential sensitivity of the attenuation on the tunnel length means that a wide range of quality factors can be engineered with relatively small changes in geometry.

Despite its simplifications, the overlap integral method that we derived from coupled-mode theory and analytic expressions converges to the numerical methods for large tunnel lengths. The method underestimates the coupling at short tunnel lengths, where the small-coupling, perturbative approximation breaks down. At these short tunnel lengths the resonator and waveguide are sufficiently strongly coupled that their eigenmodes begin to differ significantly from the case when they are infinitely separated. The perturbative approach of CMT is therefore naturally more appropriate at long tunnel lengths where there is weaker coupling.

The FDTD simulation achieves excellent agreement with the other numerical methods, particularly for shorter tunnels where the coupling is stronger. Performance in this regime demonstrates the flexibility of the FDTD method: because it directly solves the wave equation, it does not rely on any particular assumptions (e.g. small coupling) to hold. Indeed, the FDTD method is the best at simulating nonlinearity, strong coupling and transient behaviour over small timescales~\cite{romeroAcousticallyDrivenSinglefrequency2024} (see Section~\ref{sec:Coupling-FDTD} above). However, in the regime of large tunnel lengths and weak coupling we also see the key flaw of FDTD methods, which is a lower limit on the observable decay rate. This limit is caused by the finite simulation finish time, which was kept constant for each of the simulations in Fig.~\ref{fig:Coupling-yQ_diffCalcMethods}.

The fact that the transfer matrix method data plot a straight line of the expected slope is not surprising, as the propagation matrices in the tunnel regime are just exponential decay operators and remain perfectly accurate with increasing distance (to the point of floating point error). What is more surprising is the nearly-perfect agreement with the finite element model, which indicates the predicted impedance mismatch is a very good estimate of the true value. This agreement with the other numerical techniques can be attributed to the near-field correction from microwave theory, Eq.~\eqref{eqn:Coupling-impedance-montgomery-correction}. Without this correction the transfer matrix method remains qualitatively accurate but presents a systematic offset, overestimating the coupling rate by $75\%$. 

On a final note, it is worth mentioning that the transfer matrix method is by far the fastest simulation method. Because it does not rely on a discretised grid it can scale to larger geometries with negligible increases in runtime. Having benchmarked the transfer matrix method for the case of a single resonator, one can imagine using it to simulate larger devices, for example a long chain of impedance-matched resonators. This would be the phononic version of a narrowband microwave filter~\cite{montgomeryPrinciplesMicrowaveCircuits1987}.

\section{Examples of engineered coupling}
\label{sec:Coupling-examples-of-engineered-coupling}
In addition to coupling rates there are other, intuitive insights that we can take from the CMT, such as orthogonality and symmetry. Equipped with this intuition and numerical techniques we can rapidly prototype more sophisticated signal processing phononic devices. In this final section we show examples of this, using the example of power splitters and mode multiplexers incorporated into a single, compact resonator footprint.

\subsection{Resonant arbitrary ratio power splitter}
\label{sec:Coupling-arbitrary-ratio-power-splitter}
Here we detail the design process for an acoustic power splitter with tailorable operating bandwidth and splitting ratio.

\begin{figure}[ht]
    \centering
    \includegraphics[width=0.99\linewidth]{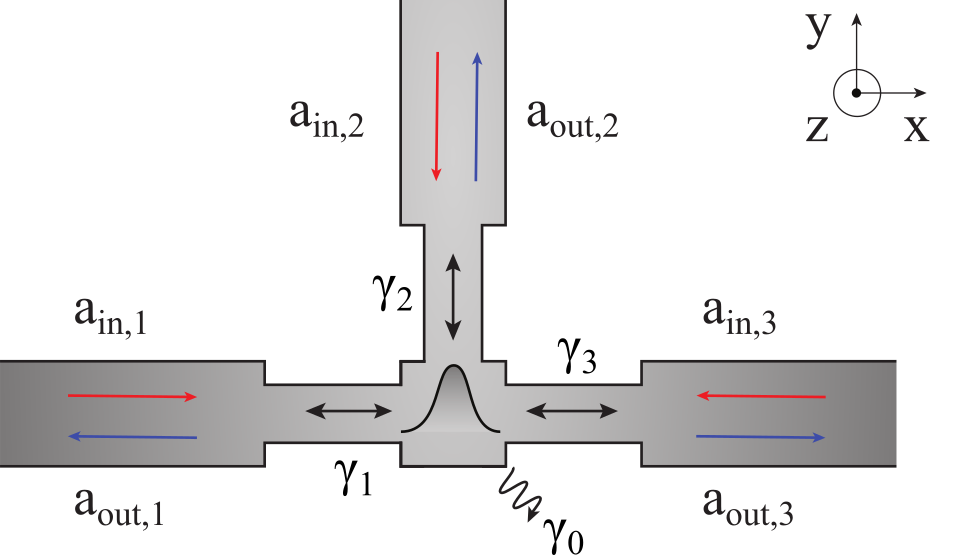}
    \caption[Schematic illustration of an acoustic power splitter]{Schematic illustration of an acoustic power splitter, outlining the relevant acoustic input ($a_{\mathrm{in},i}$) and output ($a_{\mathrm{out},i})$ fields and decay rates ($\gamma_i$). $\gamma_0$ represents internal loss such as from internal material friction or clamping losses.}
    \label{fig:Coupling-acoustic-power-splitter}
\end{figure}

This power splitter is formed by a single square resonator evanescently coupled through tunnel junctions to three waveguides, as illustrated in Fig.~\ref{fig:Coupling-acoustic-power-splitter}. The resonator size and waveguide width are chosen such that the operating frequency matches the fundamental eigenfrequency of the resonator and is in the single-mode regime of the waveguides (Section~\ref{sec:Phononic-basics-Dispersion}). 

We can achieve an arbitrary splitting ratio by precisely choosing the coupling rates. Denote the input waveguide coupling rate as $\gamma_1$ and the two output waveguide coupling rates as $\gamma_2$ and $\gamma_3$. A reflectionless device can be achieved by choosing $\gamma_1=\gamma_2+\gamma_3$, corresponding to impedance matching (see Section~\ref{sec:Coupling-input-output-impedance-matching}). In that case, the total loss or bandwidth of the resonator is $2\gamma_1$ (ignoring internal losses), so tuning $\gamma_1$ controls the bandwidth. Finally, the power splitting ratio can be set by the ratio of $\gamma_2$ to $\gamma_3$. Indeed, let us denote denote the input power as $P_1$, the power going into waveguide 2 as $P_2$, and the power going into waveguide 3 as $P_3$. At steady state the ratio of $P_2$ and $P_3$ is related to the coupling rates as: $P_2/P_3=\gamma_2/\gamma_3$. Therefore to achieve a power splitting ratio of $P_2=xP_1$ (and $P_3=(1-x)P_2$), where $x\in[0,1]$, the coupling rates must be set to $\gamma_2/\gamma_3=P_2/P_3=x/(1-x)$.

For example, consider desiring a 50:50 impedance-matched power splitter with a frequency bandwidth of $10\,\mathrm{kHz}$. Let the coupling rates $\gamma_1$, $\gamma_2$ and $\gamma_3$ represent the input waveguide and two output waveguide couplings. To achieve the desired bandwidth we should have $\gamma_1+\gamma_2+\gamma_3=2\pi\times10\,\mathrm{kHz}$. For impedance matching we also want $\gamma_1=\gamma_2+\gamma_3$, and for 50:50 power splitting we want $\gamma_2=\gamma_3$. Solving these equations gives that $\gamma_1=2\gamma_2=2\gamma_3=2\pi\times5\,\mathrm{kHz}$. Reading off of Fig.~\ref{fig:Coupling-yQ_diffCalcMethods} using the transfer matrix or FEM method data, this suggests that the input tunnel should be approximately $15\,\mathrm{\mu m}$ long, and the output tunnels should be $18\,\mathrm{\mu m}$ long. If we desire a different power splitting ratio of 90:10 we would set $\gamma_2=9\gamma_3$. This corresponds to respective tunnel lengths of $L_2=15\,\mathrm{\upmu m}$ and $L_3=25\,\mathrm{\upmu m}$. Figure~\ref{fig:Coupling-yQ_diffCalcMethods} only needs to be calculated once to be used as a reference for these different designs.

\begin{figure}[ht]
    \centering
    \includegraphics[width=0.95\linewidth]{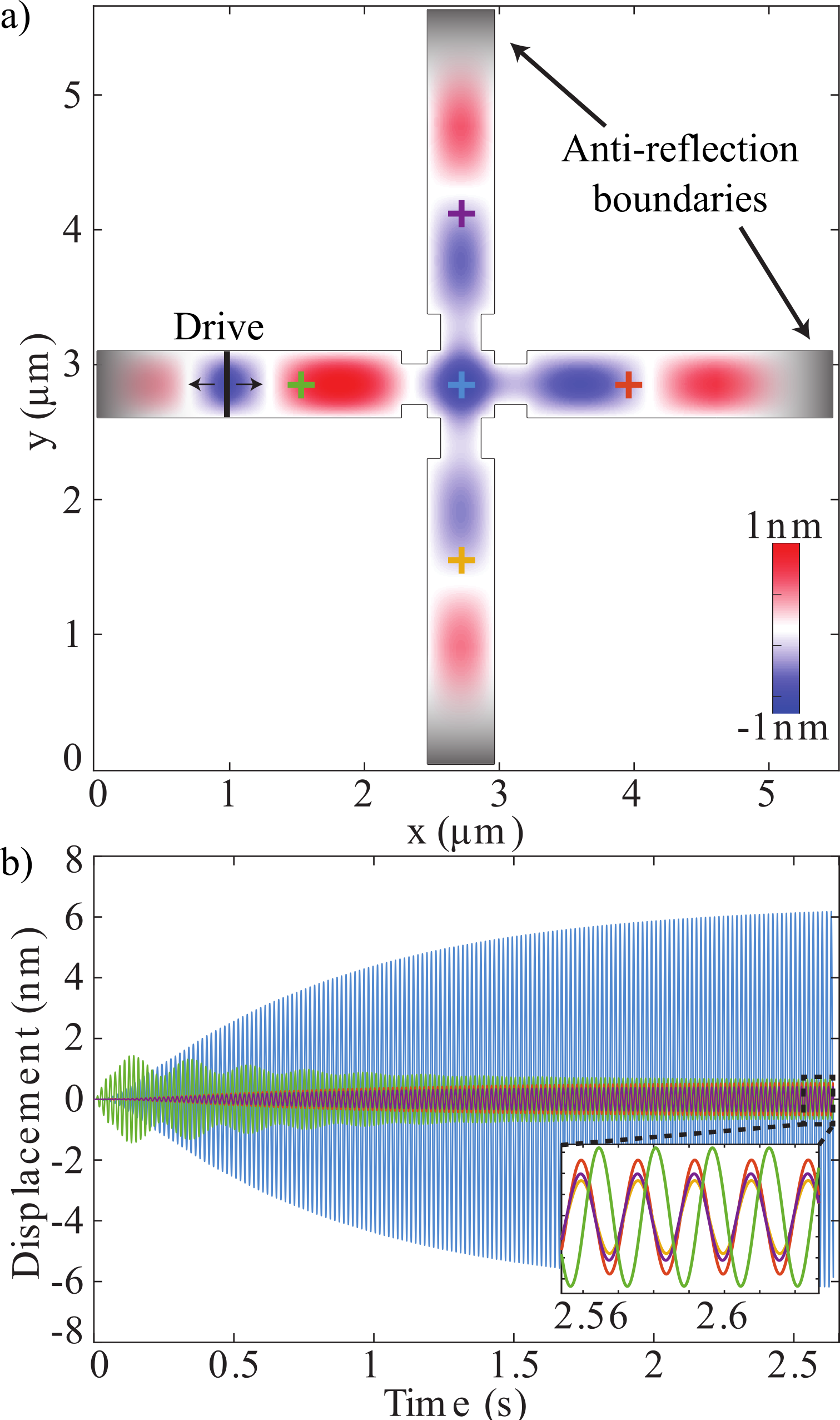}
    \caption{(a) Schematic of the FDTD simulation conditions for an impedance-matched 1:3 acoustic power splitter with different tunnel length/coupling rates for each output. The position of the drive (black line) is marked along with absorbing effective PML domains (grey shading). Coloured crosses denote where the amplitude is measured: in the center of the input waveguide (green), in the resonator (blue), and in the output waveguides (purple, orange, and red) (b) Measured amplitudes. The inset highlights the steady-state performance showing no back reflections and power splitting into each output waveguide depending on the respective coupling rate. Parameters: waveguide width=50 $\mu$m, tunnel width 30 $\mu$m, resonator dimensions $50\times50$ $\mu$m$^2$, Tunnel lengths: input $19\,\upmu$m input; right $24\,\upmu$m; top $27\,\upmu$m; bottom $30\,\upmu$m. The drive frequency is $\Omega/2\pi=6.23$ MHz.}
    \label{fig:PowerSplitter-FDTD-demo}
\end{figure}

Time-domain numerical simulations validate our design for the power splitter. For example, Fig.~\ref{fig:PowerSplitter-FDTD-demo} shows an FDTD simulation of a three-port impedance-matched power splitter. As the simulation is initialized with waveguides and resonator at rest, the resonator must first ring-up before reaching steady-state. During this transitory period, which takes on the order of $Q$ oscillations, a gradually dimishing fraction of the incident energy is backreflected until, in the steady state, acoustic energy incident in the input port flows with no back-reflections into the output ports. This is seen in Fig.~\ref{fig:PowerSplitter-FDTD-demo}, where the input amplitude (green trace) is initially beating as the resonator rings up, but then relaxes into a constant oscillation consistent with zero back reflection. 

As seen in Fig.~\ref{fig:Coupling-yQ_diffCalcMethods}, the coupling rate is exponentially sensitive to the tunnel length, so significantly different power splitter behaviours can be engineered without compromising the device footprint. Coupling rates also strongly depend on the tunnel width, providing an additional independent means of control. Tunnels with precisely different lengths or widths can be realised by using a small mesh size and the top-down release fabrication method~\cite{mauranyapinTunnelingTransverseAcoustic2021,hirschDirectionalEmissionOnchip2024}.

\subsection{Benefits and trade-offs of resonant devices}
\label{sec:Coupling-resonant-versus-non-resonant-devices}

Having seen an example of a resonator-based power splitter, it is worth discussing the benefits and disadvantages of resonant versus non-resonant power splitters and mode converters.

Non-resonant devices include waveguide couplers like we explored in Section~\ref{sec:Coupling-two-waveguides-example}. These have a high bandwidth and no transient behaviour, at the cost of a large physical footprint. The large footprint also makes it hard to include several ports, which one can see by comparison with on-chip photonics~\cite{lipsonSwitchingLightSilicon2005,sternOnchipModedivisionMultiplexing2015,tangTenPortUnitaryOptical2021}.

Resonant devices feature multiple inputs and outputs centrally coupled to a single resonator. This design is far more compact, at the cost of requiring a `ring-up' period before reaching nominal performance in the steady state. This ring up period will scale with $\gamma^{-1}=Q/\Omega$, where $\gamma$ is the total out-coupling rate of the central resonator, $Q$ is its quality factor, and $\Omega$ is the operating frequency. Larger coupling rates to input and output waveguides will produce a greater device bandwidth and shorter transient period. Conversely, bandwidth selectivity and amplitude enhancement within the resonator (if desirable) can be achieved by using lower coupling rates.

\subsection{Resonant mode converting power splitter}
\label{sec:Coupling-mode-converting-power-splitter}

\begin{figure*}[ht]
    \centering
    \includegraphics[width=0.99\linewidth]{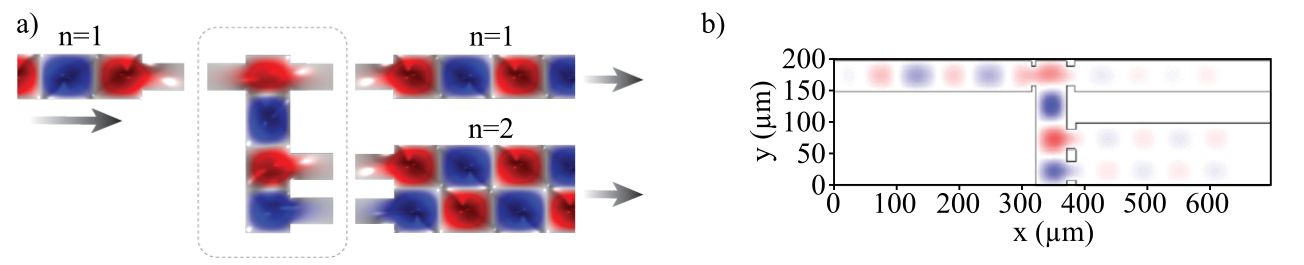}
    \caption[Phononic mode converting power splitter]{Phononic mode converting power splitter (a) Illustration of the mode converting powersplitter. A single mode waveguide provides input to a resonant cavity. One third of the incoming power exits on another single mode waveguide and the other two thirds as the second mode in a multimode waveguide. The waveguides are coupled to the resonator by reflectionless evanescent tunnels~\cite{mauranyapinTunnelingTransverseAcoustic2021} (b) Still frame from FDTD simulation. Parameters: Single mode waveguide has width 50~{\textmu}m, multimode waveguide has width 100~{\textmu}m. Tunnels have width 30~{\textmu}m; the left tunnel has length $6\,\mathrm{\upmu m}$ and the right tunnels have length $14\,\mathrm{\upmu m}$. The resonator has dimensions $50\times200$~{\textmu}m$^2$.}
    \label{fig:Coupling-powersplitter-modeconverter}
\end{figure*}

As a further example we design a $\frac{1}{3}:\frac{2}{3}$ power splitter that also performs mode conversion~\cite{zhuUnidirectionalExtraordinarySound2020}, as illustrated in Fig.~\ref{fig:Coupling-powersplitter-modeconverter}. This example three-port device accepts a single mode input and, in the steady state, splits the acoustic power between the fundamental mode of a single mode output waveguide and the second mode of a wider multimode waveguide, with no back-reflections. The $\frac{1}{3}:\frac{2}{3}$ ratio is produced when the three tunnel couplers have equal dimensions. It is important to note that back reflections will occur in the ring-up period before steady state is achieved. Lower quality factor devices with higher coupling rates will ring-up faster, as the cost of reduced frequency selectivity.

Only the second mode is excited in the wider waveguide because of symmetry constraints. The resonator is sized to have four antinodes at the operation frequency, and the tunnel junctions to the multimode waveguide are aligned to the antinodes such that they couple to opposite amplitudes of motion. Therefore by Eq.~\eqref{eqn:Coupling-FEM-overlap-integration-1} there is no coupling between the fundamental mode of the multimode waveguide and the resonator mode. At the same time there is nonzero coupling to the second order waveguide mode, so by sizing the multimode waveguide such that the third mode cutoff frequency is above the operating frequency, one can guarantee that only second mode travelling waves will be produced. Use of symmetry is essential to the device performance: our simulations show that a single tunnel junction offset from the centre of the multimode waveguide predominantly excites the second mode, but also a substantial amount of the fundamental mode.

A similar approach to Section~\ref{sec:Coupling-comparison-of-numerical-methods} is used to choose the appropriate tunnel lengths. We first compute the spatial profiles of the $n=4$ eigenmode of the resonator, the fundamental mode of the single mode waveguides, and the $n=1$ and $n=2$ modes of the multimode waveguide. Three functions of the decay rate as a function of tunnel length are then calculated for each of these coupling configurations (not shown here). The appropriate decay rates are determined by choosing the resonator bandwidth and enforcing the impedance matching condition $\gamma_\mathrm{in}=\gamma_\mathrm{out,1}+\gamma_\mathrm{out,2}.$ These predictions are again validated through acoustic FDTD simulations as shown in Fig.~\ref{fig:Coupling-powersplitter-modeconverter}(b).

\subsection{Phononic mode division multiplexer}
\label{sec:Coupling-mode-division-multiplexer}

\begin{figure*}[ht]
    \centering
    \includegraphics[width=0.95\linewidth]{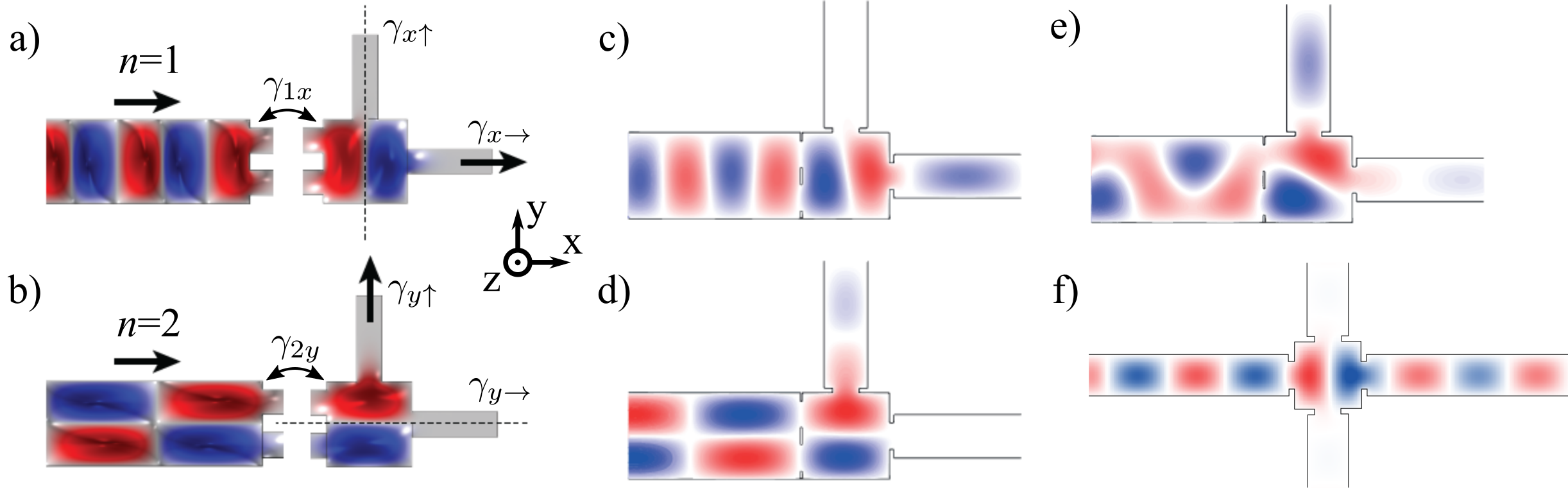}
    \caption[Phononic mode division multiplexer]{Phononic mode division multiplexer. (a) Concept: the $n=1$ mode of the multimode waveguide drives the acoustic cavity mode with mirror symmetry along the $x$ axis, and therefore exits into the right single mode output waveguide because $\gamma_{x\rightarrow}\gg\gamma_{x\uparrow}$. (b) Conversely the $n=2$ waveguide mode excites the frequency-degenerate cavity mode with mirror symmetry along the $y$ axis, and therefore exits to the top single mode output waveguide because $\gamma_{y\uparrow}\gg\gamma_{y\rightarrow}$. (c) FDTD simulation of the $n=1$ mode coupling to the $x$ resonant mode and out-coupling to the right waveguide. (d) FDTD simulation of the $n=2$ mode coupling to the $y$ resonant mode and out-coupling to the top waveguide. (e) FDTD simulation of a linear combination of $n=1$ and $n=2$ eigenmodes being demultiplexed. Parameters: Single mode waveguide has width 50~{\textmu}m, multimode waveguide has width 100~{\textmu}m. The resonator is a square with side length $100\,\mathrm{\upmu m}$. Tunnels have width 30~{\textmu}m; the short tunnels have length $2.5\,\mathrm{\upmu m}$ and the long tunnels have length $5\,\mathrm{\upmu m}$. (f) FDTD simulation of a four-way acoustic waveguide junction with minimal crosstalk. Parameters: central square resonator has side length 50~{\textmu}m, the tunnels have width 20~{\textmu}m and length 5~{\textmu}m, and the waveguides have width 30~{\textmu}m.
    } 
    \label{fig:Coupling-mode-division-multiplexer}
\end{figure*}

As a final example we consider the design of an acoustic mode division multiplexer. Mode division multiplexing is a technique to increase the bandwidth by encoding information in the different spatial modes of a multimode communication channel~\cite{zhouOpticalWavesWaveguides2024}. This requires the ability to combine multiple (preferentially single-mode) inputs, encode these into different spatial modes of a multimode waveguide, and finally separate these again into single-mode outputs.

Figure~\ref{fig:Coupling-mode-division-multiplexer}(a-b) explains the concept for a two-mode demultiplexer. (Operated in reverse, the same device functions as a mode multiplexer.) As with the previous example, symmetry is a key technique exploited here to perform the spatial filtering. The square-shaped resonator possesses two quasi-frequency degenerate acoustic resonances, one each for $x$ and $y$ mirror symmetry. Denoting them as the $x$ and $y$ modes, we can write down coupling rates $\gamma$ between each mode and the input and output modes. There are two modes in the input waveguide, $n=1$ and $n=2$, producing input coupling rates $\gamma_{1x},\,\gamma_{1y},\,\gamma_{2x},$ and $\gamma_{2y}$, and output coupling rates $\gamma_{x\uparrow},\gamma_{y\uparrow},\gamma_{x\rightarrow}$ and $\gamma_{y\rightarrow}$. Because the $n=1$ mode in the input waveguide has symmetry in the $x$ axis and antisymmetry in the $y$ axis, $\gamma_{1x}\gg\gamma_{1y}$ and it overwhelmingly couples to the $x$ resonator mode. For similar reasons the $n=2$ input mode overwhelmingly couples to the $y$ resonator mode. Another set of symmetry arguments imply that the $x$ resonator mode couples to the right output single mode waveguide ($\gamma_{x\rightarrow}\gg\gamma_{x\uparrow}$) and the $y$ resonator mode couples to the top output single mode waveguide ($\gamma_{y\uparrow}\gg\gamma_{u\rightarrow}$).

Proper operation of this device also requires that $\gamma_{1x}=\gamma_{x\rightarrow}=\gamma_{2y}=\gamma_{y\uparrow}$ for impedance matching and bandwidth matching. This can be achieved with suitable choice of tunnel barrier lengths.

The accuracy of these predictions is again validated through FDTD simulations, as shown in Fig.~\ref{fig:Coupling-mode-division-multiplexer}(c-e).

A variant of this design with four impedance-matched tunnel barriers coupled to single mode waveguides would function as a cross-talk free acoustic waveguide junction (Fig.~\ref{fig:Coupling-mode-division-multiplexer}(f)), analogous to the kind demonstrated in the photonic realm~\cite{johnsonEliminationCrossTalk1998}. To maintain a single frequency of operation, care would need be taken in the fabrication process to ensure the $x$ and $y$ modes are frequency degenerate. Specifically, any difference in the mode frequencies should be less than the linewidths of the modes---for a resonator with low intrinsic dissipation, these linewidths will be primarily determined by the waveguide coupling rates. Shorter and wider tunnel couplers will produce broader linewidths, meaning greater fabrication tolerances and higher operating bandwidth. The downside will be greater crosstalk due to direct coupling between the waveguides (against the symmetry restrictions), due to overlap of the evanescent fields of the waveguide modes~\cite{johnsonEliminationCrossTalk1998}. 

\section{Conclusion}

Suspended high tensile-stress membranes are a powerful platform for realising integrated phononic circuits. We have seen that the significant mismatch between suspended and non-suspended material produces very low radiative losses and allows single-mode waveguiding. Additionally, electrostatic actuation and geometric elongation provide two complimentary methods of eliciting Duffing nonlinearity. Perhaps the most important feature we have seen---and what separates suspended membranes apart from other phononic circuitry platforms---is the ability to evanescently couple heterogeneous membrane geometries. This ability, combined with the use of subwavelength release holes~\cite{mauranyapinTunnelingTransverseAcoustic2021}, allows a wide range of membrane geometries to be realised and coupled together~\cite{jinEngineeringErrorCorrecting2024,hirschDirectionalEmissionOnchip2024,romeroAcousticallyDrivenSinglefrequency2024} over a compact footprint.

By adapting coupled mode theory to phononics we have provided a comprehensive theoretical framework with which to understand and design evanescent couplers for membrane phononic circuits. Complimenting the analytic theory, we showed several practical ways to numerically model the coupling of membrane devices, and demonstrated their use by designing resonant signal processing devices such as tunable power splitters and mode (de)multiplexers.

In the future we anticipate far more circuit designs could be imagined and realised using the theory presented in this tutorial. For instance, one direction of future work may be creating networks of coupled resonators, using evanescent couplers to achieve precisely tailored coupling strengths, and electrostatic actuation or geometric elongation to introduce Duffing nonlinearity. The ability to fabricate nearly arbitrary network topologies with precisely tunable coupling strengths suggests membrane phononics may be a potent platform for experimenting with systems of nonlinear oscillators. Example applications include mechanical error correction~\cite{jinEngineeringErrorCorrecting2024,jinNanomechanicalErrorCorrection2025}, machine learning and neuromorphic computing~\cite{coulombeComputingNetworksNonlinear2017,erementchoukComputationalCapabilitiesNonlinear2021,kartalNanomechanicalSystemsReservoir2025} and basic research into complex networks~\cite{shimSynchronizedOscillationCoupled2007,fonComplexDynamicalNetworks2017,mathenyExoticStatesSimple2019,fonComplexDynamicalNetworks2017}.

\section*{Acknowledgements}
This research was primarily funded by the Australian Research Council and Lockheed Martin Corporation through the Australian Research Council Linkage Grant No. LP190101159. Support was also provided by the Australian Research Council Centre of Excellence for Engineered Quantum Systems (No. CE170100009). G.I.H. (No. DE210100848) and C.G.B. (No. FT240100405) acknowledge their Australian Research Council Fellowships.

The authors would like to thank Rachpon Kalra and George Brawley for helpful discussions in the development of this work.

\vspace{1cm}


\appendix

\section{Appendix: Parameters and material properties}
\label{sec:Appendix-params+properties}
Table~\ref{tab:App-material-properties} lists material properties that we use to describe high tensile-stress thin-film silicon nitride membranes.

\begin{table}[ht]
    \centering
    \begin{tabular}{c|c|c}
         Parameter & Typical value & Reference \\
         \hline
         Tensile stress, $\sigma$ & $1\,\mathrm{GPa}$ & \cite{gardeniersLPCVDSiliconrichSilicon1996,beliaevOpticalStructuralComposition2022}\\
         Volumetric density, $\rho$ & $3200\,\mathrm{kg\cdot m^{-3}}$& \cite{crcpressCRCHandbookChemistry2016}\\
         Poisson's ratio, $\nu$ & 0.23 & \cite{khanYoungsModulusSilicon2004} \\
         Young's modulus, $Y$ & $250\,\mathrm{GPa}$ & \cite{boeOnchipTestingLaboratory2009,khanYoungsModulusSilicon2004}
    \end{tabular}
    \caption{Typical material properties used in simulations, corresponding to high tensile-stress thin-film silicon nitride.}
    \label{tab:App-material-properties}
\end{table}

\bibliography{bib}

\end{document}